\documentclass[11pt,noupper,a4paper,twoside]{thesis}

\usepackage{graphics}
\usepackage{graphicx}
\usepackage{setspace}          % Used to give 1 1/2 spacing
\usepackage{dsfont}            % Used to get double line 1 and C
\usepackage[mathscr]{euscript} % Used to get script G
\usepackage{psfrag}            % Used to substitute LaTeX in figures

\chapterfont{\huge}

\newcommand{\NOT}{\textsc{not}}
\newcommand{\CNOT}{\textsc{cnot}}

\begin{document}
\ifx\href\undefined\else\hypersetup{linktocpage=true}\fi 

\pagestyle{empty}
\begin{center}

\vspace{50mm}

{\huge Towards Large-Scale Quantum Computation}

\vspace{15mm}

Austin Greig Fowler

\vspace{10mm}

Submitted in total fulfilment of the requirements \\
of the degree of Doctor of Philosophy

\vspace{10mm}

March 2005

\vspace{10mm}

School of Physics

The University of Melbourne

\end{center}

\begin{center}
{\huge Abstract}
\end{center}

\vspace{50mm}

This thesis deals with a series of quantum computer implementation
issues from the Kane $^{31}$P in $^{28}$Si architecture to Shor's
integer factoring algorithm and beyond. The discussion begins with
simulations of the adiabatic Kane \CNOT\ and readout gates,
followed by linear nearest neighbor implementations of 5-qubit
quantum error correction with and without fast measurement. A
linear nearest neighbor circuit implementing Shor's algorithm is
presented, then modified to remove the need for exponentially
small rotation gates. Finally, a method of constructing optimal
approximations of arbitrary single-qubit fault-tolerant gates is
described and applied to the specific case of the remaining
rotation gates required by Shor's algorithm.

\begin{center}
{\huge Declaration}
\end{center}

\vspace{50mm}

This is to certify that

\begin{enumerate}

\item the thesis comprises only my original work towards the PhD,

\item due acknowledgement has been made in the text to all other
material used,

\item the thesis is less than 100,000 words in length, exclusive
of tables, maps, bibliographies and appendices.

\end{enumerate}

\vspace{10mm}

\begin{center}
\rule{80mm}{0.25mm}
\end{center}

\begin{center}
{\huge Acknowledgements}
\end{center}

\vspace{50mm}

This thesis would never have been finished without the help of a
number of people. My parents, who have put up with me for more
years than I care to mention and have always been there when I
needed them. My supervisor, Lloyd Hollenberg, for the freedom he
has allowed me and his constant academic and professional support.
The DLU, and specifically Maun Suang Boey, for years of assistance
above and beyond the call of duty. I am in your debt. I have
enjoyed getting to know you as much as working with you. Adam
Healy, for his friendship, and, together with Paul Noone, turning
my life around.  And finally, Natasha, who didn't help me finish
the thesis at all, but showed me I was human.

\pagestyle{headings}
\pagenumbering{roman}
\tableofcontents
\listoffigures
\listoftables

% \pagenumbering{arabic} command must go in start of introduction.tex
\pagenumbering{arabic}
\chapter{Introduction}
\label{intro}

This chapter begins with a brief review of the quantum computing
field with a bias towards the specific topics developed in later
chapters. The main purpose of the review is to introduce the
concepts and language required to overview the aims and content of
this thesis (Section~\ref{introduction:section:overview}). The
reader familiar with quantum computing can go directly to
Section~\ref{introduction:section:overview}. Familiarity with
quantum mechanics is assumed.

In Section~\ref{introduction:section:motivation} we provide
justification and motivation for the study of quantum computing.
Section~\ref{introduction:section:models} reviews the various
models of quantum computation, namely the circuit model, adiabatic
quantum computation, cluster states, topological quantum
computation, and geometric quantum computation. The theoretical
work demonstrating that arbitrarily large quantum computations can
be performed arbitrarily reliably is gathered in
Section~\ref{introduction:section:possible}.
Section~\ref{introduction:section:overview} then overviews the
thesis.

\section{Why quantum compute?}
\label{introduction:section:motivation}

The incredible exponential growth in the number of transistors
used in conventional computers, popularly described as Moore's
Law, is expected to continue until at least the end of this decade
\cite{Inte05}. This growth is achieved through miniaturization.
Smaller transistors consume less power, can be packed more
densely, and switch faster. However, in some areas, particularly
the silicon dioxide insulating layer within each transistor,
atomic scales are already being approached \cite{Pee00}. This
fundamental barrier is one of many factors fuelling research into
radical new computing technologies.

Furthermore, certain problems simply cannot be solved efficiently
on conventional computers irrespective of their inexorable
technological and computational progress. One such problem is the
simulation of quantum systems. The amount of classical data
required to describe a quantum system grows exponentially with the
system size.  The data storage problem alone precludes the
existence of an efficient method of simulation on a conventional
computer.

The first hint of a way around this impasse was provided by
Feynman in 1982 \cite{Feyn82} when he suggested using quantum
mechanical components to store and manipulate the data describing
a quantum system. The number of quantum mechanical components
required would be directly proportional to the size of the quantum
system. This idea was built on by Deutsch in 1985 \cite{Deut85} to
form a model of computation called a quantum Turing machine ---
the quantum mechanical equivalent of the universal Turing machine
\cite{Turi36} which previously was thought to be the most powerful
and only model of computation.

That the laws of physics in principle permit the construction of
quantum computers exponentially more powerful than their classical
relatives is hugely significant. Despite this, it was not until
the publication of Shor's quantum integer factoring algorithm
\cite{Shor94a} that research into quantum computing began to
attract serious attention.  The difficulty of classically
factoring integers forms the basis of the popular RSA encryption
protocol \cite{Rive78}. RSA is used to establish secure
connections over the Internet, enabling the transmission of
sensitive data such as passwords, credit card details, and online
banking sessions. RSA also forms the heart of the popular secure
messaging utility PGP (Pretty Good Privacy) \cite{Zimm91}. Rightly
or wrongly, the prospect of rendering much of modern classical
communication insecure has arguably driven the race to build a
quantum computer.

More recently, quantum algorithm offering an exponential speed up
over their classical equivalents have been devised for problems
within group theory \cite{Kita96}, knot theory \cite{Subr02},
eigenvalue calculation \cite{Jaks03}, image processing
\cite{Schu03}, basis transformations \cite{Baco04}, and numerical
integrals \cite{Abra99}. Other promising quantum algorithms exist
\cite{Kieu04,Farh01}, but have not been thoroughly analyzed.

On the commercial front, the communication of quantum data has
been shown to enable unconditionally secure communication in
principle \cite{Lo99} resulting in the creation of companies
offering real products \cite{Magi02,idQu04a} that have already
found application in the information technology sector
\cite{idQu04b}. Despite this, it remains to be seen whether human
ingenuity is sufficient to make large-scale quantum computing a
reality.

\section{Models of quantum computing}
\label{introduction:section:models}

Classical computers have a single well defined computational model
--- the direct manipulation of bits via Boolean logic. The field of
quantum computation is too young to have settled on a single
model. This section attempts to make a brief yet complete review
of the current status of the various quantum computation models
currently under investigation. We neglect quantum Turing machines
\cite{Deut85} as they are a purely abstract rather than a
physically realisably computation model. We also neglect quantum
neural networks \cite{Fabe02,Gupt01} due to their use of nonlinear
gates, and both Type II quantum computers \cite{Yepe01} and
quantum cellular automata \cite{Lent93} due to their essentially
classical nature.

\subsection{The circuit model}
\label{introduction:subsection:circuit_model}

The most widely used model of quantum computation is called the
circuit model. Instead of the traditional bits of conventional
computing which can take the values 0 or 1, the circuit model is
based on qubits which are quantum systems with two states denoted
by $|0\rangle$ and $|1\rangle$. The power of quantum computing
lies in the fact that qubits can be placed in superpositions
$\alpha|0\rangle +\beta|1\rangle$, and entangled with one another,
eg.~$(|00\rangle + |11\rangle)/\sqrt{2}$. Manipulation of qubits
is performed via quantum gates. An $n$-qubit gate is a
$2^{n}\times 2^{n}$ unitary matrix. The most general single-qubit
gate can be written in the form
\begin{equation}
U = \left(
\begin{array}{cc}
e^{i(\alpha+\beta)/2}\cos\theta & e^{i(\alpha-\beta)/2}\sin\theta\\
-e^{i(-\alpha+\beta)/2}\sin\theta &
e^{-i(\alpha+\beta)/2}\cos\theta
\end{array}
\right).
\end{equation}
Common single-qubit gates are
\begin{displaymath}
H = \frac{1}{\sqrt{2}}\left( \begin{array}{cc}
1 & 1 \\
1 & -1 \\
\end{array} \right),\hspace{10mm}
X = \left( \begin{array}{cc}
0 & 1 \\
1 & 0 \\
\end{array} \right),\hspace{10mm}
Z = \left( \begin{array}{cc}
1 & 0 \\
0 & -1 \\
\end{array} \right),
\end{displaymath}
\begin{equation}
S = \left( \begin{array}{cc}
1 & 0 \\
0 & i \\
\end{array} \right),\hspace{10mm}
S^{\dag} = \left( \begin{array}{cc}
1 & 0 \\
0 & -i \\
\end{array} \right),\hspace{10mm}
T = \left(\begin{array}{cc}
1 & 0 \\
0 & e^{i\pi/4} \\
\end{array} \right).
\end{equation}
For example, the result of applying an $X$-gate to
$\alpha|0\rangle +\beta|1\rangle$ is
\begin{equation}
\left( \begin{array}{cc}
0 & 1 \\
1 & 0 \\
\end{array} \right)
\left( \begin{array}{c} \alpha \\
\beta \\
\end{array} \right)
= \left( \begin{array}{c} \beta \\
\alpha \\
\end{array} \right).
\end{equation}
The $X$-gate will sometimes be referred to as a \NOT\ gate or
inverter. Its action will sometimes be referred to as a bit-flip
or inversion, and that of the $Z$-gate as a phase-flip. The
$H$-gate was derived from the Walsh-Hadamard transform
\cite{Hada1893,Wals23} and first named as the Hadamard gate in
\cite{Clev97}.

Given the ability to implement arbitrary single qubit gates and
almost any multiple qubit gate, arbitrary quantum computations can
be performed \cite{DiVi95,Brem02}. The column vector form of an
arbitrary 2-qubit state $|\Psi\rangle = \alpha|00\rangle +
\beta|01\rangle + \gamma|10\rangle + \delta|11\rangle$ is
\begin{equation}
\left(
\begin{array}{c}
\alpha \\
\beta \\
\gamma \\
\delta
\end{array}
\right).
\end{equation}
Note the ordering of the computational basis states. For
convenience, such states will occasionally be denoted by
$|q_{1}q_{0}\rangle$ with $q_{1}$ ($q_{0}$) referred to as the
first (last) or left (right) qubit regardless of the actual
physical arrangement. The most common 2-qubit gate is the
controlled-\NOT\ (\CNOT)
\begin{equation}
\left(
\begin{array}{cccc}
1 & 0 & 0 & 0 \\
0 & 1 & 0 & 0 \\
0 & 0 & 0 & 1 \\
0 & 0 & 1 & 0
\end{array}
\right),
\end{equation}
which given an arbitrary 2-qubit state $|q_{1}q_{0}\rangle$,
inverts the target qubit $q_{0}$ if the control qubit $q_{1}$ is
1. Note that a \CNOT\ with target qubit $q_{1}$ and control qubit
$q_{0}$ would have the form
\begin{equation}
\left(
\begin{array}{cccc}
1 & 0 & 0 & 0 \\
0 & 0 & 0 & 1 \\
0 & 0 & 1 & 0 \\
0 & 1 & 0 & 0
\end{array}
\right).
\end{equation}

The swap gate
\begin{equation}
\left(
\begin{array}{cccc}
1 & 0 & 0 & 0 \\
0 & 0 & 1 & 0 \\
0 & 1 & 0 & 0 \\
0 & 0 & 0 & 1
\end{array}
\right)
\end{equation}
swaps the states of $q_{1}$ and $q_{0}$.  Additional 2-qubit gates
will be defined as required.

\begin{figure}
\begin{center}
\includegraphics[width=10cm]{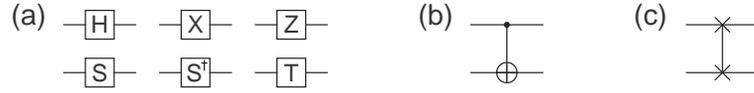}
\caption{(a) Common single-qubit gates, (b) the \CNOT\ gate with
solid dot representing the control qubit, (c) the swap
gate.}\label{introduction:figure:circuit_definitions}
\end{center}
\end{figure}

\begin{figure}
\begin{center}
\includegraphics[width=12cm]{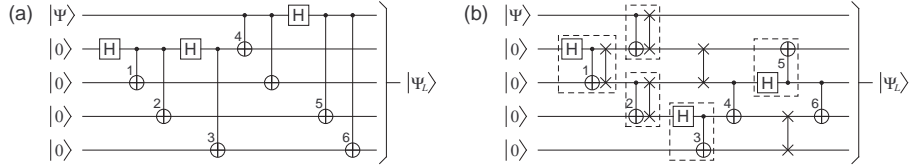}
\caption{(a) Example of a quantum circuit, (b) depth 6
rearrangement of (a). This circuit implements the encode stage of
5-qubit non-fault-tolerant quantum error
correction.}\label{introduction:figure:5qec_encode_modern}
\end{center}
\end{figure}

By representing qubits as horizontal lines, a time sequence of
quantum gates can conveniently be represented by notation that
looks like a conventional circuit. Symbols equivalent to the gates
described above are shown in
Fig.~\ref{introduction:figure:circuit_definitions}. An example of
a complete circuit is shown in
Fig.~\ref{introduction:figure:5qec_encode_modern}a. Note that the
horizontal lines represent time flowing from left to right, not
wires. We define the depth of a quantum circuit to be the number
of layers of 2-qubit gates required to implement it. Note that
multiple single-qubit gates and 2-qubit gates applied to the same
two qubits can be combined into a single 2-qubit gate. For
example, Fig.~\ref{introduction:figure:5qec_encode_modern}b is a
depth 6 rearrangement of
Fig.~\ref{introduction:figure:5qec_encode_modern}a. These circuits
are discussed in more detail in Chapter~\ref{5QEC}.

The basis $\{|0\rangle,|1\rangle\}$ is referred to as the
computational basis. The simplest representation of an $n$-qubit
state is
\begin{equation}\label{introduction:eq:nqstate}
    |\Psi\rangle =
    \frac{1}{\sqrt{2^{n}}}\sum_{k=1}^{2^{n}-1}|k\rangle,
\end{equation}
where $k$ is an $n$-bit number. The information theoretic
properties of qubits have undergone investigation by many authors
and are well reviewed in \cite{Niel00}.

One natural extension of the qubit circuit model is to use
$d$-level quantum systems (qudits). The simplest representation of
an $n$-qudit state is
\begin{equation}\label{introduction:eq:nqdstate}
    |\Psi\rangle =
    \frac{1}{\sqrt{d^{n}}}\sum_{k=1}^{d^{n}-1}|k\rangle,
\end{equation}
where $k$ is expressed in base $d$. The properties of qudit
entanglement \cite{Rung00}, qudit teleportation \cite{Gour04},
qudit error correction \cite{Knil96a,Gras02}, qudit cryptography
\cite{Cerf01,Kari02}, and qudit algorithms \cite{Brem04,Cere04}
are broadly similar to the corresponding properties of qubits, and
will not be discussed further here.

A third variant of the circuit model exists based on continuous
quantum variables such as the position and momentum eigenstates of
photons. Considerable experimental work on the entanglement of
continuous quantum variables has been performed and is reviewed in
Ref.~\cite{Adam04}. In particular, the problem of continuous
quantum variable teleportation \cite{Smol04} has received a great
deal of attention. Furthermore, methods have been devised to
perform continuous variable error correction \cite{Barn04},
cryptography \cite{Nami04}, and continuous variable versions of a
number of algorithms \cite{Pati00,Pati02,Lomo02,Lomo03}.

\subsection{Adiabatic quantum computation}

In the adiabatic model of quantum computation \cite{Farh00}, a
Hamiltonian $H_{f}$ is found such that its ground state is the
solution of the problem under consideration. This final
Hamiltonian $H_{f}$ must also be continuously deformable (for
example, by varying the strength of a magnetic field) to some
initial Hamiltonian $H_{i}$ with a ground state that is easy to
prepare. After initializing the computer in the ground state of
$H_{i}$, $H_{i}$ is adiabatically (i.e.\ sufficiently slowly to
leave the system in the ground state) deformed into $H_{f}$
yielding the ground state of $H_{f}$. The major difficulty is
determining how slowly the deformation must occur to be adiabatic.
The adiabatic model can be applied to systems of qubits, qudits,
and continuous quantum variables, and is equivalent to the circuit
model in the sense that any adiabatic algorithm can be converted
into a quantum circuit with at most polynomial overhead
\cite{Ahar04,Kemp04}. The principal advantage of the adiabatic
model is an alternative way of thinking about quantum computation
that has led to numerous new algorithms tackling problems in graph
theory \cite{Chil00}, combinatorics \cite{Farh00,Smel02},
condensed matter and nuclear physics \cite{Wu02}, and set theory
\cite{Smel01}.

\subsection{Cluster states}

Given a multi-dimensional lattice of qubits each initialized to
$(|0\rangle + |1\rangle)/\sqrt{2}$ with identical tunable nearest
neighbor interactions of form
\begin{equation}
H_{ij}(t)=\hbar g(t)(1+\sigma_{z}^{(i)})(1+\sigma_{z}^{(j)})/4,
\end{equation}
a cluster state \cite{Brie01} can be created by evolving the
system for a time such that $\int g(t)dt =\pi$. Cluster states are
highly entangled states with the remarkable property that
arbitrary quantum computations can be performed purely via
single-qubit measurements along arbitrary axes
\cite{Raus01,Raus03}. A small amount of classical computation is
required between measurements. Cluster states can also be defined
over qudits \cite{Zhou03}, and are special cases of the more
general class of graph states \cite{Hein04}. Methods have been
devised to perform quantum communication \cite{Zeng03}, error
correction \cite{Schl03b} and fault-tolerant computation
\cite{Niel04b} within the cluster state model. Entanglement
purification can be used to increase the reliability of cluster
state generation \cite{Dur03}. Unlike the adiabatic model however,
the cluster state model is yet to lead to any genuinely new
algorithms. Given the equivalence of the cluster state model to
the circuit model \cite{Schl03a}, the primary utility of the
cluster state model appears to be simpler physical implementation
in certain systems such as linear optics \cite{Niel04a,Brow04},
and possibly special cases of NV-centers in diamond, quantum dots,
and ion traps \cite{Barr04}.

\subsection{Topological quantum computation}

\begin{figure}
\begin{center}
\includegraphics[width=12cm]{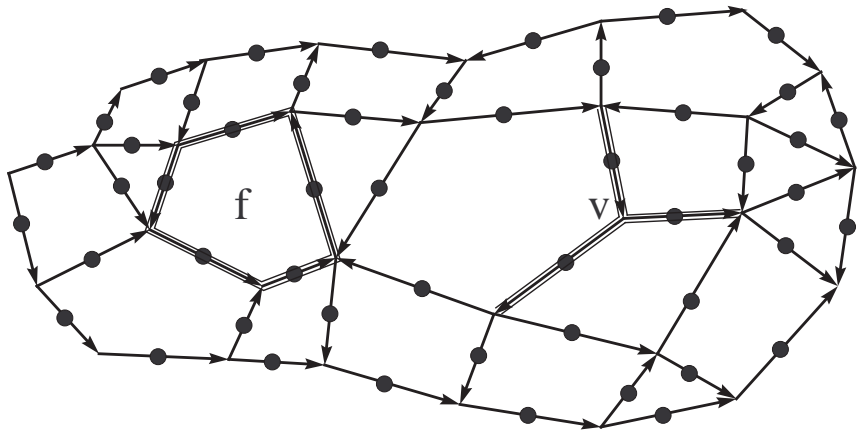}
\end{center}
\caption{Lattice of hypothetical particles used in the
construction of the anyonic model of quantum computation.}
\label{introduction:figure:lattice}
\end{figure}

\begin{figure}
\begin{center}
\includegraphics[width=8cm]{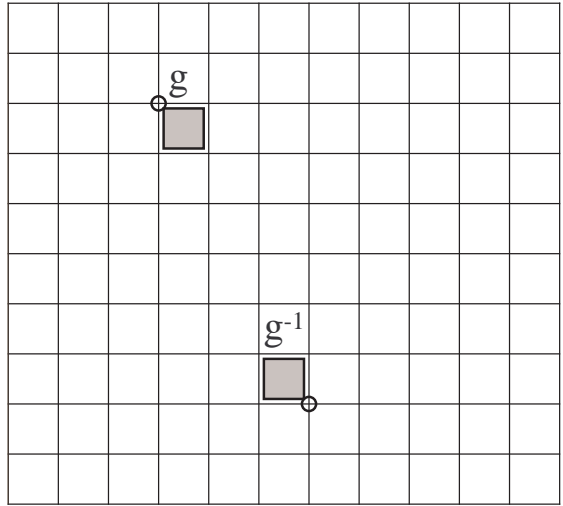}
\end{center}
\caption{A conjugate pair of non-Abelian anyons.}
\label{introduction:figure:pair}
\end{figure}

\begin{figure}
\begin{center}
\includegraphics[width=5cm]{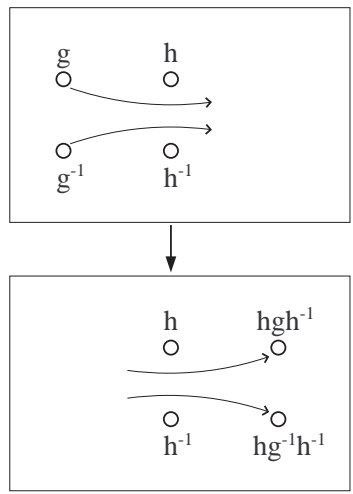}
\end{center}
\caption{Performing operations in the anyonic model of quantum
computation.} \label{introduction:figure:pull_through}
\end{figure}

The primary difficulty in building a quantum computer is
controlling data degradation through interaction with the
environment, generally called decoherence. Interaction with the
environment can in principle be eliminated by using a topological
model of computation.  Topological quantum computation was
proposed by Kitaev in 1997 \cite{Kita97b}, and developed further
in Ref.~\cite{Ogbu99}. An alternative proposal was given by
Freedman in Ref.~\cite{Free03}. Kitaev considered an oriented
2-dimensional lattice of hypothetical particles with many body
interactions as shown in Fig.~\ref{introduction:figure:lattice}.
The hypothetical particles on each lattice link are related to the
60 permutations of five distinguishable objects $P_{5}$. Certain
types of excitations of the lattice are also related to the group
$P_{5}$, and exist on lattice sites which correspond to a vertex
and face pair as shown in Fig.~\ref{introduction:figure:pair}.
These excitations are called non-Abelian anyons. Of particular
interest are pairs of excitations $|g,g^{-1}\rangle$. Non-Abelian
anyon pairs have the remarkable property that simply moving them
through one another effects computation. Given two pairs
$|g,g^{-1}\rangle, |h,h^{-1}\rangle$, moving $|g,g^{-1}\rangle$
through $|h,h^{-1}\rangle$ as shown in
Fig.~\ref{introduction:figure:pull_through} creates the state
$|hgh^{-1},hg^{-1}h^{-1}\rangle$.

Let $|0\rangle=|g,g^{-1}\rangle$ and $|1\rangle=|h,h^{-1}\rangle$.
Find $x$ such that $h=xgx^{-1}$.  A quantum inverter can be
constructed simply by moving states $|0\rangle,|1\rangle$ through
the ancilla pair $|x,x^{-1}\rangle$. More complicated operations
can be performed simply in a similar manner, using more ancilla
pairs and pull through operations.

At sufficiently low temperature, the only way data can be
corrupted in a topological quantum computer is via the spontaneous
interaction of two anyons --- an event that occurs with
probability $O(e^{-\alpha l})$, where $l$ is the minimum
separation of any pair of anyons. This probability can be made
arbitrarily small simply by keeping the anyons well separated.
While this is a desirable property, clearly a computational model
based on hypothetical particles that do not exist in nature cannot
be implemented.

Recently, significant progress has been made on the topological
quantum computation model. General schemes using anyons based on
arbitrary non-solvable groups \cite{Moch03}, and smaller,
solvable, non-nilpotent groups \cite{Moch04} have been devised. A
less robust but simpler scheme based on Abelian anyons has been
proposed \cite{Lloy00} and designs for possible experimental
realizations are emerging \cite{Aver02,Pare01}. Experimental
proposals based on non-Abelian anyons have also been constructed
\cite{Over01,Duan02}.

\subsection{Geometric quantum computation}

\begin{figure}
\begin{center}
\includegraphics[width=4cm]{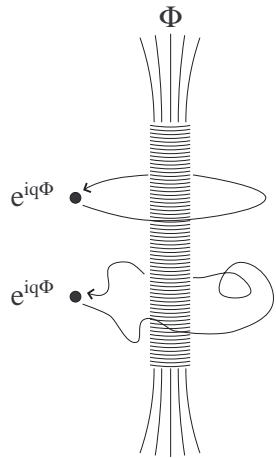}
\end{center}
\caption{Example of the path independence of the phase shift
induced by the Aharonov-Bohm effect. A particle of charge $q$
executing a loop around a perfectly insulated solenoid containing
flux $\Phi$ acquires a geometric phase $e^{iq\Phi}$.}
\label{introduction:figure:Aharonov}
\end{figure}

The basic idea of geometric quantum computation is illustrated by
the Aharonov-Bohm effect \cite{Ahar59} in which a particle of
charge $q$ executing a loop around a perfectly insulated solenoid
containing flux $\Phi$ acquires a geometric phase $e^{iq\Phi}$
\cite{Berr84}. As shown in
Fig.~\ref{introduction:figure:Aharonov}, the phase acquired is
insensitive to the exact path taken. This phase shift can be used
to build quantum gates \cite{Jone00,Falc00}. At the moment, there
are conflicting numerical and analytic calculations both
supporting and attacking the fundamental robustness of geometric
quantum computation \cite{Zhu04}.

\section{Quantum computation is possible in principle}
\label{introduction:section:possible}

For the remainder of the thesis we will focus on the qubit circuit
model. Physical realizations of this model of computation have
been proposed in the context of liquid NMR \cite{Cory96}, ion
traps \cite{Cira95} including optically \cite{Cira97,Duan04b} and
physically \cite{Cira00,Kiel02} coupled microtraps, linear optics
\cite{Knil01b}, Josephson junctions utilizing both charge
\cite{Makh99,Naka99} and flux \cite{Bock97} degrees of freedom,
quantum dots \cite{Loss98}, $^{31}$P in $^{28}$Si architectures
utilizing indirectly exchange coupled donor nuclear spins
\cite{Kane98}, exchange coupled electron spins
\cite{Vrij00,Hill04}, both spins \cite{Kane00,Skin03}, magnetic
dipolar coupled electron spins \cite{Sous04} and qubits encoded in
the charge distribution of a single electron on two donors
\cite{Holl03}, deep donors in silicon \cite{Ston03}, acceptors in
silicon \cite{Gold03}, solid-state ensemble NMR utilizing lines of
$^{29}$Si in $^{28}$Si \cite{Ladd02}, electrons floating on liquid
helium \cite{Plat99}, cavity QED \cite{Turc95,Pell95}, optical
lattices \cite{Bren99,Jaks99} and the quantum Hall effect
\cite{Priv98}. At the present time, the ion trap approach is the
closest to realizing the five basic requirements of scalable
quantum computation \cite{DiVi00}.

The mere existence of quantum computer proposals and quantum
algorithms is not sufficient to say that quantum computation is
possible in principle. Firstly, almost all quantum algorithms that
provide an exponential speed up over the best-known equivalent
classical algorithms make use of the quantum Fourier transform
which in turn uses exponentially small rotations.  For example, in
the case of Shor's algorithm, to factor an $L$-bit number, in
principle single-qubit rotations of magnitude $2\pi/2^{2L}$ are
required. Clearly this is impossible for large $L$.

Coppersmith resolved this issue by showing that most of the small
rotations in the quantum Fourier transform could simply be ignored
without significantly affecting the output of the circuit.  For
the specific case of Shor's algorithm, as described in
Chapter~\ref{ShorGates}, we took this work further to show that
rotations of magnitude $\pi/128$ implemented with accuracy
$\pm\pi/512$ were sufficient to factor integers thousands of bits
long.

More seriously, quantum systems are inherently fragile. The gamut
of relaxation processes, environmental couplings, and even
systematic errors induced by architectural imperfections are
typically grouped under the heading of decoherence. Ignoring
leakage errors in which a qubit is destroyed or placed in a state
other than those selected for computation \cite{Leun97}, Shor made
the surprising discovery that all types of decoherence could be
corrected simply by correcting unwanted bit-flips ($X$),
phase-flips ($Z$) and both simultaneously ($XZ$) \cite{Shor94b}.
Shor's scheme required each qubit of data to be encoded across
nine physical qubits. This was the first quantum error correction
code (QECC).

Later work by Laflamme gave rise to a 5-qubit QECC \cite{Lafl96}
--- the smallest code that can correct an arbitrary error to one
of its qubits \cite{Niel00}. Steane's 7-qubit code \cite{Stea96},
which also only guarantees to correct an arbitrary error to one of
its qubits, is, however, more convenient for the purposes of
quantum computation. Steane's code is part of the large class of
CSS codes (Calderbank, Shor, Steane) \cite{Cald95}, which is in
turn part of the very large class of stabilizer codes
\cite{Gott96}. Only a few examples such as permutationally
invariant codes \cite{Poll04} exists outside the class of
stabilizer codes.

To illustrate the utility of the stabilizer formalism, consider a
5-qubit QECC \cite{Gott97} in which we create logical $|0\rangle$
and $|1\rangle$ states corresponding to the superpositions
\begin{eqnarray}
|0_{L}\rangle & = & |00000\rangle+|10010\rangle+|01001\rangle+|10100\rangle \nonumber \\
& & +|01010\rangle-|11011\rangle-|00110\rangle-|11000\rangle \nonumber \\
& & -|11101\rangle-|00011\rangle-|11110\rangle-|01111\rangle \nonumber \\
& & -|10001\rangle-|01100\rangle-|10111\rangle+|00101\rangle, \\
|1_{L}\rangle & = & |11111\rangle+|01101\rangle+|10110\rangle+|01011\rangle \nonumber \\
& & +|10101\rangle-|00100\rangle-|11001\rangle-|00111\rangle \nonumber \\
& & -|00010\rangle-|11100\rangle-|00001\rangle-|10000\rangle \nonumber \\
& & -|01110\rangle-|10011\rangle-|01000\rangle+|11010\rangle.
\end{eqnarray}
These superstitions redundantly encode data in such a way that an
arbitrary error to any one qubit can be corrected. They are,
however, difficult to work with directly. In the stabilizer
formalism, $|0_{L}\rangle$ and $|1_{L}\rangle$ are instead
described as simultaneous $+1$ eigenstates of
\begin{eqnarray}
M_{1} & = & X\otimes Z\otimes Z\otimes X\otimes I \\
M_{2} & = & I\otimes X\otimes Z\otimes Z\otimes X \\
M_{3} & = & X\otimes I\otimes X\otimes Z\otimes Z \\
M_{4} & = & Z\otimes X\otimes I\otimes X\otimes Z
\end{eqnarray}
These operators are called stabilizers.  Let $S$ denote the set of
stabilizers (which is the group generated by $M_{1}$--$M_{4}$).
Any valid logical state must satisfy
$M|\Psi_{L}\rangle=|\Psi_{L}\rangle$ for all $M\in S$. This
observation allows us to determine which quantum gates can be
applied directly to the logical states. Specifically, if we wish
to apply $U$ to our logical state to obtain $U|\Psi_{L}\rangle$,
then since $UMU^{\dag}U|\Psi\rangle$ must be a logical state,
$UMU^{\dag}$ must be a stabilizer.  If we restrict our attention
to gates $U$ that are products of single-qubit gates, only two
single logical qubit gates
\begin{eqnarray}
X_{L} & = & X\otimes X\otimes X\otimes X\otimes X \\
Z_{L} & = & Z\otimes Z\otimes Z\otimes Z\otimes Z
\end{eqnarray}
can be applied directly to data encoded using this version of
5-qubit QEC. Note that by restricting our attention to products of
single-qubit gates, any error present in one of the qubits of the
code before a logical gate operation cannot be copied to other
qubits. Any circuit with the property that a single error can
cause that most one error in the output is called fault-tolerant.

So far, we have not explained how errors are corrected using a
QECC. Given a potentially erroneous state $|\Psi'\rangle$, one way
of locating any errors is to check whether $|\Psi'\rangle$ is
still a $+1$ eigenstate of each of the stabilizers. Any errors so
located can then be manually corrected.  This method is described
in some detail in Chapter~\ref{Solovay}.

\begin{figure}
\begin{center}
\includegraphics[width=8cm]{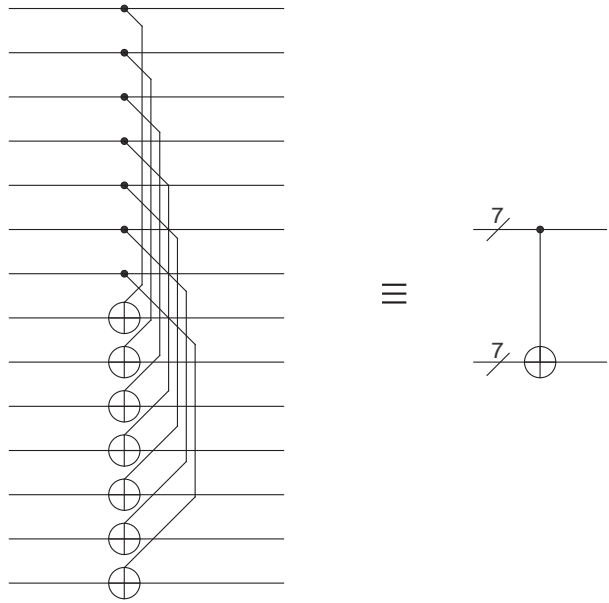}
\caption{7-qubit transversal logical \CNOT\
gate.}\label{introduction:figure:cnot_straight}
\end{center}
\end{figure}

We now have basic single logical qubit gates and error correction.
To achieve universal quantum computation we need to be able to
couple logical qubits and perform arbitrary single logical qubits
gates \cite{DiVi95,Brem02}.  Unfortunately, the 5-qubit QECC does
not readily permit multiple logical qubits to be coupled, though a
complicated three logical qubit gate does exist \cite{Gott98}. For
universal quantum computation, the 7-qubit Steane code, or indeed
any of the CSS codes, is more appropriate as they permit a simple
transversal implementation of logical \CNOT\ as shown in
Fig.~\ref{introduction:figure:cnot_straight}. The 7-qubit Steane
code also permits similar transversal single-qubit gates $H$, $X$,
$Z$, $S$ and $S^{\dag}$ (see Fig.~\ref{Solovay:figure:single}).
These are, however, insufficient to construct arbitrary
single-qubit rotations of the form
\begin{equation}
\left(
\begin{array}{cc}
\cos(\theta/2)e^{i(\alpha+\beta)/2} & \sin(\theta/2)e^{i(\alpha-\beta)/2} \\
-\sin(\theta/2)e^{i(-\alpha+\beta)/2} & \cos(\theta/2)e^{i(-\alpha-\beta)/2}
\end{array}
\right).
\end{equation}

An additional gate such as the $T$-gate is required to construct
arbitrary single-qubit gates. The simplest fault-tolerant
implementation of the $T$-gate we have been able to devise still
requires an additional 12 ancilla qubits, and at least 93 gates,
45 resets and 17 measurements arranged on a circuit of depth at
least 92 and is described in detail in Chapter~\ref{Solovay}. By
virtue of the fact that the operation $HT$ corresponds to the
rotation of a qubit by an angle that is an irrational number, and
the fact that repeated rotation by an irrational number enables
arbitrarily close approximation of a rotation of any angle, the
pair of single-qubit gates $H$, $T$ is sufficient to approximate
an arbitrary single-qubit rotation arbitrarily accurately.  In
practice, it is better to use the 23 unique combinations of $H$,
$X$, $Z$, $S$ and $S^{\dag}$ in addition to the $T$-gate since,
for example, the logical $X$-gate is vastly simpler to implement
than $HTTTTH$. The existence of efficient sequences of gates to
approximate arbitrary unitary rotations in general, and
single-qubit gates in particular, is guaranteed by the
Solovay-Kitaev theorem \cite{Solo95,Kita97,Niel00,Harr02}. The
exact length of such sequences required to achieve a given
accuracy is discussed in Chapter~\ref{Solovay}.

We now have enough machinery to consider arbitrarily large,
arbitrarily reliable quantum computation. Suppose every qubit in
our computer has probability $p$ of suffering an error per unit
time, where unit time refers to the amount of time required to
implement the slowest fundamental gate (not logical gate).
Consider a circuit consisting of the most complicated
fault-tolerant logic gate, the $T$-gate, followed by quantum error
correction. By virtue of being fault-tolerant, this circuit can
only fail to produce useful output if at least two errors occur.
For some constant $c$, and sufficiently low error rate $p$, the
probability of failure of the structure is thus $cp^{2}$. Since we
have chosen the most complicated gate, every other fault-tolerant
gate followed by error correction, including the identity $I$ (do
nothing) gate, has probability of failure at most $cp^{2}$.

Consider an arbitrary quantum circuit expressed in terms of the
fundamental gates \CNOT, $I$, $H$, $X$, $Z$, $S$, $S^{\dag}$,
their products, and $T$ (no QEC at this stage). In the worst-case,
if an error anywhere causes the circuit to fail, the probability
of success of such a circuit is $(1-p)^{qt}$ where $q$ is the
number of qubits and $t$ the number of time steps. By replacing
each gate with an error corrected fault-tolerant structure, this
can be reduced to $(1-cp^{2})^{qt}$. Note that the new circuit is
still expressed entirely in terms of the allowed gates and
measurement. We can therefore repeat the process and replace these
gates in turn with error corrected fault-tolerant structures
giving an overall reliability of $(1-c^{3}p^{4})^{qt}$. If we
repeat this $k$ times we find that the overall circuit reliability
is
\begin{equation}
\left(1-\frac{(cp)^{2^{k}}}{c}\right)^{qt}.
\end{equation}
Clearly, provided the error rate per time step is less than
$p_{th}=1/c$ and we have sufficient resources, an arbitrarily
large quantum computation can be performed arbitrarily reliably.
This is the threshold theorem of quantum computation
\cite{Knil96b}. Note that the greater the amount $p<p_{th}$, the
fewer levels of error correction that are required. Despite the
lack of a definitive reference, $p_{th}$ is frequently assumed to
be $10^{-4}$.

Substantial customization and optimization of the threshold
theorem has occurred over the years. As described, the threshold
theorem requires the ability to interact arbitrarily distant pairs
of qubits in parallel, fast measurement gates, and fast and
reliable classical computation. A detailed study of the impact of
noisy long-range communication has been performed yielding
threshold error rates $\sim$$10^{-4}$ for a variety of assumptions
\cite{Svor04}. Modifications removing the need for measurement and
classical processing but still requiring error-free long-range
interactions have been devised at the cost of greatly increased
resources \cite{Ahar99}, though subsequent work has substantially
reduce the complexity of the required quantum circuitry with a
threshold $p_{th}=2.4\times 10^{-6}$ obtained under the additional
assumption that the quantum computer consists of a single line
qubits with nearest neighbor interactions only \cite{Szko04}.

By using a much simplified error correction scheme devised by
Steane \cite{Stea02} that works on any CSS code, and using larger
QECCs and less concatenation, encouragingly high thresholds
$\sim$$10^{-3}$ \cite{Stea03b} and even $9\times 10^{-3}$
\cite{Reic04} have been calculated, although in the latter case
only under the assumption that errors occur after gates, not to
idle qubits, and in both cases long-range interactions must be
available.

An alternative approach is to perform computation by interacting
data with specially prepared ancilla states
\cite{Knil04a,Knil04b}, a method called postselected quantum
computing.  While the resources required to prepare sufficiently
reliable ancilla states are prohibitive, in principle this
approach permits arbitrarily large computations to be performed
provided $p<p_{th}=0.03$ \cite{Knil04c}.

\section{Overview}
\label{introduction:section:overview}

The primary goal of this thesis is to relax the theoretical
resource requirements required for large-scale quantum
computation. Much of our work is motivated by the Kane $^{31}$P in
$^{28}$Si architecture in which fast and reliable measurement and
classical processing is difficult to achieve, and qubits may be
limited to a single line with nearest neighbor interactions only.

In Chapters~\ref{CNOT}--\ref{readout}, the viability of the
adiabatic Kane \CNOT\ gate and readout operation are assessed.
Chapter~\ref{canonical} reviews a technique of constructing
efficient 2-qubit gates. A greatly simplified non-fault-tolerant
linear nearest neighbor implementation of 5-qubit quantum error
correction is presented in Chapter~\ref{5QEC}, and further
modified to remove the need for measurement and classical
processing in Chapter~\ref{5QECNM}. Chapter~\ref{Shor} provides a
detailed review of Shor's algorithm, followed by a linear nearest
neighbor circuit implementation in Chapter~\ref{ShorLNN}.
Chapter~\ref{ShorGates} focuses on removing the need for
exponentially small rotations in circuit implementations of Shor's
algorithm. Finally, Chapter~\ref{Solovay} presents a method of
constructing arbitrary single-qubit fault-tolerant gates, and
applies this to the specific case of the remaining rotation gates
required by Shor's algorithm. Chapter~\ref{remarks} contains
concluding remarks and summarizes the results of the thesis.
Chapters~\ref{CNOT}, \ref{readout}, \ref{5QEC}, \ref{ShorLNN} and
\ref{ShorGates} have been published in
\cite{Fowl03a,Holl04,Fowl04,Fowl04b,Fowl03b} respectively.

\chapter{The adiabatic Kane \CNOT\ gate}
\label{CNOT}

The spins of the $^{31}$P nucleus and donor electron in bulk
$^{28}$Si have extremely long coherence times
\cite{Fehe59,Tyry03}. Consequently, a number of quantum computer
proposals have been based on this system
\cite{Kane98,Skin03,Sous04,Hill04}. In this chapter and
Chapter~\ref{readout}, we focus on Kane's 1998 proposal
\cite{Kane98} which calls for a line of single $^{31}$P atoms
spaced approximately 20nm apart \cite{Kett04,Well04}. The spin of
each phosphorus nucleus is used as a qubit and the donor electron
used to couple to neighboring qubits. Neglecting readout
mechanisms for the moment, each qubit requires at least two
electrodes to achieve single- and 2-qubit gates. The extent to
which the presence and operation of these electrodes will reduce
the system's coherence times is unknown. We therefore study the
error rate of the Kane \CNOT\ gate as a function of the coherence
times to determine their approximate minimum acceptable values. We
find that the coherence times required to achieve a \CNOT\ error
rate of $10^{-4}$ are a factor of 6 less than those already
observed experimentally.

The chapter is organized as follows.  In
Section~\ref{CNOT:section:background}, the Kane architecture is
briefly described followed by the method of performing a \CNOT\ in
Section~\ref{CNOT:section:cnot}. In
Section~\ref{CNOT:section:deph}, the technique we used to model
finite coherence times is presented along with contour plots of
the \CNOT\ error rate as a function of the coherence times. In
Section~\ref{CNOT:section:conc}, we discuss the implications of
our results.

\section{The Kane architecture}
\label{CNOT:section:background}

\begin{figure}[h]
\begin{center}
\includegraphics[width=12cm]{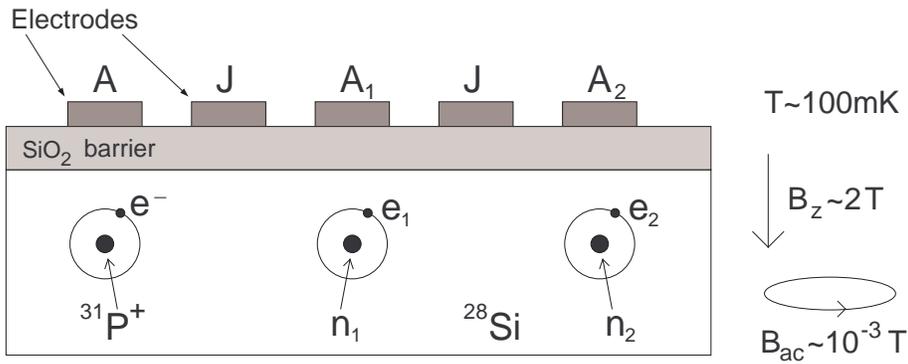}
\end{center}
\caption{Schematic of the Kane architecture. The rightmost two
qubits show the notation to be used when discussing the \CNOT\
gate.} \label{CNOT:figure:3q}
\end{figure}

Ignoring readout mechanisms, which we discuss in
Chapter~\ref{readout}, the basic layout of the adiabatic Kane
phosphorus in silicon solid-state quantum computer
\cite{Kane98,Kane00} is shown in Fig.~\ref{CNOT:figure:3q}. The
phosphorous donor electrons are used primarily to mediate
interactions between neighboring nuclear spin qubits. As such, the
donor electrons are polarized to remove their spin degree of
freedom. This can be achieved by maintaining a steady $B_{z}=2$T
at around $T=100$mK \cite{Goan00}. Techniques for relaxing the
high field and low temperature requirements such as spin
refrigeration are under investigation \cite{Kane00}.

In addition to the potential to build on the vast expertise
acquired during the last 50 years of silicon integrated circuit
development, the primary attraction to the Kane architecture, and
$^{31}$P in $^{28}$Si architectures in general, is their
extraordinarily long spin coherence times \cite{Fehe59,Tyry03}.
Four quantities are of interest --- the relaxation ($T_{1}$) and
dephasing ($T_{2}$) times of both the donor electron and nucleus.
Both times only have meaning when the system is in a steady
magnetic field. Assuming the field is parallel with the z-axis,
the relaxation time refers to the time taken for $1/e$ of the
spins in the sample to spontaneously flip whereas the dephasing
time refers to the time taken for the $x$ and $y$ components of a
single spin to decay by a factor of $1/e$. Existing experiments
cannot measure $T_{2}$ directly, but instead a third quantity
$T^{*}_{2}$ which is the time taken for the $x$ and $y$ components
of an ensemble of spins to decay by a factor of $1/e$. Since
$T^{*}_{2}\leq T_{2}$ \cite{Hu01}, we can use experimental values
$T^{*}_{2}$ as a lower bound for $T_{2}$.

In natural silicon containing 4.7\% $^{29}$Si, relaxation times
$T_{1}$ in excess 1 hour have been observed for the donor electron
at $T=1.25$K and $B\sim 0.3$T \cite{Fehe59}. The nuclear
relaxation time has been estimated at over 80 hours in similar
conditions \cite{Honi60}. These times are so long that we will
ignore relaxation in our simulations of gate reliability. The
donor electron dephasing time $T^{*}_{2}$ in enriched $^{28}$Si
containing less than 50ppm $^{29}$Si, at 7K and donor
concentration $0.8\times 10^{15}$ has been measured to be 14ms
\cite{Tyry03}. Extrapolation to a single donor suggests
$T_{2}=60$ms at 7K. An even longer $T_{2}$ is expected at lower
temperatures. At the time of writing, to the authors' knowledge,
no experimental data relating to the nuclear dephasing time has
been obtained. However, considering the much greater isolation
from the environment of the nuclear spin, this time is expected to
be much larger than the electron dephasing time.

Control of the nuclear spin qubits is achieved via electrodes
above and between each phosphorus atom and global transverse
oscillating fields of magnitude $\sim$$10^{-3}$T. To selectively
manipulate a single qubit, the $A$-electrode above it is biased. A
positive/negative bias draws/drives the donor electron away from
the nucleus, in both cases reducing the magnitude of the hyperfine
interaction. This in turn reduces the energy difference between
nuclear spin up ($|0\rangle$) and down ($|1\rangle$) allowing this
transition to be brought into resonance with a globally applied
oscillating magnetic field. Depending on the timing of the
$A$-electrode bias, an arbitrary rotation about an axis in the
$x$-$y$ plane can be implemented \cite{Hill03}. By utilizing up to
three such rotations, a qubit can be rotated into an arbitrary
superposition $\alpha|0\rangle+\beta|1\rangle$.

Biasing an $A$-electrode above a particular donor will also effect
neighboring and more distant donors. It is likely that
compensatory biasing of nearby electrodes will be required to
ensure non-targeted qubits remain off resonant.  Subject to this
restriction, there is no limit to the number of simultaneous
single-qubit rotations that can be implemented throughout a Kane
quantum computer.

Interactions between neighboring qubits are governed by the
$J$-electrodes. A positive bias encourages greater overlap of the
donor electron wave functions leading to indirect coupling of
their associated nuclei. In analogy to the single-qubit case, this
allows multiple 2-qubit gates to be performed selectively between
arbitrary neighbors. A discussion of the electrode pulses required
to implement a \CNOT\ is given in the next section.

\section{Adiabatic \CNOT\ pulse profiles}
\label{CNOT:section:cnot}

Performing a \CNOT\ gate on an adiabatic Kane QC is an involved
process described in detail in \cite{Goan00}. Given the high field
(2T) and low temperature (100mK) operating conditions, we can
model the behavior of the system with a spin Hamiltonian. Only two
qubits are required to perform a \CNOT, so for the remainder of
the chapter we will restrict our attention to a computer with just
two qubits. The basic notation is shown in the right half of
Fig.~\ref{CNOT:figure:3q}. Furthermore, let
\mbox{$\sigma^{z}_{n1}\equiv\sigma^{z}\otimes I\otimes I\otimes
I$}, \mbox{$\sigma^{z}_{e1}\equiv I\otimes\sigma^{z}\otimes
I\otimes I$}, \mbox{$\sigma^{z}_{n2}\equiv I\otimes
I\otimes\sigma^{z}\otimes I$} and \mbox{$\sigma^{z}_{e2}\equiv
I\otimes I\otimes I\otimes\sigma^{z}$} where $I$ is the $2\times
2$ identity matrix, $\sigma^{z}$ is the usual Pauli matrix and
$\otimes$ denotes the matrix outer product. With these definitions
the meaning of terms such as $\sigma^{y}_{n2}$ and
$\vec{\sigma}_{e1}$ should be self evident.

Let $g_{n}$ be the g-factor for the phosphorus nucleus, $\mu_{n}$
the nuclear magneton and $\mu_{B}$ the Bohr magneton. The
Hamiltonian can be broken into three parts
\begin{equation}
\label{CNOT:eq:H} H = H_{\rm Z}+H_{\rm int}(t)+H_{\rm ac}(t).
\end{equation}
The Zeeman energy terms are contained in $H_{\rm Z}$
\begin{equation}
\label{CNOT:eq:HZ} H_{\rm
Z}=-g_{n}\mu_{n}B_{z}(\sigma^{z}_{n1}+\sigma^{z}_{n2})+\mu_{B}B_{z}(\sigma^{z}_{e1}+\sigma^{z}_{e2}).
\end{equation}
The contact hyperfine and exchange interaction terms, both of
which can be modified via the electrode potentials are
\begin{equation}
\label{CNOT:eq:Hint} H_{\rm
int}(t)=A_{1}(t)\vec{\sigma}_{n1}\cdot\vec{\sigma}_{e1}+A_{2}(t)\vec{\sigma}_{n2}\cdot\vec{\sigma}_{e2}+J(t)\vec{\sigma}_{e1}\cdot\vec{\sigma}_{e2},
\end{equation}
where $A_{i}(t) = 8\pi\mu_{B}g_{n}\mu_{n}|\Phi_{i}(0,t)|^{2}/3$,
$|\Phi_{i}(0,t)|$ is the magnitude of the wavefunction of donor
electron $i$ at phosphorous nucleus $i$ at time $t$, and $J(t)$
depends on the overlap of the two donor electron wave functions.
The dependance of these quantities on their associated electrode
voltages is a subject of ongoing research
\cite{Koil02,Lari00,Vali99,Kett03,Kett04,Well04}, though it
appears atomic precision placement of the phosphorus donors is
required due to strong oscillatory dependence of the exchange
interaction strength on the distance and direction of separation
of donors. For our purposes, it is sufficient to ignore the exact
voltage required and assume that the hyperfine and exchange
interaction energies $A_{i}$ and $J$ are directly manipulable.

The last part of the Hamiltonian contains the coupling to a
globally applied oscillating field of magnitude $B_{\rm ac}(t)$.
\begin{equation}
\label{CNOT:eq:Hac}
\begin{array}{rcl}
H_{\rm ac}(t) & = & B_{\rm ac}(t)\cos(\omega
t)[-g_{n}\mu_{n}(\sigma^{x}_{n1}+\sigma^{x}_{n2})
                                             +\mu_{B}(\sigma^{x}_{e1}+\sigma^{x}_{e2})] \\
          & + & B_{\rm ac}(t)\sin(\omega t)[-g_{n}\mu_{n}(\sigma^{y}_{n1}+\sigma^{y}_{n2})
                                             +\mu_{B}(\sigma^{y}_{e1}+\sigma^{y}_{e2})].
\end{array}
\end{equation}
Using the above definitions, only the quantities $A_{1}$, $J$ and
$B_{\rm ac}$ need to be manipulated to perform a \CNOT\ gate.

\begin{figure}[p]
\begin{center}
\includegraphics[width=12cm]{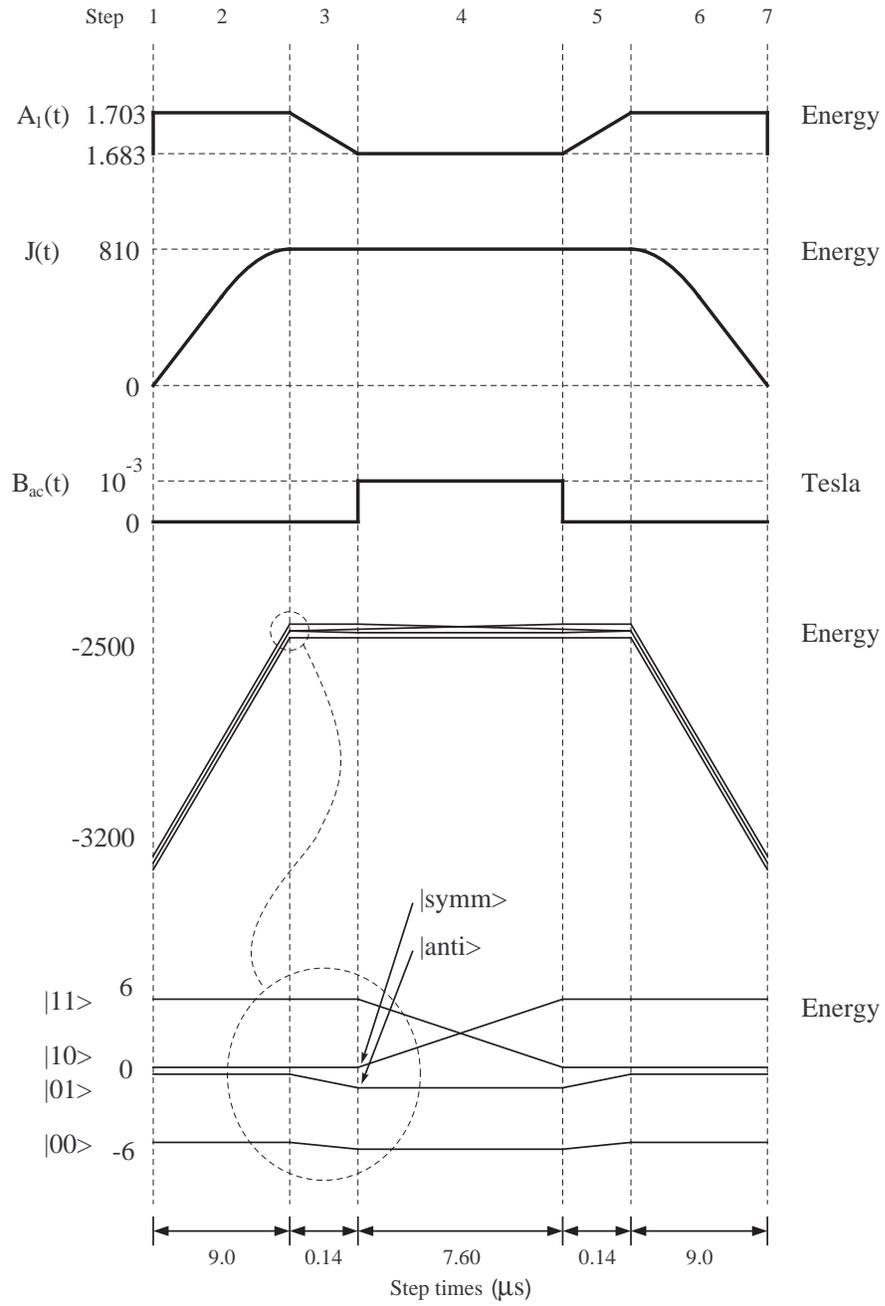}
\end{center}
\caption{Gate profiles and state energies during a \CNOT\ gate in
units of $g_{n}\mu_{n}B_{z}=7.1\times10^{-5}$meV.}
\label{CNOT:figure:pulses}
\end{figure}

For clarity, assume the computer is initially in one of the states
$|00\rangle$, $|01\rangle$, $|10\rangle$ or $|11\rangle$ and that
we wish to perform a \CNOT\ gate with qubit 1 (the left qubit) as
the control. The necessary profiles are shown in
Fig.~\ref{CNOT:figure:pulses}. Step one is to break the degeneracy
of the two qubits' energy levels to allow the control and target
qubits to be distinguished. To make qubit 1 the control, the value
of $A_{1}$ is increased (qubit 1 will be assumed to be the control
qubit for the remainder of the chapter).

Step two is to gradually apply a positive potential to the
$J$-electrode in order to force greater overlap of the donor
electron wave functions and hence greater (indirect) coupling of
the underlying nuclear qubits. The rate of this change is limited
so as to be adiabatic --- qubits initially in energy eigenstates
remain in energy eigenstates throughout this step.  This point
shall be discussed in more detail shortly.

Let $|$symm$\rangle$ and $|$anti$\rangle$ denote the standard
symmetric and antisymmetric superpositions of $|10\rangle$ and
$|01\rangle$. Step three is to adiabatically reduce the $A_{1}$
coupling back to its initial value once more. During this step,
anti-level-crossing behavior changes the input states
$|10\rangle\rightarrow|$symm$\rangle$ and
$|01\rangle\rightarrow|$anti$\rangle$.

Step four is the application of an oscillating field $B_{\rm ac}$
resonant with the $|$symm$\rangle\leftrightarrow|11\rangle$
transition. This oscillating field is maintained until these two
states have been interchanged. Steps five to seven are the time
reverse of steps one to three. Note that steps one and seven (the
increasing and decreasing of $A_{1}$) appear instantaneous in
Fig.~\ref{CNOT:figure:pulses} as the only limit to their speed is
that they be done in a time much greater than
$\hbar/0.01$eV$\sim0.1$ps where 0.01eV is the orbital excitation
energy of the donor electron.

In principle, the adiabatic steps should be performed slowly to
achieve maximum fidelity. In practice, slow gates are more
vulnerable to decoherence. To resolve this conflict, consider the
degree to which the evolution of a given $H(t)$ deviates from
perfect adiabaticity \cite{Bran89}
\begin{equation}
\Theta(t) \equiv {\rm Max_{a\neq b}} \left[ {\hbar | \langle
\psi_{\rm a}(t)| {\partial \over \partial t}(H(t)) | \psi_{\rm
b}(t)\rangle | \over (\langle \psi_{\rm a}(t)|H(t)| \psi_{\rm
a}(t)\rangle - \langle \psi_{\rm b} (t) |H(t)| \psi_{\rm
b}(t)\rangle )^2} \right]. \label{CNOT:eq:adiabaticcond}
\end{equation}
Ignoring decoherence for the moment, for high fidelity it is
necessary that $\Theta(t)\ll 1$. The states ${|\psi_a(t)\rangle}$
are the eigenstates of $H(t)$. To a certain extent, it is possible
to reduce $\Theta(t)$ without increasing the duration of a step by
optimizing the profiles of the adiabatically varying parameters in
$H(t)$. In the case of the adiabatic Kane \CNOT, this means
optimizing the profiles of $A_{1}(t)$ and $J(t)$.

\begin{figure}[h]
\begin{center}
\includegraphics[width=8cm]{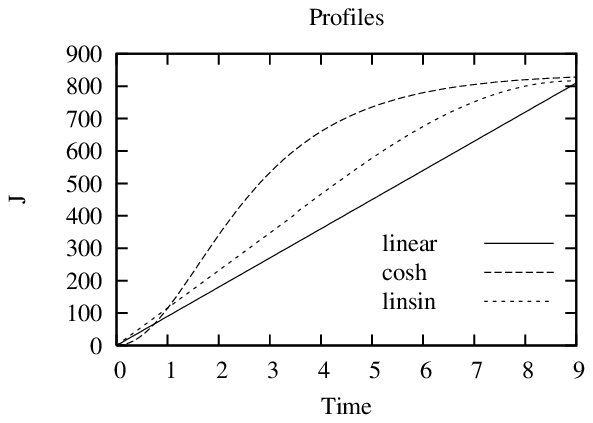}
\end{center}
\caption{Possible forms of the $J(t)$ profile for step 2 of the
adiabatic \CNOT\ gate. $J(t)$ is in units of
$g_{n}\mu_{n}B_{z}=7.1\times10^{-5}$meV.}
\label{CNOT:figure:profiles}
\end{figure}

\begin{figure}[h]
\begin{center}
\includegraphics[width=8cm]{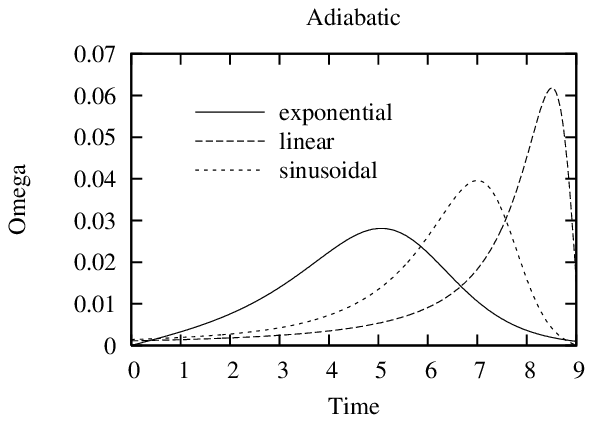}
\end{center}
\caption{The adiabatic measure $\Theta(t)$ for each $J(t)$
profile.} \label{CNOT:figure:adiabatic}
\end{figure}

Various profiles for the adiabatic steps in the \CNOT\ procedure
have been investigated in \cite{WelPhD}. In
Fig.~\ref{CNOT:figure:profiles}, we have plotted three possible
$J(t)$ profiles for step two of the \CNOT\ gate. The function
$\Theta(t)$ for each profile is shown in
Fig.~\ref{CNOT:figure:adiabatic}. Profile 1 is a simple linear
pulse. Profile 2 can be seen to be the best of the three and is
described by $J(t) = 810\alpha (1 - {\rm sech}(5 t /\tau ))$ where
$\tau=9\mu s$ is the duration of the pulse and $\alpha = 1.0366$
is a factor introduced to ensure that $J(\tau) = 810$. The third
profile
\begin{equation}
J(t)=\left\{
\begin{array}{ll}
\frac{J_{max}}{2}\frac{t(1+\pi /2)}{\tau}, & 0<t<\tau/(1+\pi /2)  \\
\frac{J_{max}}{2} \left[ 1+\sin \left(\frac{\pi}{2}\frac{t-\tau/(1+\pi /2)}{\tau/(1+2/\pi )} \right) \right], & \tau/(1+\pi /2)<t<\tau
\end{array}
\right.
\label{CNOT:eq:linsin}
\end{equation}
is a composite linear-sinusoidal profile that was used in the
calculations presented in this chapter due to numerical
difficulties in solving the Schr\"odinger equation for profile 2.
The advantage of the second two profiles over the linear one is
that they flatten out as $J$ approaches 810. At $J=816.65$, the
system undergoes a level crossing. To maintain adiabatic
evolution, $J(t)$ needs to change more slowly near this value.
Note that the reason it is desirable to make $J(t)$ so large is to
ensure that there is a large energy difference between
$|$symm$\rangle$ and $|$anti$\rangle$ during step 4 (the
application of $B_{\rm ac}$). This difference is given by
\begin{equation}
\delta E = 2 A^2({1 \over \mu_B B_z + g_n \mu_n B_z} - {1 \over \mu_B B_z + g_n \mu_n B_z-2J}).
\end{equation}
Without a large energy difference, the oscillating field $B_{\rm
ac}$ which is set to resonate with the transition $|$symm$\rangle
\leftrightarrow |11\rangle$ will also be very close to resonant
with $|$anti$\rangle \leftrightarrow |11\rangle$ causing a large
error during the operation of the \CNOT\ gate.  This source of
error can be further reduced by using a weaker $B_{\rm ac}$ at the
cost of slower gate operation.

Step 3 (the decreasing of $A_{1}$) could be performed without degrading
the overall fidelity of the gate in a time of less than a
micro-second with a linear pulse profile.

The above steps were simulated using an adaptive Runge-Kutta
routine to solve the density matrix form of the Schr\"odinger
equation
\begin{equation}
\label{CNOT:eq:rho dot}
\dot{\rho}(t)=\frac{1}{i\hbar}[H(t),\rho(t)]
\end{equation}
in the computational basis $|n_{1}e_{1}n_{2}e_{2}\rangle$. The
times used for each stage were as follows

\begin{center}
\begin{tabular}{|c|c|} \hline
\emph{stage} & \emph{duration ($\mu s$)} \\ \hline
2 & 9.0000 \\
3 & 0.1400 \\
4 & 7.5989 \\
5 & 9.0000 \\
6 & 0.1400 \\ \hline
\end{tabular}
\end{center}

Note that the precision of the duration of stage 4 is required as
the oscillating field $B_{\rm ac}$ induces the states $|11\rangle$
and $|$symm$\rangle$ to swap smoothly back and forth. The duration
7.5989$\mu s$ is the time required for one swap at $B_{\rm
ac}=10^{-3}$T.

The other step times were obtained by first setting them to
arbitrary values ($\sim$5$\mu s$) and increasing them until the
gate fidelity ceased to increase. The step times were then
decreased one by one until the fidelity started to decrease. As
such, the above times are the minimum time in which the maximum
fidelity can be achieved. This maximum fidelity was found to be
$5\times 10^{-5}$ for all computation basis states.

\section{Intrinsic dephasing and fidelity}
\label{CNOT:section:deph}

In this chapter and Chapter~\ref{readout}, dephasing is modelled
as exponential decay of the off diagonal components of the density
matrix. While a large variety of more detailed dephasing models
exist \cite{Baue02,Burk99,Kimm02,Mozy02,Sous03,Thor02}, the chosen
method is consistent with the observed experimental behavior of
dephasing in solid-state systems \cite{Hu01}. The donor electrons
and phosphorous nuclei are assumed to dephase at independent
rates. With the inclusion of dephasing terms,
Eq.~(\ref{CNOT:eq:rho dot}) becomes
\begin{eqnarray}
\label{CNOT:eq:rho dot dephase}
\dot{\rho} & = & \frac{1}{i\hbar}[H,\rho] \nonumber \\
           &   & -\Gamma_{e}[\sigma^{z}_{e_{1}},[\sigma^{z}_{e_{1}},\rho]]-\Gamma_{e}[\sigma^{z}_{e_{2}},[\sigma^{z}_{e_{2}},\rho]] \nonumber \\
           &   & -\Gamma_{n}[\sigma^{z}_{n_{1}},[\sigma^{z}_{n_{1}},\rho]]-\Gamma_{n}[\sigma^{z}_{n_{2}},[\sigma^{z}_{e_{2}},\rho]].
\end{eqnarray}
To understand the effect of each double commutator, it is instructive to
consider the following simple mathematical example :
\begin{eqnarray}
\label{CNOT:eq:simp dephase}
\dot{M} & = & -\Gamma[\sigma^{z},[\sigma^{z},M]] \nonumber \\
\left( \begin{array}{cc}
\dot{m}_{11} & \dot{m}_{12}  \\
\dot{m}_{21} & \dot{m}_{22}
\end{array} \right) & = &
\left( \begin{array}{cc}
0 & -4\Gamma m_{12} \\
-4\Gamma m_{21} & 0
\end{array} \right) \nonumber \\
\left( \begin{array}{cc}
m_{11}(t) & m_{12}(t)  \\
m_{21}(t) & m_{22}(t)
\end{array} \right) & = &
\left( \begin{array}{cc}
m_{11}(0) & m_{12}(0)e^{-4\Gamma t} \\
m_{21}(0)e^{-4\Gamma t} & m_{22}(0)
\end{array} \right).
\end{eqnarray}
Thus each double commutator in Eq.~(\ref{CNOT:eq:rho dot dephase})
exponentially decays its associated off diagonal elements with
characteristic time $\tau_{e}=1/4\Gamma_{e}$ or
$\tau_{n}=1/4\Gamma_{n}$.

\begin{figure}[p]
\begin{center}
\includegraphics[width=9cm]{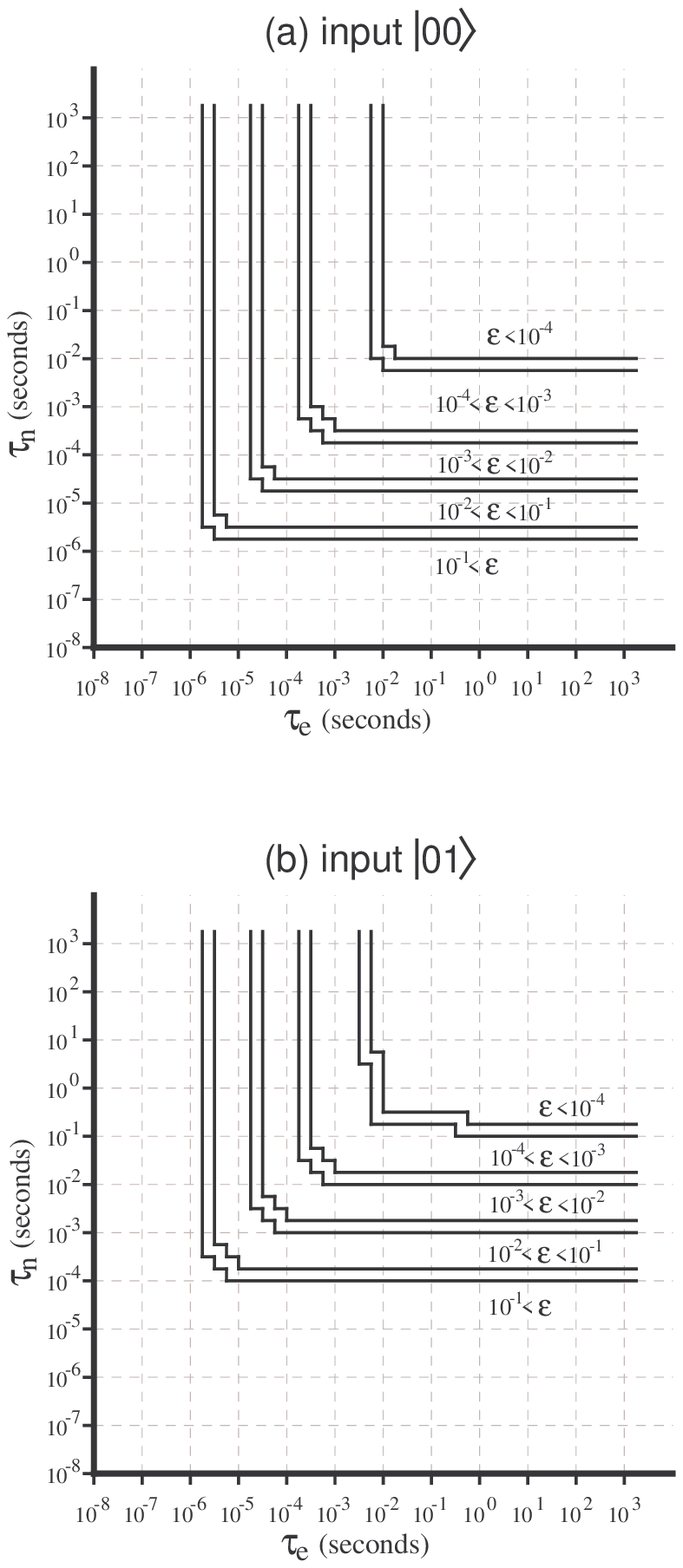}
\end{center}
\caption{Probability of error $\varepsilon$ during a \CNOT\ gate
as a function of $\tau_{e}$ and $\tau_{n}$ for input state (a)
$|00\rangle$ and (b) $|01\rangle$. The first qubit is the
control.} \label{CNOT:figure:00_01}
\end{figure}

\begin{figure}[p]
\begin{center}
\includegraphics[width=9cm]{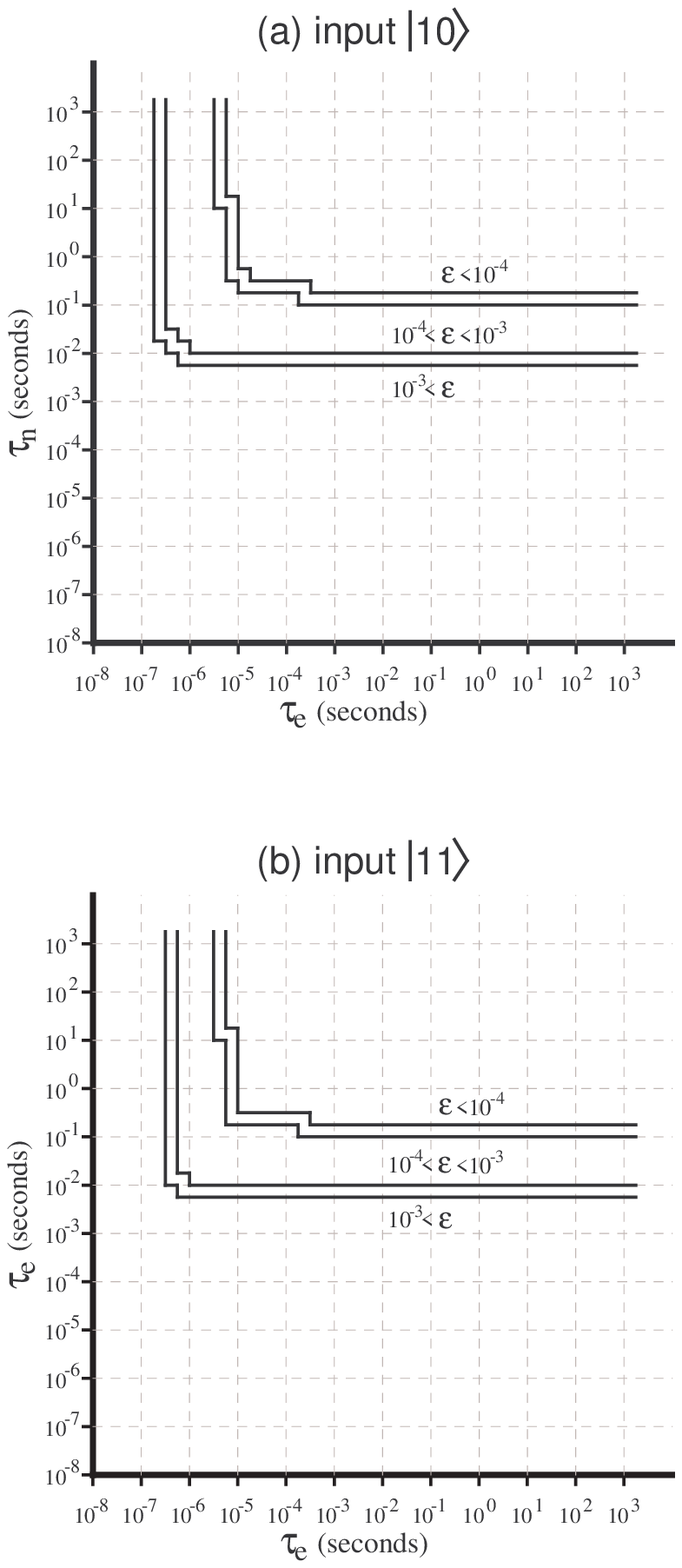}
\end{center}
\caption{Probability of error $\varepsilon$ during a \CNOT\ gate
as a function of $\tau_{e}$ and $\tau_{n}$ for input state (a)
$|10\rangle$ and (b) $|11\rangle$. The first qubit is the
control.} \label{CNOT:figure:10_11}
\end{figure}

\begin{figure}[h]
\begin{center}
\includegraphics[width=10cm]{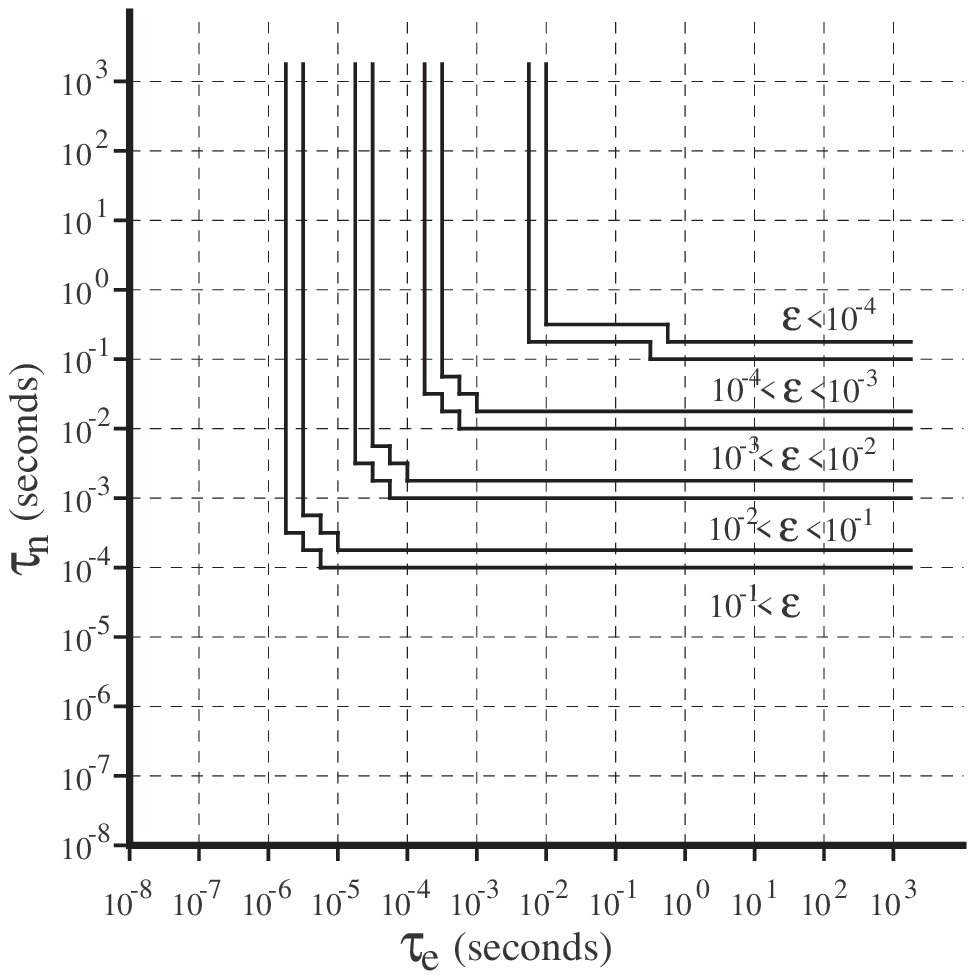}
\end{center}
\caption{The worst case probability of error $\varepsilon$ during
a \CNOT\ gate as a function of $\tau_{e}$ and $\tau_{n}$ for all
input states.} \label{CNOT:figure:all}
\end{figure}

For each initial state $|00\rangle$, $|01\rangle$, $|10\rangle$
and $|11\rangle$, Eq.~(\ref{CNOT:eq:rho dot dephase}) was solved
for a range of values of $\tau_{e}$ and $\tau_{n}$ using the pulse
profiles described in Section~\ref{CNOT:section:cnot} allowing a
contour plot of the gate error versus $\tau_{e}$ and $\tau_{n}$ to
be constructed
(Figs~\ref{CNOT:figure:00_01}--\ref{CNOT:figure:10_11}). Note that
each contour is a double line as each run of the simulation
required considerable computational time and the data available
does not allow finer delineation of exactly where each contour is.
The worst case error of all input states as a function of
$\tau_{e}$ and $\tau_{n}$ is shown in Fig.~\ref{CNOT:figure:all}.

\section{Conclusion}
\label{CNOT:section:conc}

Fig.~\ref{CNOT:figure:all} suggests that it would be acceptable
for the dephasing times of the phosphorus donor electron and
nuclei to be 10ms and 0.5s respectively if a \CNOT\ reliability of
$10^{-4}$ was desired. Given that the current best estimate of the
donor electron dephasing time is 60ms \cite{Tyry03}, and assuming
that the nuclear dephasing time is at least a factor of 80 longer
again, as for the case of the relaxation times, it would be
acceptable for the presence of the silicon dioxide barrier, gate
electrodes, and other control structures to reduce the dephasing
times by a factor of 6 without impacting on the desired
reliability of the gate.

Since the publication of this work \cite{Fowl03a}, simpler and
faster methods of implementing gates in the Kane architecture have
been devised \cite{Hill03}, though they are slightly more
vulnerable to decoherence \cite{Hill04b}. An interesting avenue of
further work would be to similarly analyze 2-qubit gates in the
context of three electron spin encoded qubits which enable
arbitrary computations to be performed utilizing the exchange
interaction only \cite{DiVi00b}, thereby eliminating the need for
oscillating magnetic fields and resulting in much faster gates.

\chapter{Adiabatic Kane single-spin readout}
\label{readout}

Meaningful quantum computation can not occur without qubit
readout. In this chapter, we assess the viability of the adiabatic
Kane single-spin readout proposal \cite{Kane98} when dephasing and
other effects such as finite exchange coupling control are taken
into account. We find that there are serious barriers to the
implementation of the proposal, and briefly review alternatives
under active investigation.

The evolution of the hyperfine and exchange interaction strengths
required to implement readout is reviewed in
Section~\ref{readout:section:pulse}. The performance of the scheme
is described in Section~\ref{readout:section:performance}.
Section~\ref{readout:section:conc} summaries our results and
points to alternative approaches to readout.

\section{Adiabatic readout pulse profiles}
\label{readout:section:pulse}

\begin{figure}
\begin{center}
\includegraphics[width=12cm]{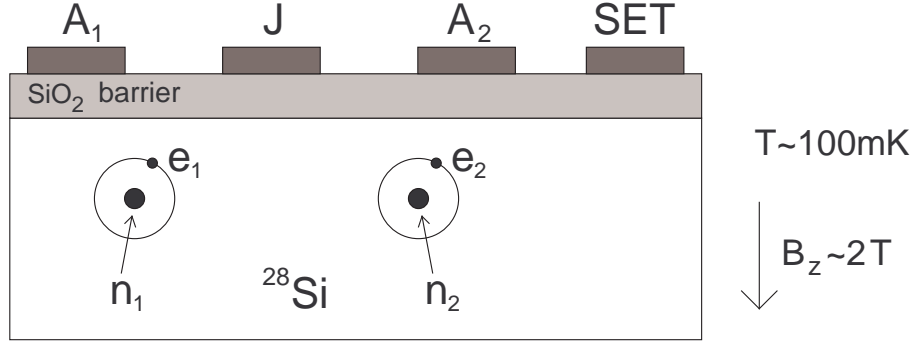}
\end{center}
\caption{Geometry of adiabatic Kane readout.}
\label{readout:figure:readout}
\end{figure}

\begin{figure}
\begin{center}
\includegraphics[width=12cm]{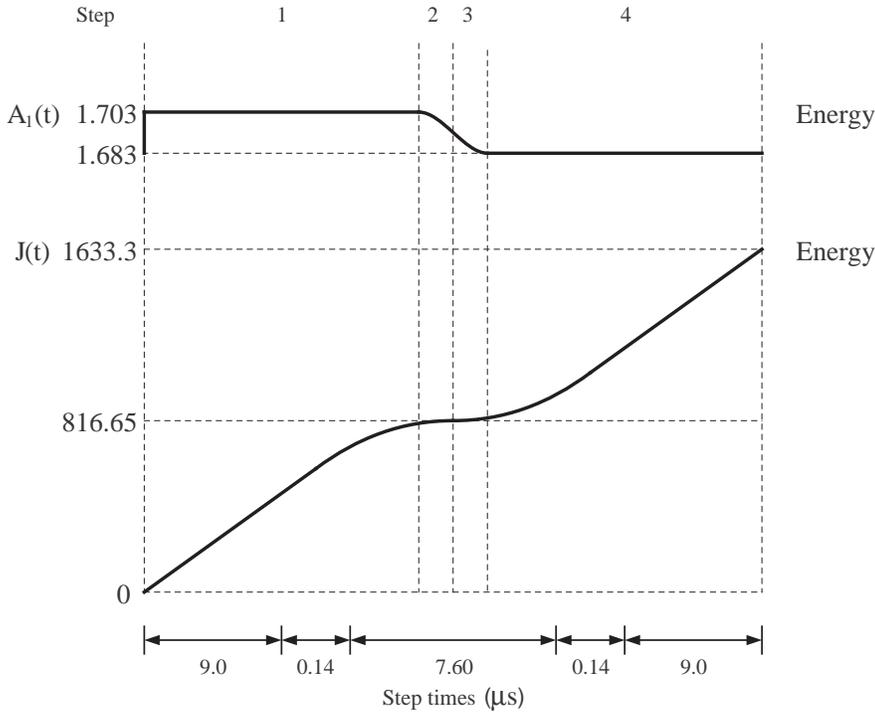}
\end{center}
\caption{Evolution of the hyperfine and exchange interaction
strengths required to convert nuclear spin information into
electron spin information, as shown in
Eqs.~(\ref{readout:eq:conv1}--\ref{readout:eq:conv4}).}
\label{readout:figure:readout_pulses}
\end{figure}

The geometry of the adiabatic Kane readout proposal is shown in
Fig.~\ref{readout:figure:readout}. The basic idea is to raise
$A_{1}$ to distinguish the qubits, and apply appropriate voltages
to induce the evolution of the hyperfine and exchange interaction
strengths as shown in Fig.~\ref{readout:figure:readout_pulses}.
This evolution passes through a level crossing resulting in the
conversion of states \cite{Goan00}
\begin{eqnarray}
\label{readout:eq:conv1} \left|\downarrow\downarrow\right\rangle|11\rangle & \rightarrow & \left|\downarrow\downarrow\right\rangle|11\rangle, \\
\label{readout:eq:conv2} \left|\downarrow\downarrow\right\rangle|10\rangle & \rightarrow & \left|\downarrow\downarrow\right\rangle|s_{n}\rangle, \\
\label{readout:eq:conv3} \left|\downarrow\downarrow\right\rangle|01\rangle & \rightarrow & |a_{e}\rangle|11\rangle, \\
\label{readout:eq:conv4} \left|\downarrow\downarrow\right\rangle|00\rangle & \rightarrow &
|a_{e}\rangle|a_{n}\rangle,
\end{eqnarray}
where $\left|\downarrow\right\rangle$ denotes a spin-down
electron, $|a_{e}\rangle$ denotes the antisymmetric superposition
of the two electrons, and similarly for $|s_{n}\rangle$ and
$|a_{n}\rangle$. Note that if qubit 1 is in state $|1\rangle$
($|0\rangle$) the final electron state will be
$\left|\downarrow\downarrow\right\rangle$ ($|a_{e}\rangle$).

By converting the nuclear spin information into electron spin
information in this manner, in principle we can apply a potential
difference to the $A_{1}$ and $A_{2}$-electrodes and, by virtue of
the Pauli exclusion principle, use the SET (single electron
transistor) to observe tunnelling of electron 1 onto donor 2 if
and only if the nuclear spin was $|0\rangle$.

% readout_pulses figure was here

\section{Readout performance}
\label{readout:section:performance}

\begin{figure}
\begin{center}
\includegraphics[width=9cm]{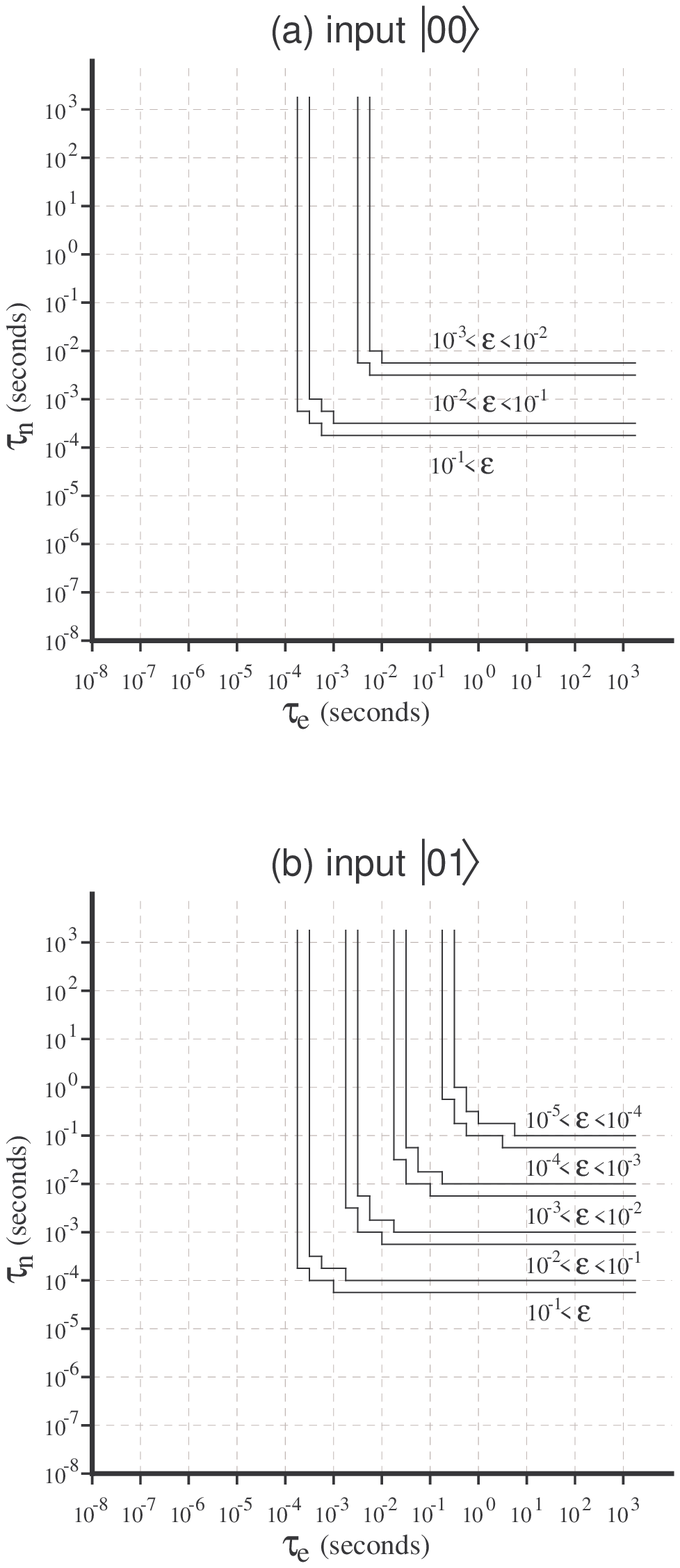}
\end{center}
\caption{Probability of error $\varepsilon$ during readout state
preparation as a function of $\tau_{e}$ and $\tau_{n}$ for input
state (a) $|00\rangle$ and (b) $|01\rangle$.}
\label{readout:figure:Read00_01}
\end{figure}

\begin{figure}
\begin{center}
\includegraphics[width=9cm]{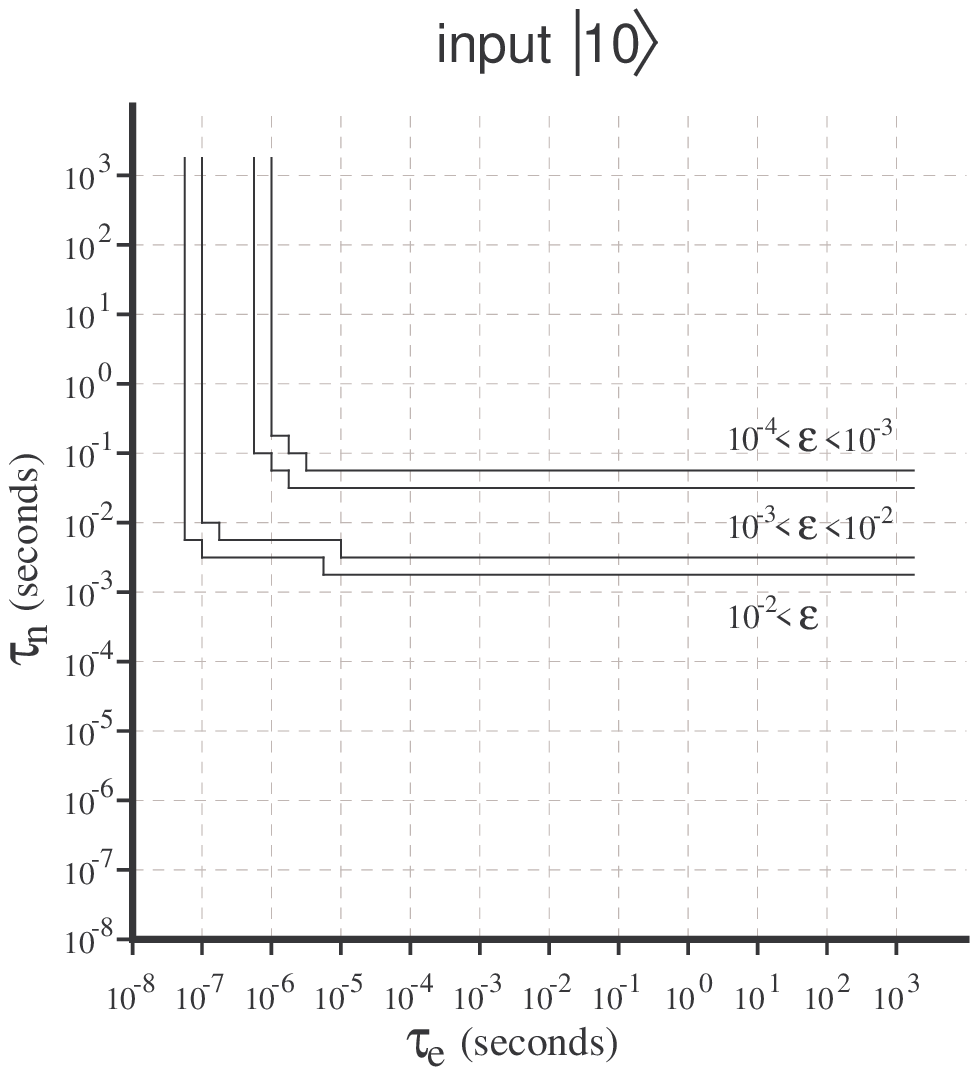}
\end{center}
\caption{Probability of error $\varepsilon$ during readout state
preparation as a function of $\tau_{e}$ and $\tau_{n}$ for input
state $|10\rangle$.} \label{readout:figure:Read10}
\end{figure}

In exactly the same manner as Chapter~\ref{CNOT}, the performance
of the adiabatic state conversion stage of readout was simulated
with variable nuclear and electronic dephasing times $\tau_{n}$
and $\tau_{e}$. The results of these simulations are shown in
Figs.~\ref{readout:figure:Read00_01}--\ref{readout:figure:Read10}
and show strong dependence on the initial computational basis
state. Indeed, we found the basis state $|11\rangle$ to be immune
to
dephasing, and, since it is far from any level crossings,
perfectly preserved during the adiabatic evolution. Consequently,
no figure has been included for $|11\rangle$.

At the other extreme, even in the absence of dephasing, the
nuclear state $|00\rangle$ is converted into electron state
$|a_{e}\rangle$ with high probability of error $\epsilon\sim
10^{-3}$. With dephasing, state $|00\rangle$ also embodies the
worst-case fidelity of the readout operation. For realistic
dephasing times, this suggests a fidelity of state preparation
(before tunnelling an actual readout) of $10^{-2}$.

While this low fidelity is not ideal, it could probably be
tolerated.  There are, however, more serious concerns.  Firstly,
even with donors spaced 20nm apart and 1V applied to the
$J$-electrode, simulations suggest that the exchange coupling
could be as weak as 0.03meV, or 420 in units of
$g_{n}\mu_{n}B_{z}$ \cite{Well03}. This is insufficient to access
the desired level crossing.

Furthermore, the $D^{-}$ state of two electrons on one donor has a
very weak binding energy of 1.7meV. Calculations suggest that a DC
field designed to encourage electron 1 to tunnel onto donor 2
would also be sufficient to ionize the $D^{-}$ state
\cite{Holl04}.

\section{Conclusion}
\label{readout:section:conc}

Current simulations suggests that there are serious concerns with
regard to the accessibility, fidelity, and stability of the states
used in the adiabatic Kane readout proposal. Modifications to the
scheme such as using an rf field to Rabi flip electron 1 dependent
on nuclear spin 1 instead of the adiabatic process and using
resonant fields to induce Rabi oscillations of electron 1 onto
donor 2 provided the transition is permitted by spin have been
suggested as a way around these problems \cite{Holl04}. A
completely new scheme involving three donors, one ionized, and
avoiding the need for double occupancy has also been proposed
\cite{Gree04}. Further theoretical and experimental work is
required to develop these ideas.

\chapter{Implementing arbitrary 2-qubit gates}
\label{canonical}

Efficiently implementing arbitrary 2-qubit gates using a given
physical architecture is an important part of practical quantum
computation. In this chapter, we combine techniques from
Refs~\cite{Makh02,Krau01,Zhan03,Chil03} to present an efficient,
but not necessarily time-optimal implementation of an arbitrary
2-qubit gate using at most three periods of free evolution of a
certain class of 2-qubit system interspersed with at most eight
single-qubit gates. The construction requires that the
architecture be able to isolate qubits from one another. This
chapter should be considered clarification and review rather than
original research. A number of examples of gates built using the
method detailed in this chapter can be found in
Ref.~\cite{Hill03}.

In Section~\ref{canonical:section:background}, prior work on
implementing quantum gates is reviewed.
Section~\ref{canonical:section:terminology} contains essential
terminology and notation. Given an arbitrary 2-qubit gate $G$,
Section~\ref{canonical:section:decomposition} details how to
construct a canonical decomposition comprised of four single-qubit
gates and one purely non-local 2-qubit gate.
Section~\ref{canonical:section:unique} describes how to make the
canonical decomposition unique.
Section~\ref{canonical:section:Zhang} uses the unique canonical
decompositions of $G$ and a given 2-qubit evolution operator
$U(t)$ to obtain a physical implementation of $G$ using at most
three periods of free evolution of $U(t)$ and eight single-qubit
gates. Section~\ref{canonical:section:conc} concludes with a
discussion of the implications of canonically decomposed gates,
and describes further work.

\section{Background}
\label{canonical:section:background}

Implementing arbitrary gates given an arbitrary architecture is a
nontrivial task. General methods have been developed for two
special classes of physical architecture --- those in which
single-qubit gates can be approximated as instantaneous due to
their speed relative to multiple qubit interactions, and those
with the ability to isolate qubits from one another. When
single-qubit gates are much faster than multiple-qubit
interactions, time-optimal implementations can be found
\cite{Khan01,Hamm02,Chil03}. In the Kane architecture, single
qubit gates cannot be approximated as instantaneous, but qubits
can be isolated from one another \cite{Well04}, so we focus on the
efficient, but not necessarily time-optimal method described in
\cite{Zhan03}.

\begin{figure}
\begin{center}
\includegraphics[width=12cm]{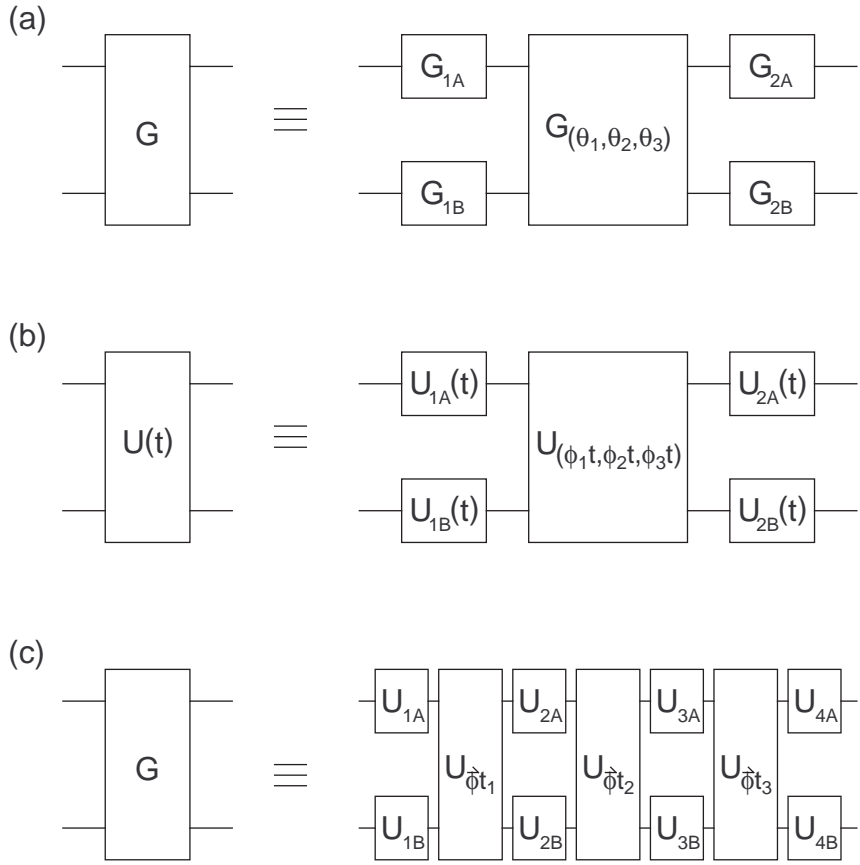}
\end{center}
\caption{(a) Circuit equivalent to an arbitrary gate constructed
via the canonical decomposition. (b) Similar equivalent circuit
which exists for a restricted class of 2-qubit evolution
operators. (c) Arbitrary gate expressed as at most three periods
evolution of the 2-qubit evolution operator and eight single-qubit
gates.} \label{canonical:figure:canonical}
\end{figure}

In both cases, the canonical decomposition
\cite{Makh02,Krau01,Zhan03,Chil03} forms the starting point of the
construction. Given an arbitrary 2-qubit gate $G$, the canonical
decomposition is a circuit, equivalent up to global phase,
involving up to four single-qubit gates $G_{1A}, G_{1B}, G_{2A},
G_{2B}$ and a purely non-local gate
$G_{\vec{\theta}}=e^{i(\theta_{1}X\otimes X+\theta_{2}Y\otimes
Y+\theta_{3}Z\otimes Z)}$
(Fig.~\ref{canonical:figure:canonical}a). Note that this
decomposition is only unique if certain restrictions are placed on
$\vec{\theta}$. A complete discussion of the many symmetries of
the canonical decomposition is given in \cite{Zhan03}.

Not all 2-qubit time evolution operators $U(t)$ admit such a
simple decomposition. The most general time-dependent
decomposition has a 2-qubit term
\begin{equation}
U_{(\phi_{1}(t),\phi_{2}(t),\phi_{3}(t))}=e^{i(\phi_{1}(t)X\otimes
X+\phi_{2}(t)Y\otimes Y+\phi_{3}(t)Z\otimes Z)}.
\end{equation}
This case is dealt with in \cite{Zhan03}, but is not required
here. Instead, we focus on time evolution operators that can be
decomposed as $U_{\vec{\phi}t}$
(Fig.~\ref{canonical:figure:canonical}b). Comparing $\vec{\theta}$
and $\vec{\phi}$, $G$ can be expressed as at most three
applications of $U_{\vec{\phi}t}$ interspersed with at most eight
single-qubit gates (Fig.~\ref{canonical:figure:canonical}c).

\section{Terminology and notation}
\label{canonical:section:terminology}

This section contains the terminology and notation used in this
chapter. Unless otherwise specified, all quantities in this
chapter are expressed in the computational basis
$\{|00\rangle,|01\rangle,|10\rangle,|11\rangle\}$. Heavy use will
also be made of the magic basis \cite{Hill97}
\begin{eqnarray}
|\Phi_{1}\rangle & = & \frac{1}{\sqrt{2}}(|00\rangle + |11\rangle) \nonumber \\
|\Phi_{2}\rangle & = & \frac{-i}{\sqrt{2}}(|00\rangle - |11\rangle) \nonumber \\
|\Phi_{3}\rangle & = & \frac{1}{\sqrt{2}}(|01\rangle - |10\rangle) \nonumber \\
|\Phi_{4}\rangle & = & \frac{-i}{\sqrt{2}}(|01\rangle + |10\rangle).
\end{eqnarray}
The transformation matrix from the magic basis to the
computational basis is
\begin{equation}\label{canonical:eq:magtocomp}
Q = \frac{1}{\sqrt{2}}\left(
\begin{array}{cccc}
1 & -i & 0 & 0 \\
0 & 0 & 1 & -i \\
0 & 0 & -1 & -i \\
1 & i & 0 & 0
\end{array}\right).
\end{equation}
States $|\Psi\rangle$ and matrices $G$ expressed in the magic
basis will be denoted by
$|\tilde{\Psi}\rangle=Q^{\dag}|\Psi\rangle$ and
$\tilde{G}=Q^{\dag}GQ$ respectively.

Given an arbitrary 2-qubit gate $G$, a canonical decomposition is
a set of four single-qubit gates $G_{1A}, G_{1B}, G_{2A}, G_{2B}$
and a purely non-local gate
$G_{\vec{\theta}}=e^{i(\theta_{1}X\otimes X+\theta_{2}Y\otimes
Y+\theta_{3}Z\otimes Z)}$ such that
\begin{equation}
G \cong (G_{2A}\otimes G_{2B})G_{\vec{\theta}}(G_{1A}\otimes
G_{1B}),
\end{equation}
where $\cong$ denotes equality up to global phase. Two sets of
non-local parameters $\vec{\theta}$, $\vec{\phi}$ are called
locally equivalent if there exist matrices $U_{1}, U_{2}, U_{3},
U_{4} \in U(2)$ such that $G_{\vec{\phi}} = (U_{3}\otimes
U_{4})G_{\vec{\theta}}(U_{1}\otimes U_{2})$. The canonical
decomposition can be made unique by requiring
$\theta_{1}\geq\theta_{2}$, $\pi/2-\theta_{1}\geq\theta_{2}$,
$\theta_{2}\geq\theta_{3}$, $\theta_{3}\geq 0$, and
$\pi/2-\theta_{1}\geq \theta_{1}$ if $\theta_{3}=0$.

\section{Constructing a canonical decomposition}
\label{canonical:section:decomposition}

In this section, we take a 2-qubit gate $G$ and construct a
canonical decomposition $G \cong (G_{2A}\otimes
G_{2B})G_{\vec{\theta}}(G_{1A}\otimes G_{1B})$. Firstly, let
$G_{S}=e^{-i\delta}G$, $\delta=-\arg(\det(G))/4$, so that
$\det(G_{S})=1$. Since $\arg$ only determines values up to
$2n\pi$, $\delta$ is only determined up to $n\pi/2$. Arbitrarily
choose $\delta \in (-\pi/4,\pi/4]$. As we shall see,
$\det(G_{S})=1$ is required to enable the construction of
$G_{\vec{\theta}}$.

Next, we determine the eigensystem of
$\tilde{G}_{S}^{T}\tilde{G}_{S}$ to obtain eigenvalues
$\{e^{2i\epsilon_{k}}\}$ and eigenvectors
$\{|\tilde{\Psi}_{k}\rangle\}$. Note that while
$\tilde{G}_{S}^{T}\tilde{G}_{S}$ is symmetric and hence possesses
a real, orthonormal eigenbasis, standard analytic and numerical
methods of obtaining the eigenvectors in general only yield a
linearly independent set. Hence, for the moment, we only assume
that $\{|\tilde{\Psi}_{k}\rangle\}$ is linearly independent.

In Ref.~\cite{Makh02} it was shown that
\begin{equation}
(\tilde{G}_{S}^{T}\tilde{G}_{S})^{*}|\tilde{\Psi}_{k}\rangle =
(\tilde{G}_{S}^{T}\tilde{G}_{S})^{-1}|\tilde{\Psi}_{k}\rangle =
e^{-2i\epsilon_{k}}|\tilde{\Psi}_{k}\rangle,
\end{equation}
implying
$\tilde{G}_{S}^{T}\tilde{G}_{S}|\tilde{\Psi}_{k}\rangle^{*} =
e^{2i\epsilon_{k}}|\tilde{\Psi}_{k}\rangle^{*}$. By considering
states $(|\tilde{\Psi}_{k}\rangle \pm
|\tilde{\Psi}_{k}\rangle^{*})/2$ it can be seen that both the real
and imaginary parts of $\{|\tilde{\Psi}_{k}\rangle\}$ are
themselves eigenvectors with the same eigenvalue. It can be shown
that a subset of $\{{\rm Re}(|\tilde{\Psi}_{k}\rangle),{\rm
Im}(|\tilde{\Psi}_{k}\rangle)\}$ always exists that is linearly
independent and forms a real eigenbasis. Using the Gram-Schmidt
procedure then gives us a real, orthonormal eigenbasis which we
redefine $\{|\tilde{\Psi}_{k}\rangle\}$ to equal. Let
$\tilde{O}_{1}$ be the matrix with rows equal to
$|\tilde{\Psi}_{k}\rangle$. Note that $\tilde{O}_{1}$ is
orthogonal and hence $|\det(\tilde{O}_{1})|=1$. If
$\det(\tilde{O}_{1})=-1$, redefine $|\tilde{\Psi}_{1}\rangle$ to
equal $-|\tilde{\Psi}_{1}\rangle$ so that $\det(\tilde{O}_{1})=1$.
Let
\begin{equation}
\label{canonical:eq:G2eps}
\tilde{G}_{2\vec{\epsilon}} =
\left(
\begin{array}{cccc}
e^{2i\epsilon_{1}} & 0 & 0 & 0 \\
0 & e^{2i\epsilon_{2}} & 0 & 0 \\
0 & 0 & e^{2i\epsilon_{3}} & 0 \\
0 & 0 & 0 & e^{2i\epsilon_{4}}
\end{array}
\right).
\end{equation}
Note that as defined $\tilde{O}_{1}$ diagonalizes
$\tilde{G}_{S}^{T}\tilde{G}_{S}$ meaning
$\tilde{G}_{S}^{T}\tilde{G}_{S} =
\tilde{O}_{1}^{T}G_{2\vec{\epsilon}}\tilde{O}_{1}$.

Compute $\epsilon_{k} \in (-\pi/2,\pi/2]$ from the eigenvalues
$e^{2i\epsilon_{k}}$ of $\tilde{G}_{S}^{T}\tilde{G}_{S}$. Since
$\det(\tilde{G}_{S}^{T}\tilde{G}_{S})=1$, $\sum_{k} \epsilon_{k} =
n\pi$. If $n>0$, subtract $\pi$ from the $n$ largest values of
$\epsilon_{k}$ as we shall later use $\sum\epsilon_{k}=0$ to
eliminate $\epsilon_{3}$ and allow $\theta_{1}$, $\theta_{2}$,
$\theta_{3}$ to be expressed in closed form in terms of
$\epsilon_{1}$, $\epsilon_{2}$, $\epsilon_{4}$. Note that the
eigenvalues $e^{2i\epsilon_{k}}$ are not changed by this
convenient redefinition. Similarly, if $n<0$, add $\pi$ to the $n$
most negative values of $\epsilon_{k}$. Let
$\tilde{O}_{2}=\tilde{G}_{S}\tilde{O}_{1}^{T}\tilde{G}_{\vec{\epsilon}}^{*}$.
It can be directly verified that $\tilde{O}_{2}$ is special
orthogonal. We now have
$\tilde{G}_{S}=\tilde{O}_{2}\tilde{G}_{\vec{\epsilon}}\tilde{O}_{1}$,
and are close to obtaining a canonical decomposition.

In \cite{Zhan03} it was shown that $SO(4)=Q^{\dag}SU(2)\otimes
SU(2)Q$, where $Q$ is the transformation matrix of
Eq.~(\ref{canonical:eq:magtocomp}). The generic form of an element
of $SU(2)\otimes SU(2)$ is
\begin{equation}
\label{canonical:eq:SU2SU2}
\left(
\begin{array}{cccc}
a\alpha & a\beta & b\alpha & b\beta \\
-a\beta^{*} & a\alpha^{*} & -b\beta^{*} & b\alpha^{*} \\
-b^{*}\alpha & -b^{*}\beta & a^{*}\alpha & a^{*}\beta \\
b^{*}\beta^{*} & -b^{*}\alpha^{*} & -a^{*}\beta^{*} & a^{*}\alpha^{*}
\end{array}
\right)
=
\left(
\begin{array}{cc}
a & b \\
-b^{*} & a^{*} \\
\end{array}
\right)
\otimes
\left(
\begin{array}{cc}
\alpha & \beta \\
-\beta^{*} & \alpha^{*} \\
\end{array}
\right),
\end{equation}
where $a,b,\alpha,\beta \in \mathds{C}$, $|a|^{2}+|b|^{2}=1$, and
$|\alpha|^{2}+|\beta|^{2}=1$. After calculating $O_{1} =
Q\tilde{O}_{1}Q^{\dag} \in SU(2)\otimes SU(2)$, two matrices
$G_{1A},G_{1B} \in SU(2)$ such that $O_{1}=G_{1A}\otimes G_{1B}$
can be readily constructed from Eq.~(\ref{canonical:eq:SU2SU2}).
For example, $a^{2}=(a\alpha)(a\alpha^{*}) -
(a\beta)(-a\beta^{*})$. Matrices $G_{1A}$, $G_{1B}$ are unique up
to a possible mutual sign flip. Matrices $G_{2A},G_{2B} \in SU(2)$
such that $O_{2}=G_{2A}\otimes G_{2B}$ can similarly be
constructed.

All that remains to do is convert $\tilde{G}_{\vec{\epsilon}}$ into
$\tilde{G}_{\vec{\theta}} = Q^{\dag}G_{\vec{\theta}}Q$, which is
\begin{equation}
\left(
\begin{array}{cccc}
e^{i(\theta_{1}-\theta_{2}+\theta_{3})} & 0 & 0 & 0 \\
0 & e^{i(-\theta_{1}+\theta_{2}+\theta_{3})} & 0 & 0 \\
0 & 0 & e^{-i(\theta_{1}+\theta_{2}+\theta_{3})} & 0 \\
0 & 0 & 0 & e^{i(\theta_{1}+\theta_{2}-\theta_{3})}
\end{array}
\right).
\label{canonical:eq:magicGtheta}
\end{equation}
We therefore need
\begin{eqnarray}
\label{canonical:eq:theta_epsilon1}
\epsilon_{1}&=&\theta_{1}-\theta_{2}+\theta_{3},
\\
\label{canonical:eq:theta_epsilon2}
\epsilon_{2}&=&-\theta_{1}+\theta_{2}+\theta_{3},
\\
\label{canonical:eq:theta_epsilon3}
\epsilon_{3}&=&-\theta_{1}-\theta_{2}-\theta_{3},
\\
\label{canonical:eq:theta_epsilon4}
\epsilon_{4}&=&\theta_{1}+\theta_{2}-\theta_{3}.
\end{eqnarray}
Since we have ensured that $\sum_{k}\epsilon_{k}=0$,
Eqs~(\ref{canonical:eq:theta_epsilon1}--\ref{canonical:eq:theta_epsilon4})
are consistent and can be inverted to give
\begin{eqnarray}
\theta_{1}&=&(\epsilon_{1}+\epsilon_{4})/2 \\
\theta_{2}&=&(\epsilon_{2}+\epsilon_{4})/2 \\
\theta_{3}&=&(\epsilon_{1}+\epsilon_{2})/2.
\end{eqnarray}
We now have a canonical decomposition $G\cong G_{S} =
(G_{2A}\otimes G_{2B})G_{\vec{\theta}}(G_{1A}\otimes G_{1B})$.

\section{Making a canonical decomposition unique}
\label{canonical:section:unique}

In this section, we explain how to modify $G_{1A}$, $G_{2A}$,
$G_{2A}$, $G_{2B}$ to make $\vec{\theta}$ unique. As it stands,
$\vec{\theta} \in (-\pi,\pi]^{3}$. By direct multiplication it can
be seen that
\begin{eqnarray}
G_{(\theta_{1}\pm\pi/2,\theta_{2},\theta_{3})}&=&\pm i(X\otimes
X)G_{(\theta_{1},\theta_{2},\theta_{3})}
\label{canonical:eq:pio2shift1}\\
G_{(\theta_{1},\theta_{2}\pm\pi/2,\theta_{3})}&=&\pm i(Y\otimes
Y)G_{(\theta_{1},\theta_{2},\theta_{3})}
\label{canonical:eq:pio2shift2}\\
G_{(\theta_{1},\theta_{2},\theta_{3}\pm\pi/2)}&=&\pm i(Z\otimes
Z)G_{(\theta_{1},\theta_{2},\theta_{3})}.
\label{canonical:eq:pio2shift3}
\end{eqnarray}
By appropriately redefining $G_{1A}$, $G_{1B}$, $G_{2A}$, $G_{2B}$,
 $\vec{\theta}$ can be restricted
to the range $[0,\pi/2)$.

As explained in beautiful detail in \cite{Zhan03}, there are 24
locally equivalent sets of values of $\vec{\theta} \in
[0,\pi/2)^{3}$. Given $(\theta_{1},\theta_{2},\theta_{3})$, the
locally equivalent values are
$(\theta_{i},\theta_{j},\theta_{k})$,
$(\pi/2-\theta_{i},\pi/2-\theta_{j},\theta_{k})$,
$(\pi/2-\theta_{i},\theta_{j},\pi/2-\theta_{k})$, and
$(\theta_{i},\pi/2-\theta_{j},\pi/2-\theta_{k})$ where $i$, $j$,
$k$, are permutations of $1$, $2$, $3$. To be more constructive,
let
\begin{eqnarray}
P_{12}&=&e^{i\pi/4 Z}\otimes e^{i\pi/4 Z}\\
P_{13}&=&e^{i\pi/4 Y}\otimes e^{i\pi/4 Y}\\
P_{23}&=&e^{i\pi/4 X}\otimes e^{i\pi/4 X}\\
M_{12}&=&e^{i\pi/4 Z}\otimes e^{-i\pi/4 Z}\\
M_{13}&=&e^{i\pi/4 Y}\otimes e^{-i\pi/4 Y}\\
M_{23}&=&e^{i\pi/4 X}\otimes e^{-i\pi/4 X}.
\end{eqnarray}
Given $G_{(\theta_{1},\theta_{2},\theta_{3})}$, we find that
\begin{eqnarray}
G_{(\theta_{2},\theta_{1},\theta_{3})}&=&P_{12}G_{(\theta_{1},\theta_{2},\theta_{3})}P_{12}^{\dag}
\label{canonical:eq:perm_and_sign1}\\
G_{(\theta_{3},\theta_{2},\theta_{1})}&=&P_{13}G_{(\theta_{1},\theta_{2},\theta_{3})}P_{13}^{\dag}
\label{canonical:eq:perm_and_sign2}\\
G_{(\theta_{1},\theta_{3},\theta_{2})}&=&P_{23}G_{(\theta_{1},\theta_{2},\theta_{3})}P_{23}^{\dag}
\label{canonical:eq:perm_and_sign3}\\
G_{(-\theta_{2},-\theta_{1},\theta_{3})}&=&M_{12}G_{(\theta_{1},\theta_{2},\theta_{3})}M_{12}^{\dag}
\label{canonical:eq:perm_and_sign4}\\
G_{(-\theta_{3},\theta_{2},-\theta_{1})}&=&M_{13}G_{(\theta_{1},\theta_{2},\theta_{3})}M_{13}^{\dag}
\label{canonical:eq:perm_and_sign5}\\
G_{(\theta_{1},-\theta_{3},-\theta_{2})}&=&M_{23}G_{(\theta_{1},\theta_{2},\theta_{3})}M_{23}^{\dag}
\label{canonical:eq:perm_and_sign6}
\end{eqnarray}
Using
Eqs~(\ref{canonical:eq:perm_and_sign1}--\ref{canonical:eq:perm_and_sign6})
and
(\ref{canonical:eq:pio2shift1}--\ref{canonical:eq:pio2shift3}), it
is always possible to obtain $\vec{\theta}$ such that
$\theta_{1}\geq\theta_{2}$, $\pi/2-\theta_{1}\geq\theta_{2}$,
$\theta_{2}\geq\theta_{3}$, $\theta_{3}\geq 0$, and
$\pi/2-\theta_{1}\geq \theta_{1}$ if $\theta_{3}=0$. To obtain
this unique $\vec{\theta}$, use
Eqs~(\ref{canonical:eq:perm_and_sign1}--\ref{canonical:eq:perm_and_sign3})
to order $\theta_{k}$ such that
$\theta_{1}\geq\theta_{2}\geq\theta_{3}\geq 0$. If
$\theta_{1}>\pi/4$ and $\theta_{2}\geq \pi/2-\theta_{1}$, use
Eq.~(\ref{canonical:eq:perm_and_sign4}) and
Eqs~(\ref{canonical:eq:pio2shift1}--\ref{canonical:eq:pio2shift2})
to obtain $(\pi/2-\theta_{2},\pi/2-\theta_{1},\theta_{3})$.
Finally, again use
Eqs~(\ref{canonical:eq:perm_and_sign1}--\ref{canonical:eq:perm_and_sign3})
to order $\theta_{k}$. This gives the unique $\vec{\theta}$ with
the desired properties.

For the remainder of the thesis, when talking about a canonical
decomposition of a gate $G$, we mean
$G_{1A},G_{1B},G_{2A},G_{2B}\in U(2)$ and the unique
$\vec{\theta}$ described above such that $G\cong (G_{2A}\otimes
G_{2B})G_{\vec{\theta}}(G_{1A}\otimes G_{1B})$ up to global phase.
We have not restricted the single-qubit unitaries to be special as
none of the standard single-qubit gates $H$, $X$, $Z$, $S$ and $T$
are special and we wish to express $G_{1A}$, $G_{1B}$, $G_{2A}$
and $G_{2B}$ in terms of standard gates whenever possible.

\section{Building gates out of physical interactions}
\label{canonical:section:Zhang}

To complete the construction of an arbitrary gate $G$ in terms of
a 2-qubit evolution operator $U(t)$ admitting a canonical
decomposition with 2-qubit term $U_{\vec{\phi}t}$ and single-qubit
gates, let
\begin{eqnarray}
G & \cong & (G_{2A}\otimes G_{2B})G_{\vec{\theta}}(G_{1A}\otimes G_{1B}), \\
U(t) & \cong & (U_{2A}(t)\otimes U_{2B}(t))U_{\vec{\phi}t}(U_{1A}(t)\otimes U_{1B}(t)),
\end{eqnarray}
and following \cite{Zhan03}, consider
\begin{equation}
\label{canonical:eq:indep_rearrange}
\begin{tabular}{rl}
& $M_{23}P_{23}P_{13}e^{i(\phi_{1}X\otimes X + i\phi_{2}Y\otimes Y
+ i\phi_{3}Z\otimes
Z)t_{3}}P_{13}^{\dag}P_{23}^{\dag}M_{23}^{\dag}$\\
$\times$ & $M_{12}P_{23}e^{i(\phi_{1}X\otimes X + i\phi_{2}Y\otimes
Y
+ i\phi_{3}Z\otimes Z)t_{2}}P_{23}^{\dag}M_{12}^{\dag}$\\
$\times$ & $e^{i(\phi_{1}X\otimes X + i\phi_{2}Y\otimes Y + i\phi_{3}Z\otimes Z)t_{1}}$\\
= & $e^{i(\phi_{1}t_{1} - \phi_{3}t_{2} + \phi_{3}t_{3})X\otimes X
     + i(\phi_{2}t_{1} - \phi_{1}t_{2} - \phi_{2}t_{3})Y\otimes Y
     + i(\phi_{3}t_{1} + \phi_{2}t_{2} - \phi_{1}t_{3})Z\otimes
     Z}$.
\end{tabular}
\end{equation}
With the conditions on $\vec{\phi}$, the equations
\begin{equation}
\label{canonical:eq:phi_theta}
\left(
\begin{array}{ccc}
  \phi_{1} & -\phi_{3} & \phi_{3} \\
  \phi_{2} & -\phi_{1} & -\phi_{2} \\
  \phi_{3} & \phi_{2} & -\phi_{1}
\end{array}
\right)
\left(
\begin{array}{c}
  t_{1} \\
  t_{2} \\
  t_{3}
\end{array}
\right)
=
\left(
\begin{array}{c}
  \theta_{1} \\
  \theta_{2} \\
  \theta_{3}
\end{array}
\right)
\end{equation}
are invertible. Note that this system of equations is neither unique nor special,
as other matrices from Eqs~(\ref{canonical:eq:perm_and_sign1}--\ref{canonical:eq:perm_and_sign6})
could have been used in Eq.~(\ref{canonical:eq:indep_rearrange}). For
example
\begin{equation}
\label{canonical:eq:indep_rearrange2}
\begin{tabular}{rl}
& $M_{12}P_{12}e^{i(\phi_{1}X\otimes X + i\phi_{2}Y\otimes Y +
i\phi_{3}Z\otimes Z)t_{3}}P_{12}^{\dag}M_{12}^{\dag}$\\
$\times$ & $M_{23}M_{13}e^{i(\phi_{1}X\otimes X + i\phi_{2}Y\otimes
Y + i\phi_{3}Z\otimes Z)t_{2}}M_{13}^{\dag}M_{23}^{\dag}$\\
$\times$ & $e^{i(\phi_{1}X\otimes X + i\phi_{2}Y\otimes Y + i\phi_{3}Z\otimes Z)t_{1}}$\\
= & $e^{i(\phi_{1}t_{1} - \phi_{3}t_{2} - \phi_{1}t_{3})X\otimes X
     + i(\phi_{2}t_{1} + \phi_{1}t_{2} - \phi_{2}t_{3})Y\otimes Y
     + i(\phi_{3}t_{1} - \phi_{2}t_{2} + \phi_{3}t_{3})Z\otimes
     Z}$
\end{tabular}
\end{equation}
also leads to an invertible system of equations.

After solving Eq.~(\ref{canonical:eq:phi_theta}) for $t_{1}$, $t_{2}$, $t_{3}$, let

\begin{eqnarray}
U_{1}&=& U_{1A}^{\dag}(t_{1}) G_{1A} \\
U_{2}&=& U_{1B}^{\dag}(t_{1}) G_{1B} \\
U_{3}&=& U_{1A}^{\dag}(t_{2}) e^{ i\pi/4 X} e^{ i\pi/4 Z} U_{2A}^{\dag}(t_{1}) \\
U_{4}&=& U_{1B}^{\dag}(t_{2}) e^{ i\pi/4 X} e^{-i\pi/4 Z} U_{2B}^{\dag}(t_{1}) \\
U_{5}&=& U_{1A}^{\dag}(t_{3}) e^{ i\pi/4 Y} e^{ i\pi/4 X} e^{ i\pi/4 X} e^{-i\pi/4 Z} e^{-i\pi/4 X} U_{2A}^{\dag}(t_{2}) \\
U_{6}&=& U_{1B}^{\dag}(t_{3}) e^{ i\pi/4 Y} e^{ i\pi/4 X} e^{-i\pi/4 X} e^{ i\pi/4 Z} e^{-i\pi/4 X} U_{2B}^{\dag}(t_{2}) \\
U_{7}&=& G_{2A} e^{-i\pi/4 X} e^{-i\pi/4 X} e^{-i\pi/4 Y} U_{2A}^{\dag}(t_{3}) \\
U_{8}&=& G_{2B} e^{ i\pi/4 X} e^{-i\pi/4 X} e^{-i\pi/4 Y} U_{2B}^{\dag}(t_{3}).
\end{eqnarray}

With these definitions, it can be verified by direct
multiplication that
\begin{equation}
G\cong (U_{7}\otimes U_{8})U(t_{3})(U_{5}\otimes U_{6})U(t_{2})(U_{3}\otimes U_{4})U(t_{1})(U_{1}\otimes U_{2}).
\end{equation}
Note that arbitrary choices have entered the calculation at a
number of points meaning the above construction is not unique and
may not be optimal. It can, however, be shown that no general
construction of this form can use fewer than three periods of
evolution of $U(t)$ and eight single qubit gates in the worst case
\cite{Zhan03}. This suggests that the above implementation is
close to optimal, and at worst efficient.

\section{Conclusion}
\label{canonical:section:conc}

The construction of this chapter enables arbitrary 2-qubit gates
to be expressed as a simple circuit involving at most eight
single-qubit gates and three periods of evolution of a restricted
class of 2-qubit interactions. This implies that if multiple
single and 2-qubit gates are applied to the same pair of qubits,
they should all be combined into a single compound gate to save
time and reduce circuit complexity. Chapters \ref{5QEC},
\ref{5QECNM} and \ref{ShorLNN} rely heavily on this technique to
absorb swap gates into neighboring useful gates, dramatically
simplifying the circuits described therein.

\chapter{5-qubit QEC on an LNN QC}
\label{5QEC}

The question has been raised as to how well quantum error
correction (QEC) can be implemented on a linear nearest neighbor
(LNN) quantum computer \cite{Gott00} due to the expectation that
numerous swap gates will be required. Working out a way around
this is important due to the large number of LNN architectures
currently under investigation
\cite{Kane98,Loss98,Vrij00,Gold03,Holl03,Tian03,Feng03,Pach03,Vand02,Soli03,Jeff02,Petr02,Ladd02,Vyur00,Kame03}.
In this chapter, a quantum circuit implementing 5-qubit QEC on an
LNN architecture is described. Our goal is to keep the error
correction scheme as simple as possible to facilitate physical
realization. In particular, fault-tolerance has not been built
into the circuit to minimize its complexity and the required
number of qubits. Despite the lack of fault-tolerance, we show
that, for both a discrete and continuous error model, a threshold
physical error rate exists below which the circuit reduces the
probability of error in the protected logical qubit. We also
determine the required physical error rate for the logical qubit
to be 10 times and 100 times as reliable as a single unprotected
qubit.

This chapter is organized as follows. Firstly, explicit examples
of canonically decomposed compound gates incorporating the swap
gates required on an LNN architecture are given in
Section~\ref{5QEC:section:compound}. In
Section~\ref{5QEC:section:5qecLNN}, the non-fault-tolerant 5-qubit
QEC scheme is described and the LNN circuit presented. Simulations
of the performance of the LNN scheme when subjected to both
discrete and continuous errors are discussed in
Section~\ref{5QEC:section:sims}. Section~\ref{5QEC:section:conc}
concludes with a summary of all results and a description of
further work.

\section{Compound gates}
\label{5QEC:section:compound}

As discussed in detail in Chapter~\ref{canonical}, the canonical
decomposition enables any 2-qubit gate $G$ to be expressed
(non-uniquely) in the form
\begin{equation}
(G_{2A}\otimes G_{2B})G_{\vec{\theta}}(G_{1A}\otimes G_{1B})
\end{equation}
where $G_{1A}, G_{1B}, G_{2A}, G_{2B} \in U(2)$ and
\begin{equation}
G_{\vec{\theta}}=e^{i(\theta_{1}X\otimes X+\theta_{2}Y\otimes
Y+\theta_{3}Z\otimes Z)}.
\end{equation}
Provided a quantum computer allows qubits to be isolated, and has
a 2-qubit evolution operator $U(t)$ admitting a canonical
decomposition with 2-qubit term $U_{\vec{\phi}t}$, an
implementation of $G$ exists using at most three periods of free
evolution of $U(t)$ and eight single-qubit gates.

\begin{figure*}
\begin{center}
\includegraphics[width=12cm]{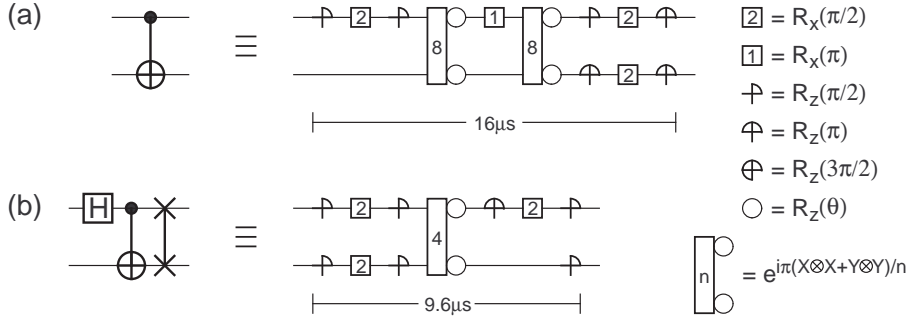}
\end{center}
\caption{Decomposition into physical operations of (a) \CNOT\ and
(b) Hadamard, \CNOT\ then swap. Note that the Kane architecture
has been used for illustrative purposes.}
\label{5QEC:figure:cnot-hcnots3}
\end{figure*}

Fig.~\ref{5QEC:figure:cnot-hcnots3}a shows the form of a
canonically decomposed \CNOT\ on a Kane quantum computer
\cite{Kane98,Hill03}. $Z$-rotations have been represented by
quarter, half and three-quarter circles corresponding to
$R_{z}(\pi/2)$, $R_{z}(\pi)$, and $R_{z}(3\pi/2)$ respectively,
where
\begin{equation}
R_{z} = e^{i\theta Z/2}.
\end{equation}
Full circles represent $Z$-rotations of angle dependent on the
physical construction of the computer (static magnetic field,
phosphorus donor placement etc). The details of obtaining the
canonical decomposition of the Kane 2-qubit evolution operator
contained in \cite{Hill03}. Up to a couple of $Z$-rotations, the
2-qubit interaction corresponds to $\phi_{1} = \phi_{2} = \pi/n$,
and $\phi_{3} = 0$. Square gates 1 and 2 correspond to X-rotations
$R_{x}(\pi)$ and $R_{x}(\pi/2)$.
Fig.~\ref{5QEC:figure:cnot-hcnots3}b shows an implementation of
the composite gate Hadamard followed by \CNOT\ followed by swap.
Note that the total time of the compound gate is significantly
less than the \CNOT\ on its own. This fact has been used to
minimize the total execution time of the LNN circuit of
Fig.~\ref{5QEC:figure:5qecboth2}b.

The above implies that the swaps inevitably required in an LNN
architecture to bring qubits together to be interacted can, in
some cases, be incorporated into other gates without additional
cost. On any architecture, canonically decomposed compound gates
should be used whenever multiple single and 2-qubit gates are
applied to the same two qubits.

\section{5-qubit LNN QEC}
\label{5QEC:section:5qecLNN}

\begin{figure*}
\begin{center}
\includegraphics[width=12cm]{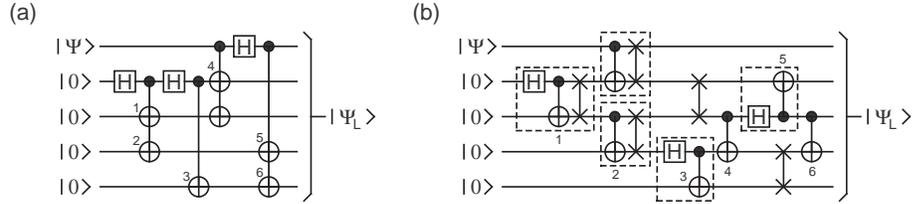}
\end{center}
\caption{(a) 5-qubit encoding circuit for general architecture,
(b) equivalent circuit for linear nearest neighbor architecture
with dashed boxes indicating compound gates. \CNOT\ gates that
must be performed sequentially are numbered.}
\label{5QEC:figure:5qecboth2}
\end{figure*}

\begin{figure*}
\begin{center}
\includegraphics[width=12cm]{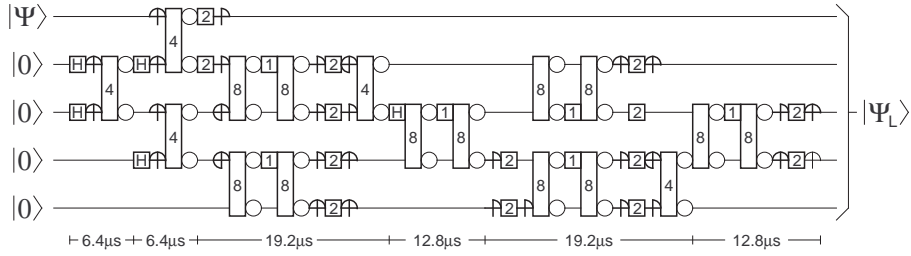}
\end{center}
\caption{A sequence of physical gates implementing the circuit of
Fig.~\ref{5QEC:figure:5qecboth2}b. Note the Kane architecture has
been used for illustrative purposes.} \label{5QEC:figure:qec5dec}
\end{figure*}

\begin{figure*}
\begin{center}
\includegraphics[width=12cm]{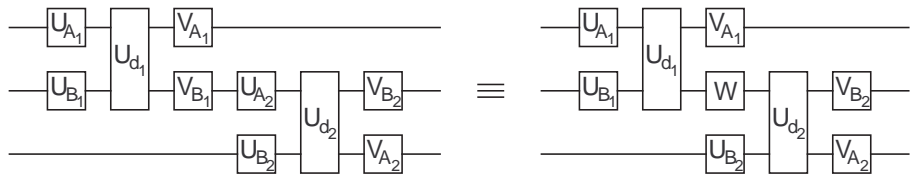}
\end{center}
\caption{Circuit equivalence used to reduce the number of physical
gates in Fig.~\ref{5QEC:figure:qec5dec}. $W=U_{A_{2}}V_{B_{1}}$}
\label{5QEC:figure:equiv}
\end{figure*}

\begin{figure*}
\begin{center}
\includegraphics[width=12cm]{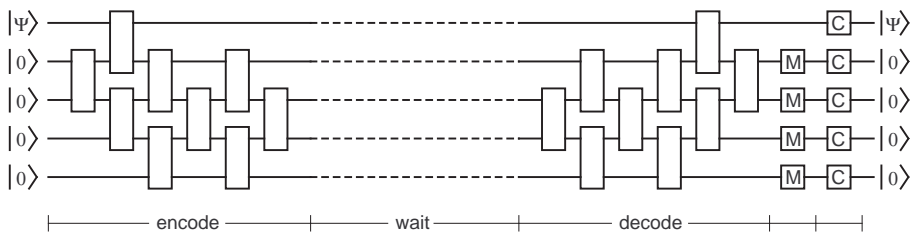}
\end{center}
\caption{A complete encode-wait-decode-measure-correct QEC cycle.}
\label{5QEC:figure:cycle}
\end{figure*}

5-qubit QEC schemes are designed to correct a single arbitrary
error. No QEC scheme designed to correct a single arbitrary error
can use less than five qubits \cite{Niel00}. A number of 5-qubit
QEC proposals exist \cite{Brau97,Knil01,Lafl96,Niwa02,Benn96}.
Fig.~\ref{5QEC:figure:5qecboth2}b shows a non-fault-tolerant
circuit appropriate for an LNN architecture implementing the
encode stage of the QEC scheme proposed in \cite{Brau97}. For
reference, the original circuit is shown in
Fig.~\ref{5QEC:figure:5qecboth2}a. Note that the LNN circuit uses
exactly the same number of \CNOT s and achieves minimal depth
since the \CNOT\ gates numbered 1--6 in
Fig.~\ref{5QEC:figure:5qecboth2}a must be performed sequentially
on any architecture that can only interact pairs of qubits (not
three or more at once). The two extra ``naked'' swaps in
Fig.~\ref{5QEC:figure:5qecboth2}b do not significantly add to the
total time of the circuit. Fig.~\ref{5QEC:figure:qec5dec} shows an
equivalent circuit broken into physical operations for a Kane
quantum computer. Note that this circuit uses the fact that if two
2-qubit gates share a qubit then two single-qubit unitaries can be
combined as shown in Fig.~\ref{5QEC:figure:equiv}. The decode
circuit is simply the encode circuit run backwards. 5-qubit QEC
schemes are primarily useful for data storage due to the
impossibility of fault-tolerantly interacting two logical qubits
\cite{Gott98}, though with some effort it is possible to
nontrivially interact three logical qubits.
Fig.~\ref{5QEC:figure:cycle} shows a full
encode-wait-decode-measure-correct data storage cycle.
Table~\ref{5QEC:table:one} shows the range of possible
measurements and the action required in each case.

\begin{table}
\begin{center}
\begin{tabular}{c|c}
Measurement & Action \\
\hline
\rule[-1ex]{0pt}{3.7ex}%
$\Psi'\otimes$0000 & \texttt{I$\otimes$IIII} \\
$\Psi'\otimes$0001 & \texttt{I$\otimes$IIIX} \\
$\Psi'\otimes$0010 & \texttt{I$\otimes$IIXI} \\
$\Psi'\otimes$0011 & \texttt{Z$\otimes$IIXX} \\
$\Psi'\otimes$0100 & \texttt{I$\otimes$IXII} \\
$\Psi'\otimes$0101 & \texttt{X$\otimes$IXIX} \\
$\Psi'\otimes$0110 & \texttt{Z$\otimes$IXXI} \\
$\Psi'\otimes$0111 & \texttt{X$\otimes$IXXX} \\
$\Psi'\otimes$1000 & \texttt{Z$\otimes$XIII} \\
$\Psi'\otimes$1001 & \texttt{I$\otimes$XIIX} \\
$\Psi'\otimes$1010 & \texttt{X$\otimes$XIXI} \\
$\Psi'\otimes$1011 & \texttt{X$\otimes$XIXX} \\
$\Psi'\otimes$1100 & \texttt{Z$\otimes$XXII} \\
$\Psi'\otimes$1101 & \texttt{X$\otimes$XXIX} \\
$\Psi'\otimes$1110 & \texttt{XZ$\otimes$XXXI} \\
$\Psi'\otimes$1111 & \texttt{Z$\otimes$XXXX}
\end{tabular}
\caption{Action required to correct the data qubit $\Psi'$ vs
measured value of ancilla qubits. Note that the X-operations
simply reset the ancilla.} \label{5QEC:table:one}
\end{center}
\end{table}

\section{Simulation of performance}
\label{5QEC:section:sims}

When simulating the QEC cycle, the LNN circuit of
Fig.~\ref{5QEC:figure:5qecboth2}b was used to keep the analysis
independent of the specific architecture used. Each compound gate
was modelled as taking the same time, allowing the time $T$ to be
made an integer such that each gate takes one time step. Gates
were furthermore simulated as though perfectly reliable and errors
applied to each qubit (including idle qubits) at the end of each
time step. The rationale for including idle qubits is that, in an
LNN architecture, active physical manipulation of some description
is frequently required to decouple neighboring qubits. Both the
manipulation itself and the degree of decoupling are likely to be
imperfect, leading to errors. Furthermore, in schemes utilizing
global electromagnetic fields to manipulate active qubits,
supposedly idle qubits may not be sufficiently off resonant.

Two error models were used --- discrete and continuous. In the
discrete model, a qubit can suffer either a bit-flip (X),
phase-flip (Z) or both simultaneously (XZ). Each type of error is
equally likely with total probability of error $p$ per qubit per
time step. The continuous error model involves applying
single-qubit unitary operations of the form
\begin{equation}
\label{5QEC:eq:continuous} U_{\sigma} =\left(
\begin{array}{cc}
\cos(\theta/2)e^{i(\alpha+\beta)/2} & \sin(\theta/2)e^{i(\alpha-\beta)/2} \\
-\sin(\theta/2)e^{i(-\alpha+\beta)/2} &
\cos(\theta/2)e^{i(-\alpha-\beta)/2}
\end{array} \right)
\end{equation}
where $\alpha$, $\beta$, and $\theta$ are normally distributed
about 0 with standard deviation $\sigma$.

Both the single-qubit and single logical qubit (five qubit)
systems were simulated. The initial state
\begin{equation}
\label{5QEC:eq:state}
|\Psi\rangle=\sin(\pi/8)|0\rangle+\cos(\pi/8)|1\rangle
\end{equation}
was used in both cases since $|\langle\Psi|X|\Psi\rangle|^{2} =
0.5$, $|\langle\Psi|Z|\Psi\rangle|^{2} = 0.5$, and
$|\langle\Psi|XZ|\Psi\rangle|^{2}= 0$ thus allowing each type of
error to be detected (but not necessarily distinguished). Simpler
states such as $|0\rangle$, $|1\rangle$,
$(|0\rangle+|1\rangle)/\sqrt{2}$, and
$(|0\rangle-|1\rangle)/\sqrt{2}$ do not have this property. For
example, the states $|0\rangle$ and $|1\rangle$ are insensitive to
phase errors, whereas the other two states are insensitive to bit
flip errors.

Let $T_{wait}$ denote the duration of the wait stage. Note that
the total duration of the encode, decode, measure and correct
stages is 14. In the QEC case the total time $T=T_{wait}+14$ of
one QEC cycle was varied to determine the time that minimizes the
error per time step
\begin{equation}
\label{5QEC:eq:step error}
\epsilon_{step}=1-\sqrt[T]{1-\epsilon_{final}}
\end{equation}
where $\epsilon_{final}=1-|\langle\Psi'|\Psi\rangle|^{2}$ and
$|\Psi'\rangle$ is the final data qubit state. An optimal time
$T_{opt}$ exists since the logical qubit is only protected during
the wait stage and the correction process can only cope with one
error. If the wait time is zero, extra complexity has been added
but no corrective ability. Similarly, if the wait time is very
large, it is almost certain that more than one error will occur,
resulting in the qubit being destroyed during the correction
process. Somewhere between these two extremes is a wait time that
minimizes $\epsilon_{step}$. This property of non-fault-tolerant
QEC has been noted previously \cite{Bare97}.

Table~\ref{5QEC:table:two} shows $T_{opt}$, $\epsilon_{step}$ and
the reduction in error $\epsilon_{step}/p$ versus $p$ for discrete
errors. Table~\ref{5QEC:table:three} shows the corresponding data
for continuous errors. Note that, in the continuous case, the
single qubit $p$ has been obtained via 1-qubit simulations using
the indicated $\sigma$ and wait time $T=T_{opt}+14$ and a 1-qubit
version of Eq.~(\ref{5QEC:eq:step error})
\begin{equation}
\label{5QEC:eq:step error_1}
p=1-\sqrt[T]{1-\epsilon_{final}}
\end{equation}
where $\epsilon_{final}=1-|\langle\Psi'|\Psi\rangle|^{2}$ and
$|\Psi'\rangle$ is the final single-qubit state. In this context,
$p$ is the discrete error rate yielding the same final error
probability as the corresponding $\sigma$ over time $T$.

\begin{table} \begin{center}
\begin{tabular}{c|c|c|c}
$p$ & $T_{opt}$ & $\epsilon_{step}$ & $\epsilon_{step}/p$ \\
\hline
\rule[-1ex]{0pt}{3.7ex}%
$10^{-2}$ & 25 & $1.7\times 10^{-2}$ & $1.7\times 10^{0}$ \\
$1.6\times 10^{-3}$ & 40 & $1.6\times 10^{-3}$ & $1.0\times 10^{0}$ \\
$10^{-3}$ & 50 & $8.4\times 10^{-4}$ & $8.4\times 10^{-1}$ \\
$10^{-4}$ & 150 & $3.1\times 10^{-5}$ & $3.1\times 10^{-1}$ \\
$10^{-5}$ & 500 & $1.0\times 10^{-6}$ & $1.0\times 10^{-1}$ \\
$10^{-6}$ & 1500 & $3.2\times 10^{-8}$ & $3.2\times 10^{-2}$ \\
$10^{-7}$ & 5000 & $1.0\times 10^{-9}$ & $1.0\times 10^{-2}$ \\
$10^{-8}$ & 10000 & $2.0\times 10^{-11}$ & $2.0\times 10^{-3}$
\end{tabular}
\caption{Probability per time step $\epsilon_{step}$ of the
logical qubit being destroyed when using 5-qubit QEC vs physical
probability $p$ per qubit per time step of a discrete error.}
\label{5QEC:table:two}
\end{center} \end{table}

\begin{table} \begin{center}
\begin{tabular}{c|c|c|c|c}
$\sigma$ & $T_{opt}$ & $p$ & $\epsilon_{step}$ & $\epsilon_{step}/p$ \\
\hline
\rule[-1ex]{0pt}{3.7ex}%
$10^{-1}$ & $2.5\times 10^{1}$ & $4.4\times 10^{-3}$ & $6.9\times 10^{-3}$ & $1.6\times 10^{0}$ \\
$4.7\times 10^{-2}$ & $5.5\times 10^{1}$ & $1.1\times 10^{-3}$ & $1.1\times 10^{-3}$ & $1.0\times 10^{0}$ \\
$10^{-2}$ & $2.5\times 10^{2}$ & $4.9\times 10^{-5}$ & $1.4\times 10^{-5}$ & $2.9\times 10^{-1}$ \\
$3.6\times 10^{-3}$ & $5.5\times 10^{2}$ & $6.4\times 10^{-6}$ & $6.4\times 10^{-7}$ & $1.0\times 10^{-1}$ \\
$10^{-3}$ & $2.5\times 10^{3}$ & $5.0\times 10^{-7}$ & $1.3\times 10^{-8}$ & $2.6\times 10^{-2}$ \\
$4.0\times 10^{-4}$ & $5.0\times 10^{3}$ & $8.0\times 10^{-8}$ & $8.0\times 10^{-10}$ & $1.0\times 10^{-2}$ \\
$10^{-4}$ & $2.5\times 10^{4}$ & $5.0\times 10^{-9}$ & $1.0\times 10^{-11}$ & $2.0\times 10^{-3}$ \\
$10^{-5}$ & $2.5\times 10^{5}$ & $5.0\times 10^{-11}$ & $7.2\times 10^{-15}$ & $1.4\times 10^{-4}$
\end{tabular}
\caption{Probability per time step $\epsilon_{step}$ of the
logical qubit being destroyed when using 5-qubit QEC vs standard
deviation $\sigma$ of continuous errors.} \label{5QEC:table:three}
\end{center} \end{table}

The threshold $p = 1.6\times 10^{-3}$ shown in
Table~\ref{5QEC:table:two} is comparable to some of the highest
thresholds of fault-tolerant quantum computation described in
Chapter~\ref{intro}, which were obtained using weaker noise models
and architectures able to interact arbitrary pairs of qubits. If
an error rate improvement of a factor of 10 or 100 is desired when
using our scheme, then $p = 10^{-5}$ or $p = 10^{-7}$ is required
respectively. Note that unlike fault-tolerant schemes, the error
rate of the logical qubit does not scale as $cp^{2}$.

For continuous errors, the threshold standard deviation is $\sigma
= 4.7\times 10^{-2}$. The logical qubit is a factor of 10 more
reliable than a single physical qubit for $\sigma = 3.6\times
10^{-3}$. A factor of 100 improvement is achieved when $\sigma =
4.0\times 10^{-4}$.

\section{Conclusion}
\label{5QEC:section:conc}

To summarize, we have presented a non-fault-tolerant circuit
implementing 5-qubit QEC on an LNN architecture that achieves the
same depth as the current least depth circuit \cite{Brau97}, and
simulated its effectiveness against both discrete and continuous
errors. For the discrete error model, if error correction is to
provide an error rate reduction of a factor of 10 or 100, the
physical error rate $p$ must be $10^{-5}$ or $10^{-7}$
respectively. The corresponding figures for the continuous error
model are $\sigma = 3.6\times 10^{-3}$ and $4.0\times 10^{-4}$.

Further work is required to determine whether the discrete or
continuous error model or some other model best describes errors
in physical quantum computers. The relationship between the two
error models also warrants further investigation. Further
simulation is required to determine the error thresholds and
scaling associated with single and 2-qubit LNN QEC protected
gates.

\chapter{QEC without measurement}
\label{5QECNM}

In Chapter~\ref{5QEC}, we described and analyzed an explicit
5-qubit non-fault-tolerant LNN QEC scheme. In this chapter, we
wish to relax the engineering requirements of this scheme further.
In particular, for many architectures the most difficult aspect of
quantum error correction is measuring qubits quickly and/or
reliably. Interactive classical processing of measurement results
can also be problematic. We therefore present an explicit 5-qubit
QEC scheme that only requires slow resetting and no classical
processing at the cost of halving the performance of the original
approach and potentially requiring an additional 4 ancilla qubits.
Prior work exists on removing measurement from fault-tolerant QEC
\cite{Ahar99,Boyk99}, but this approach requires many more qubits
than the scheme presented here. A similar no measurement
non-fault-tolerant LNN QEC scheme has been devised and physically
implemented using liquid NMR technology \cite{Knil01}, but is not
directly applicable to other technologies, and only permits a
single cycle of QEC to be performed.

The discussion is organized as follows. In
Section~\ref{5QECNM:section:reset}, the concept of resetting as
distinct from measurement is explained in more detail. In
Section~\ref{5QECNM:section:5qecnm}, a QEC scheme requiring fast
resetting and no classical processing is described and its
performance simulated. In Section~\ref{5QECNM:section:5qecslow},
additional qubits are added to the circuit to enable the use of a
slow reset operation. Section~\ref{5QECNM:section:conc} concludes
with a summary of our results and a description of further work.

\section{Resetting}
\label{5QECNM:section:reset}

\begin{figure}
\begin{center}
\includegraphics[width=70mm]{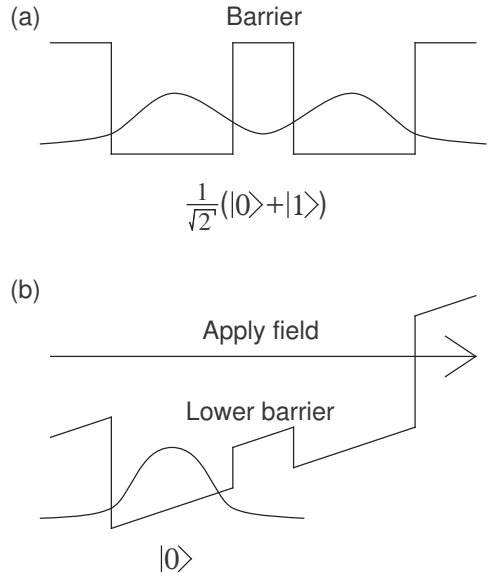}
\end{center}
\caption{Example of a physical model of resetting through
relaxation. Given a double potential well with left and right
occupancy representing $|0\rangle$ and $|1\rangle$ respectively,
resetting to $|0\rangle$ can be achieved by lowering the barrier
and applying a bias.} \label{5QECNM:figure:reset}
\end{figure}

Resetting is distinct from measurement in that a given qubit is in
a known state after resetting but no information about its state
beforehand is provided. A physical example of resetting is
provided by a double quantum dot system separated by a potential
barrier in which $|0\rangle$ is represented by an electron in the
left dot, and $|1\rangle$ by an electron in the right dot. By
applying an electric field across the double dot system and
lowering the barrier potential, the electron can be encouraged to
relax into the $|0\rangle$ state. This process is illustrated in
Fig.~\ref{5QECNM:figure:reset}.

\section{5-qubit QEC without measurement}
\label{5QECNM:section:5qecnm}

\begin{figure}
\begin{center}
\includegraphics[width=70mm]{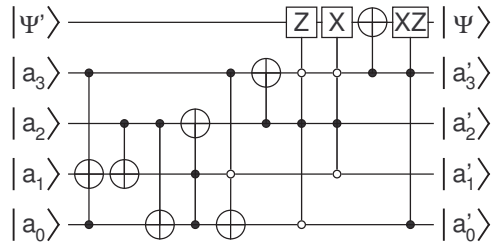}
\end{center}
\caption{Quantum circuit acting on a recently decoded logical
qubit to correct the data qubit based on the value of the ancilla
qubits. Hollow dots represent control qubits that must be
$|0\rangle$ for the attached gate to be applied.}
\label{5QECNM:figure:correct}
\end{figure}

To eliminate measurement from Fig.~\ref{5QEC:figure:cycle}, the
implicit classical logic that converts the measured result into a
corrective action must be converted into quantum logic gates.
Fig.~\ref{5QECNM:figure:correct} shows a quantum circuit
performing the necessary logic. The first half of the circuit
rearranges the states 0000 to 1111 such that the required
corrective action is as shown in column 3 of
Table~\ref{5QECNM:table:syndrome}.  This rearrangement of actions
leads to the relatively simple corrective logic of the second half
of Fig.~\ref{5QECNM:figure:correct}.

\begin{table} \begin{center}
\begin{tabular}{c|c|c}
Ancilla & Action & Action\\
\hline
\rule[-1ex]{0pt}{3.7ex}%
0000 & \texttt{I} & \texttt{I} \\
0001 & \texttt{I} & \texttt{I} \\
0010 & \texttt{I} & \texttt{I} \\
0011 & \texttt{Z} & \texttt{I} \\
0100 & \texttt{I} & \texttt{XZ} \\
0101 & \texttt{X} & \texttt{X} \\
0110 & \texttt{Z} & \texttt{Z} \\
0111 & \texttt{X} & \texttt{I} \\
1000 & \texttt{Z} & \texttt{X} \\
1001 & \texttt{I} & \texttt{Z} \\
1010 & \texttt{X} & \texttt{X} \\
1011 & \texttt{X} & \texttt{Z} \\
1100 & \texttt{Z} & \texttt{X} \\
1101 & \texttt{X} & \texttt{Z} \\
1110 & \texttt{XZ} & \texttt{X} \\
1111 & \texttt{Z} & \texttt{Z}
\end{tabular}
\caption{Second column shows the action required to correct the
data qubit given a certain ancilla value immediately after
decoding. Third column shows the action required to correct the
data qubit given a certain ancilla value after the application of
the first half of Fig.~\ref{5QECNM:figure:correct}.}
\label{5QECNM:table:syndrome}
\end{center} \end{table}

\begin{figure*}
\begin{center}
\includegraphics[width=12cm]{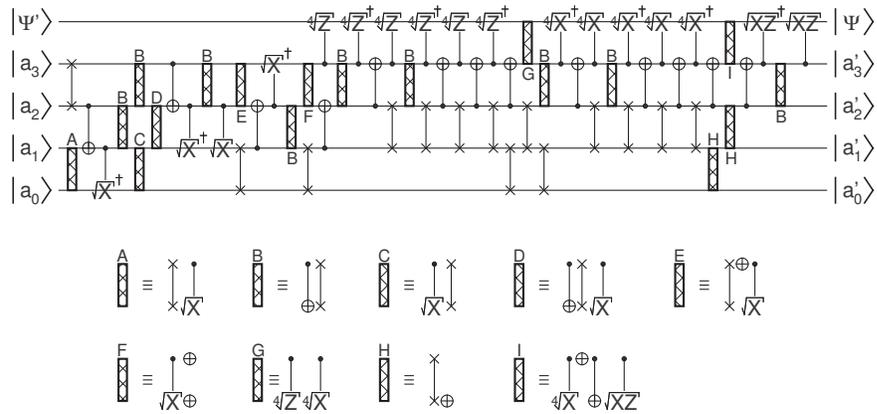}
\end{center}
\caption{Linear nearest neighbor version of
Fig.~\ref{5QECNM:figure:correct}.}
\label{5QECNM:figure:correctLNN}
\end{figure*}

In keeping with Chapter~\ref{5QEC}, an LNN version of
Fig.~\ref{5QECNM:figure:correct} has been devised and is shown in
Fig.~\ref{5QECNM:figure:correctLNN}. The LNN circuit was used in
all simulations.

Given the increased complexity of
Fig.~\ref{5QECNM:figure:correctLNN} compared with
Fig.~\ref{5QEC:figure:cycle}, it is expected that the threshold
error rates for both the discrete and continuous error models will
be lower.  Furthermore, the optimal wait time between correction
cycles it is expected to be longer to balance the need for
correction against the longer period of vulnerability during
correction. Both of these effects can be observed in
Tables~\ref{5QECNM:table:two} and \ref{5QECNM:table:three}.

\begin{table} \begin{center}
\begin{tabular}{c|c|c|c}
$p$ & $T_{opt}$ & $\epsilon_{step}$ & $\epsilon_{step}/p$ \\
\hline
\rule[-1ex]{0pt}{3.7ex}%
$10^{-3}$ & 120 & $1.4\times 10^{-3}$ & $1.4\times 10^{0}$ \\
$3.7\times 10^{-4}$ & 190 & $3.7\times 10^{-4}$ & $1.0\times 10^{0}$ \\
$10^{-4}$ & 320 & $6.0\times 10^{-5}$ & $6.0\times 10^{-1}$ \\
$10^{-5}$ & 1100 & $2.1\times 10^{-6}$ & $2.1\times 10^{-1}$ \\
$2.2\times 10^{-6}$ & 2300 & $2.2\times 10^{-7}$ & $1.0\times 10^{-1}$ \\
$10^{-6}$ & 3000 & $7.1\times 10^{-8}$ & $7.1\times 10^{-2}$ \\
$10^{-7}$ & 15000 & $1.9\times 10^{-9}$ & $1.9\times 10^{-2}$ \\
$2.1\times 10^{-8}$ & 40000 & $2.1\times 10^{-10}$ & $1.0\times 10^{-2}$ \\
\end{tabular}
\caption{Probability per time step $\epsilon_{step}$ of the
logical qubit being destroyed when using no measurement 5-qubit
QEC vs physical probability $p$ per qubit per time step of a
discrete error.} \label{5QECNM:table:two}
\end{center} \end{table}

\begin{table} \begin{center}
\begin{tabular}{c|c|c|c|c}
$\sigma$ & $T_{opt}$ & $p$ & $\epsilon_{step}$ & $\epsilon_{step}/p$ \\
\hline
\rule[-1ex]{0pt}{3.7ex}%
$3.1\times 10^{-2}$ & $1.0\times 10^{2}$ & $4.6\times 10^{-4}$ & $4.6\times 10^{-4}$ & $1.0\times 10^{0}$ \\
$10^{-2}$ & $3.2\times 10^{2}$ & $4.9\times 10^{-5}$ & $2.2\times 10^{-5}$ & $4.5\times 10^{-1}$ \\
$2.0\times 10^{-3}$ & $1.8\times 10^{3}$ & $2.0\times 10^{-6}$ & $2.0\times 10^{-7}$ & $1.0\times 10^{-1}$ \\
$10^{-3}$ & $3.7\times 10^{3}$ & $5.0\times 10^{-7}$ & $2.4\times 10^{-8}$ & $4.8\times 10^{-2}$ \\
$2.1\times 10^{-4}$ & $2.4\times 10^{4}$ & $2.1\times 10^{-8}$ & $2.1\times 10^{-10}$ & $1.0\times 10^{-2}$ \\
$10^{-4}$ & $5.0\times 10^{4}$ & $5.0\times 10^{-9}$ & $1.8\times 10^{-11}$ & $3.6\times 10^{-3}$ \\
$10^{-5}$ & $4.0\times 10^{5}$ & $5.1\times 10^{-11}$ & $1.5\times 10^{-14}$ & $2.9\times 10^{-4}$
\end{tabular}
\caption{Probability per time step $\epsilon_{step}$ of the
logical qubit being destroyed when using no measurement 5-qubit
QEC vs standard deviation $\sigma$ of continuous errors.}
\label{5QECNM:table:three}
\end{center} \end{table}

For discrete errors, the new threshold error rate, and the error
rates at which a factor of 10 and 100 improvement in reliability
are achieved are $p=3.7\times 10^{-4}$, $2.2\times 10^{-6}$, and
$2.1\times 10^{-8}$ respectively. Note that these are
approximately a factor of 5 less than the corresponding results
obtained in Chapter~\ref{5QEC}. For continuous errors, the
pertinent standard deviations are $\sigma=3.1\times 10^{-2}$,
$2.0\times 10^{-3}$, and $2.1\times 10^{-4}$.  These are very
comparable to the results obtained in Chapter~\ref{5QEC}, being at
most a factor of 2 less.

\section{5-qubit QEC with slow resetting}
\label{5QECNM:section:5qecslow}

\begin{figure*}[p]
\begin{center}
\includegraphics[height=18cm]{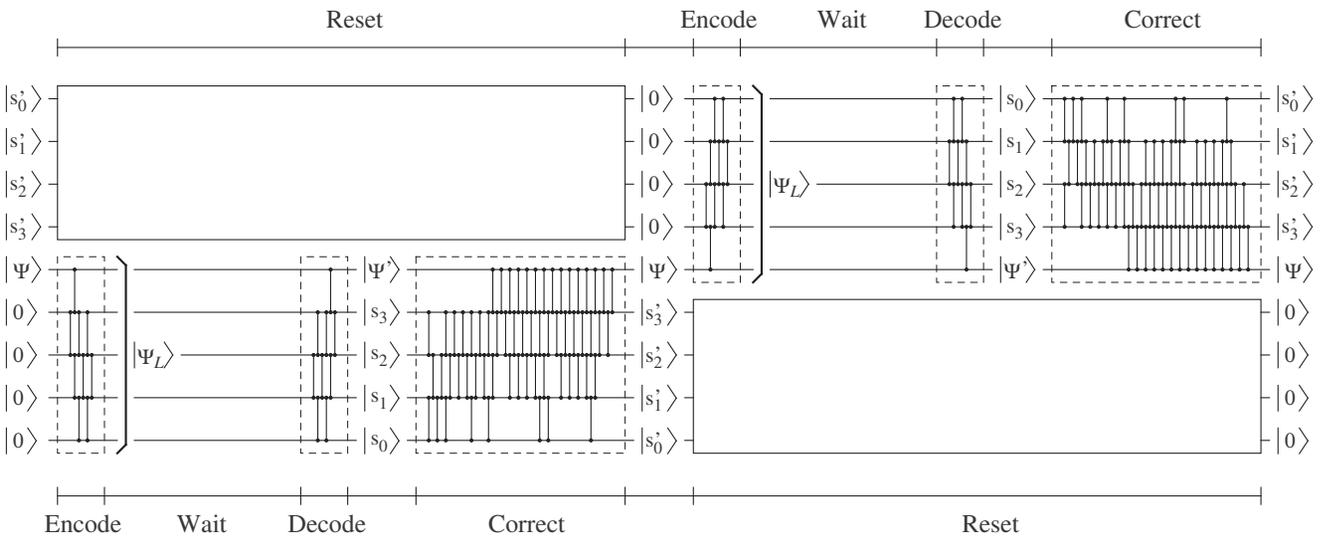}
\end{center}
\caption{5-qubit quantum error correction scheme not requiring
fast measurement or classical processing.}
\label{5QECNM:figure:9qubitcycle}
\end{figure*}

The use of Fig.~\ref{5QECNM:figure:correctLNN} in a 5-qubit system
assumes the reset operation is fast (comparable to the time
required to implement a single quantum gate). This requirement can
be eliminated with the addition of four ancilla qubits as shown in
Fig.~\ref{5QECNM:figure:9qubitcycle}. By re-encoding with fresh
ancilla, the reset operation may now take an amount of time equal
to an entire QEC cycle.  From Tables~\ref{5QECNM:table:two} and
\ref{5QECNM:table:three}, depending on the physical error rate,
this can be thousands of times longer than a single gate
operation.

\section{Conclusion}
\label{5QECNM:section:conc}

To summarize, we have shown that even without fault-tolerance,
fast measurement, classical processing, and large numbers of
qubits, it is still possible to construct a general quantum error
correction scheme with a reasonably high ($p=3.7\times 10^{-4}$ or
$\sigma=3.1\times 10^{-2}$) threshold error rate.

\chapter{Shor's algorithm}
\label{Shor}

Chapters~\ref{ShorLNN}, \ref{ShorGates} and to a lesser extent
Chapter~\ref{Solovay} deal with aspects of implementing Shor's
algorithm \cite{Shor94a,Shor97}. This chapter provides a detailed
review of the algorithm. In addition to Shor's papers, we draw
heavily on Ref.~\cite{Niel00}.

When Shor's algorithm was published in 1994, it was greeted with
great excitement due to its potential to break the popular RSA
encryption protocol \cite{Rive78}. RSA is used in all aspects of
e-commerce from Internet banking to secure online payment and can
also be used to facilitate secure message transmission. The
security of RSA is conditional on large integers being difficult
to factorize, which has so far proven to be the case when using
classical computers. Given a quantum computer, Shor's algorithm
renders the integer factoring problem tractable.

To be precise, let $N = N_{1}N_{2}$ be a product of prime numbers.
Let $L = \ln_{2}N$ be the binary length of $N$. Given $N$, Shor's
algorithm enables the determination of $N_{1}$ and $N_{2}$ in a
time polynomial in $L$.  This is achieved indirectly by finding
the period $r$ of $f(k) = m^{k} \bmod N$, where $1<m<N$, ${\rm
gcd}(m,N)=1$ and ${\rm gcd}$ denotes the greatest common divisor.
Provided $r$ is even and $f(r/2)\neq N-1$, the factors are
$N_{1}={\rm gcd}(f(r/2)+1,N)$ and $N_{2}={\rm gcd}(f(r/2)-1,N)$.
Note that the greatest common divisor can be computed in a time
linear in $L$ using a classical computer. For odd $N$ and randomly
selected $m$ such that $1<m<N$ and ${\rm gcd}(m,N)=1$, the
probability that $f(k)$ has a suitable $r$ is at least $0.75$
\cite{Niel00}. Thus on average very few values of $m$ need to be
tested to factor $N$.

\begin{figure}
\begin{center}
\includegraphics[width=70mm]{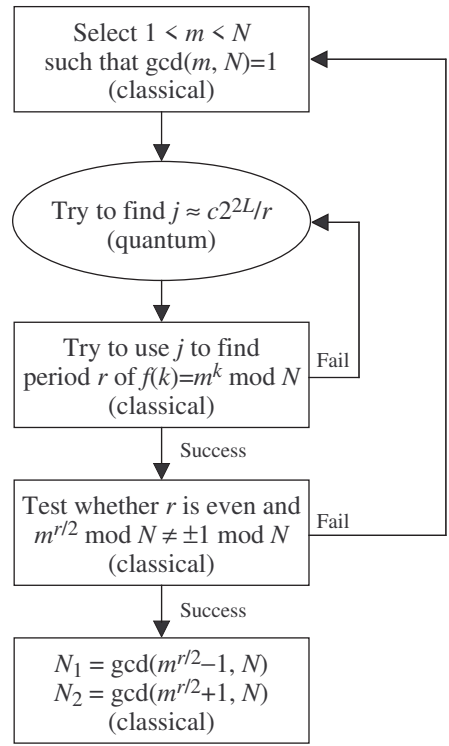}
\end{center}
\caption{The complete Shor's algorithm including classical pre-
and postprocessing.  The first branch is highly likely to fail,
resulting in many repetitions of the quantum heart of the
algorithm, whereas the second branch is highly likely to succeed.}
\label{Shor:figure:flowchart}
\end{figure}

The quantum part of Shor's algorithm can be viewed as a subroutine
that generates numbers of the form $j \simeq c 2^{2L}/r$. To
distinguish this from the necessary classical pre- and
post-processing, this subroutine will be referred to as quantum
period finding (QPF). Due to decoherence and imprecise gates, the
probability $s$ that QPF will successfully generate useful data
(defined precisely below) may be quite low with many repetitions
required to work out the period $r$ of a given $f(k)=m^{k}\bmod
N$. Using this terminology, Shor's algorithm consists of classical
preprocessing, potentially many repetitions of QPF with classical
postprocessing and possibly a small number of repetitions of the
entire cycle followed by more classical post-processing
(Fig.~\ref{Shor:figure:flowchart}).

The efficiency of QPF can only be quantified with reference to a
specific quantum circuit implementation. To date, the most
thorough work on different quantum circuit implementations has
been performed by Van Meter \cite{VanM04} drawing on work by
Vedral \cite{Vedr96}, Beckman \cite{Beck96}, Gossett
\cite{Goss98}, Draper \cite{Drap04} and Zalka \cite{Zalk98}.
Table~\ref{Shor:table:one} gives representative examples of the
variety of circuits in existence and their qubit counts and depths
as a function of $L$. Note that generally speaking time can be
saved at the cost of more qubits.

\begin{table} \begin{center}
\begin{tabular}{c|c|c}
Circuit & Qubits & Depth \\
\hline
\rule[-1ex]{0pt}{3.7ex}%
Beauregard \cite{Beau03} & $\sim 2L$ & $\sim 32L^{3}$ \\
Vedral \cite{Vedr96} & $\sim 5L$ & $240L^{3}$ \\
Zalka 1 \cite{Zalk98}& $\sim 5L$ & $\sim 3000L^{2}$ \\
Zalka 2 \cite{Zalk98}& $\sim 50L$ & $\sim 2^{19}L^{1.2}$ \\
Van Meter \cite{VanM04} & $O(L^2)$ & $O(L\log^{2}L)$
\end{tabular}
\caption{Required number of qubits and circuit depth of different
implementations of the quantum part of Shor's algorithm. Where
possible, figures are accurate to leading order in $L$.}
\label{Shor:table:one}
\end{center} \end{table}

An underlying procedure common to all implementations does exist.
The first common step involves initializing the quantum computer
to a single pure state $|0\rangle_{2L}|0\rangle_{L}$. Note that
for clarity the computer state has been broken into a $2L$ qubit
$k$ register and an $L$ qubit $f$ register.  The meaning of this
will become clearer below.

Step two is to Hadamard transform each qubit in the $k$-register
yielding
\begin{equation}
\label{Shor:eq:two}
\frac{1}{2^{L}}\sum_{k=0}^{2^{2L}-1}|k\rangle_{2L}|0\rangle_{L}.
\end{equation}

Step three is to calculate and store the corresponding values of
$f(k)$ in the $f$-register
\begin{equation}
\label{Shor:eq:three}
\frac{1}{2^{L}}\sum_{k=0}^{2^{2L}-1}|k\rangle_{2L}|f(k)\rangle_{L}.
\end{equation}
Note that this step requires additional ancilla qubits. The exact
number depends heavily on the circuit used.

Step four can actually be omitted but it explicitly shows the
origin of the period $r$ being sought. Measuring the $f$-register
yields
\begin{equation}
\label{Shor:eq:four} \frac{\sqrt{r}}{2^{L}}\sum_{n=0}^{
2^{2L}/r-1}|k_{0}+nr\rangle_{2L}|f_{M}\rangle_{L}
\end{equation}
where $k_{0}$ is the smallest value of $k$ such that $f(k)$ equals
the measured value $f_{M}$.

Step five is to apply the quantum Fourier transform
\begin{equation}
\label{Shor:eq:qft1} |k\rangle \rightarrow
\frac{1}{2^{L}}\sum_{j=0}^{2^{2L}-1}\mathrm{exp}\Big(\frac{2\pi
i}{2^{2L}} jk\Big)|j\rangle
\end{equation}
to the $k$-register resulting in
\begin{equation}
\label{Shor:eq:fivesum}
\frac{\sqrt{r}}{2^{2L}}\sum_{j=0}^{2^{2L}-1}\sum_{p=0}^{
2^{2L}/r-1}\mathrm{exp}\Big(\frac{2\pi i}{2^{2L}}
j(k_{0}+pr)\Big)|j\rangle_{2L}|f_{M}\rangle_{L}.
\end{equation}
The meaning of equation \ref{Shor:eq:fivesum} is best illustrated by
reversing the order of the summation
\begin{equation}
\label{Shor:eq:sumreverse}
\frac{\sqrt{r}}{2^{2L}}\sum_{j=0}^{2^{2L}-1}\sum_{p=0}^{
2^{2L}/r-1}\mathrm{exp}\Big(\frac{2\pi i}{2^{2L}}
j(k_{0}+pr)\Big)|j\rangle_{2L}|f_{M}\rangle_{L}.
\end{equation}
The probability of measuring a given value of $j$ is thus
\begin{equation}
\label{Shor:eq:prj} {\rm
Pr}(j,r,L)=\left|\frac{\sqrt{r}}{2^{2L}}\sum_{p=0}^{
2^{2L}/r-1}\mathrm{exp}\Big(\frac{2\pi i}{2^{2L}}
jpr\Big)\right|^{2}.
\end{equation}

\begin{figure}
\begin{center}
\includegraphics[width=90mm]{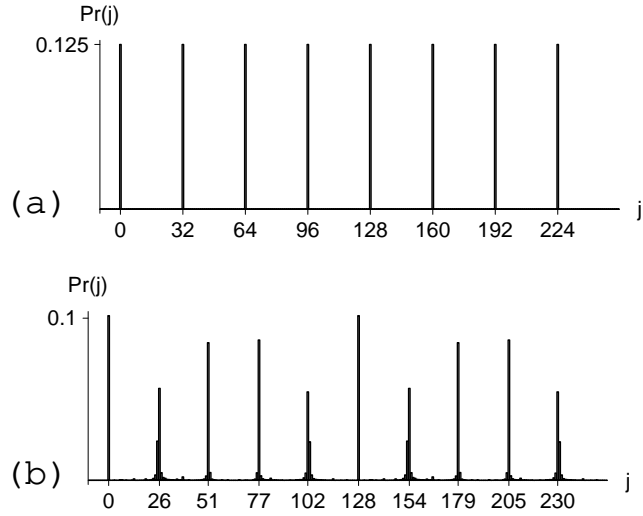}
\end{center}
\caption{Probability of different measurements $j$ at the end of
quantum period finding with total number of states $2^{2L}=256$
and (a) period $r=8$, (b) period $r=10$.}
\label{Shor:figure:period}
\end{figure}

If $r$ divides $2^{2L}$, Eq.~(\ref{Shor:eq:prj}) can be evaluated
exactly. In this case the probability of observing $j=c2^{2L}/r$
for some integer $0\leq c<r$ is $1/r$ whereas if $j\neq c2^{2L}/r$
the probability is 0. This situation is illustrated in
Fig.~\ref{Shor:figure:period}a. However if $r$ divides $2^{2L}$
exactly a quantum computer is not needed as $r$ would then be a
power of 2 and easily calculable. When $r$ is not a power of 2 the
perfect peaks of Fig.~\ref{Shor:figure:period}a become slightly
broader as shown in Fig.~\ref{Shor:figure:period}b. All one can
then say is that with high probability the value $j$ measured will
satisfy $j\simeq c2^{2L}/r$ for some $0\leq c<r$.

Given a measurement $j\simeq c2^{2L}/r$ with $c\neq 0$, classical
postprocessing is required to extract information about $r$. The
process begins with a continued fraction expansion. To illustrate,
consider factoring 143 ($L=8$). Suppose we choose $m$ equal 2 and
the output $j$ of QPF is 31674. The relation $j\simeq c2^{2L}/r$
becomes $31674\simeq c65536/r$. The continued fraction expansion
of $c/r$ is
\begin{equation}
\label{Shor:eq:contfrac854}
\frac{31674}{65536}=\frac{1}{\frac{32768}{15837}}=\frac{1}{2+\frac{1094}{15837}}=\frac{1}{2+\frac{1}{14+\frac{1}{2+\frac{1}{10+1/52}}}}.
\end{equation}
The continued fraction expansion of any number between 0 and 1 is
completely specified by the list of denominators which in this
case is $\{2,14,2,10,52\}$. The $n$th convergent of a continued
fraction expansion is the proper fraction equivalent to the first
$n$ elements of this list.
\begin{eqnarray}
\label{Shor:eq:convergeants427}
\{2\} & = & \frac{1}{2} \nonumber \\
\{2,14\} & = & \frac{14}{29} \nonumber \\
\{2,14,2\} & = & \frac{29}{60} \nonumber \\
\{2,14,2,10\} & = & \frac{304}{629} \nonumber \\
\{2,14,2,10,52\} & = & \frac{15837}{32768}
\end{eqnarray}
An introductory exposition and further properties of continued
fractions are described in Ref.~\cite{Niel00}. The period $r$ can
be sought by substituting each denominator into the function
$f(k)= 2^{k} \bmod 143$. With high probability only the largest
denominator less than $2^{L}$ will be of interest. In this case
$2^{60}\bmod 143=1$ and hence $r=60$.

Two modifications to the above are required. Firstly, if $c$ and
$r$ have common factors, none of the denominators will be the
period but rather one will be a divisor of $r$. After repeating
QPF a number of times, let $\{j_{m}\}$ denote the set of measured
values. Let $\{c_{mn}/d_{mn}\}$ denote the set of convergents
associated with each measured value $\{j_{m}\}$. If a pair
$c_{mn}$, $c_{m'n'}$ exists such that ${\rm gcd}(c_{mn},
c_{m'n'})=1$ and $d_{mn}$, $d_{m'n'}$ are divisors of $r$ then
$r={\rm lcm}(d_{mn}, d_{m'n'})$, where ${\rm lcm}$ denotes the
least common multiple. It can be shown that given any two divisors
$d_{mn}$, $d_{m'n'}$ with corresponding $c_{mn}$, $c_{m'n'}$ the
probability that ${\rm gcd}(c_{mn},c_{m'n'})=1$ is at least $1/4$
\cite{Niel00}. Thus, on average, only a small number of different
divisors are required. In practice, it will not be known which
denominators are divisors so every pair $d_{mn}$, $d_{m'n'}$ with
${\rm gcd}(c_{mn}, c_{m'n'})=1$ must be tested.

The second modification is simply allowing for the possibility
that the output $j$ of QPF may be useless. Let $s$ denote the
probability that $j=\lfloor c2^{2L}/r\rfloor$ or $\lceil
c2^{2L}/r\rceil$ for some $0<c<r$ where $\lfloor \rfloor$, $\lceil
\rceil$ denote rounding down and up respectively. Such values of
$j$ will be called useful as the denominators of the associated
convergents are guaranteed to include a divisor of $r$
\cite{Niel00}. To obtain a divisor of $r$, $O(1/s)$ runs of QPF
must be performed.

To summarize, as each new value $j_{m}$ is measured, the
denominators $d_{mn}$ less than $2^{L}$ of the convergents of the
continued fraction expansion of $j_{m}/2^{2L}$ are substituted
into $f(k)=m^{k} \bmod N$ to determine whether any $f(d_{mn})=1$
which would imply that $r=d_{mn}$. If not, every pair $d_{mn}$,
$d_{m'n'}$ with associated numerators $c_{mn}$, $c_{m'n'}$
satisfying  ${\rm gcd}(c_{mn}, c_{m'n'})=1$ is tested to see
whether $r={\rm lcm}(d_{mn}, d_{m'n'})$. Note that as shown in
Fig.~\ref{Shor:figure:flowchart}, if $r$ is even or $m^{r/2}\bmod
N = \pm 1\bmod N$ then the entire process needs to be repeated
$O(1)$ times. Thus Shor's algorithm always succeeds provided
$O(1/s)$ runs of QPF can be performed. Note that if $s$ is too
small, it may not be possible to repeat QPF $O(1/s)$ times in a
practical amount of time.

\chapter{Shor's algorithm on an LNN QC}
\label{ShorLNN}

Implementing Shor's factorization algorithm \cite{Shor94a,Shor97}
is arguably the ultimate goal of much experimental quantum
computer research. A necessary test of any quantum computer
proposal is therefore whether or not it can implement quantum
period finding (QPF) as described in Chapter~\ref{Shor}. While
several different quantum circuits implementing QPF have been
designed (Table~\ref{Shor:table:one}), most tacitly assume that
arbitrary pairs of qubits within the computer can be interacted.
As discussed in Chapter~\ref{5QEC}, a large number of promising
proposals, including the Kane quantum computer, are best suited to
realizing a single line of qubits with nearest neighbor
interactions only. Determining whether these linear nearest
neighbor (LNN) architectures can implement QPF in particular and
quantum algorithms in general in a practical manner is a
nontrivial and important question. In this chapter we present a
circuit implementing QPF designed for an LNN QC. As the rest of
Shor's algorithm is classical, this implies that Shor's algorithm
can be implemented on an LNN QC. Despite the interaction
restrictions, the circuit presented uses just $2L+4$ qubits and to
leading order requires $8L^{4}$ gates arranged in a circuit of
depth $32L^{3}$ --- identical to leading order to the Beauregard
circuit \cite{Beau03} upon which this work is based. Note that the
original Beauregard's circuit used just $2L+3$ qubits, but with
the extra qubit repeated Toffoli gates can be implemented more
quickly, reducing the overall depth of the circuit by a factor of
4. The precise differences between the LNN and Beauregard circuit
are detailed throughout the chapter.

As controlling large numbers of qubits has so far proven to be
extraordinarily difficult, this work places emphasis firstly on
minimizing the required number of qubits. Secondly, the depth has
been minimized over the total gate count in an effort to reduce
the need for quantum error correction assuming the primary source
of error will be decoherence rather than the quantum gates
themselves. If gate errors dominate, a higher depth but lower gate
count circuit would be preferable \cite{Vedr96}.

The chapter is structured as follows. In
Section~\ref{ShorLNN:section:shor_decomp} Shor's algorithm is
broken into a series of simple tasks appropriate for direct
translation into circuits. Sections~\ref{ShorLNN:section:qft} to
\ref{ShorLNN:section:comp_circ} then present, in order of
increasing complexity, the LNN quantum circuits that together
comprise the LNN Shor quantum circuit.  The LNN quantum Fourier
transform (QFT) is presented first, followed by a modular
addition, the controlled swap, modular multiplication, and finally
the complete circuit. Section~\ref{ShorLNN:section:conc} contains
a summary of all results, and a description of further work.

\section{Decomposing Shor's algorithm}
\label{ShorLNN:section:shor_decomp}

The purpose of this section is to break Shor's algorithm into a
series of steps that can be easily implemented as quantum
circuits.  Neglecting the classical computations and optional
measurement step described in the previous chapter, Shor's
algorithm has already been broken into four steps.
\begin{enumerate}
\item Hadamard transform.
\item Modular exponentiation.
\item Quantum Fourier transform.
\item Measurement.
\end{enumerate}
The modular exponentiation step is the only one that requires
further decomposition.

The calculation of $f(k) = m^{k} \bmod N$ is firstly broken up
into a series of controlled modular multiplications.
\begin{equation}
f(k) = \prod_{i=0}^{2L-1}(m^{2^{i}k_{i}} \bmod N),
\label{ShorLNN:eq:mult_series}
\end{equation}
where $k_{i}$ denotes the $i$th bit of $k$. If $k_{i}=1$ the
multiplication $m^{2^{i}} \bmod N$ occurs, and if $k_{i}=0$
nothing happens.

There are many different ways to implement controlled modular
multiplication (Table~\ref{Shor:table:one}). The methods of
\cite{Beau03} require the fewest qubits and will be used here. To
illustrate how each controlled modular multiplication proceeds,
let $a(i)=m^{2^{i}} \bmod N$ and
\begin{equation}
\label{ShorLNN:eq:xdef}
x(i) = \prod_{j=0}^{i-1}(m^{2^{j}k_{j}}
\bmod N).
\end{equation}
$x(i)$ represents a partially completed modular exponentiation and
$a(i)$ the next term to multiply by. Let $|x(i),0\rangle$ denote a
quantum register containing $x(i)$ and another of equal size
containing 0. Firstly, add $a(i)$ modularly multiplied by the
first register to second register if and only if (iff) $k_{i}=1$.
\begin{eqnarray}
|x(i),0\rangle & \mapsto & |x(i),0+a(i)x(i) \bmod N \rangle
\nonumber \\
& = & |x(i),x(i+1)\rangle.
\end{eqnarray}
Secondly, swap the registers iff $k_{i}=1$.
\begin{equation}
|x(i),x(i+1)\rangle\mapsto|x(i+1),x(i)\rangle
\end{equation}
Thirdly, subtract $a(i)^{-1}$ modularly multiplied by the first
register from the second register iff $k_{i}=1$.
\begin{eqnarray}
& & |x(i+1),x(i)\rangle \nonumber \\
& \mapsto & |x(i+1),x(i)-a(i)^{-1}x(i+1) \bmod
N\rangle \nonumber \\
& = & |x(i+1),0\rangle.
\end{eqnarray}
Note that while nothing happens if $k_{i}=0$, by the definition of
$x(i)$ the final state in this case will still be
$|x(i+1),0\rangle$.

The first and third steps described in the previous paragraph are
further broken up into series of controlled modular additions and
subtractions respectively.
\begin{eqnarray}
0+a(i)x(i) & = & 0+\sum_{j=0}^{L-1}a(i)2^{j}x(i)_{j} \bmod N, \label{ShorLNN:eq:add_series_1} \\
x(i)-a(i)^{-1}x(i+1) & = &
x(i)-\sum_{j=0}^{L-1}a(i)^{-1}2^{j}x(i+1)_{j} \bmod N,
\hspace{10mm} \label{ShorLNN:eq:add_series_2}
\end{eqnarray}
where $x(i)_{j}$ and $x(i+1)_{j}$ denote the $j$th bit of $x(i)$
and $x(i+1)$ respectively. Note that the additions associated with
a given $x(i)_{j}$ can only occur if $x(i)_{j}=1$ and similarly
for the subtractions. Given that these additions and subtractions
form a multiplication that is conditional on $k_{i}$, it is also
necessary that $k_{i}=1$.

Further decomposition will be left for subsequent sections.

\section{Quantum Fourier Transform}
\label{ShorLNN:section:qft}

The first circuit that needs to be described, as it will be used
in all subsequent circuits, is the QFT.
\begin{equation}
\label{ShorLNN:eq:qft2} |k\rangle \rightarrow
\frac{1}{\sqrt{2^{L}}}\sum_{j=0}^{2^{L}-1}\exp(2\pi
ijk/2^{L})|j\rangle
\end{equation}

\begin{figure}
\begin{center}
\includegraphics[width=90mm]{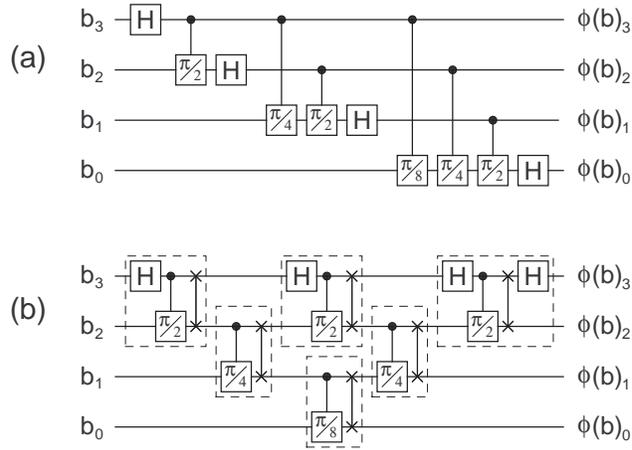}
\end{center}
\caption{(a) Standard quantum Fourier transform circuit. (b) An
equivalent linear nearest neighbor circuit.}
\label{ShorLNN:figure:serial_parallel_qft_no_ket}
\end{figure}

\begin{figure}
\begin{center}
\includegraphics[width=90mm]{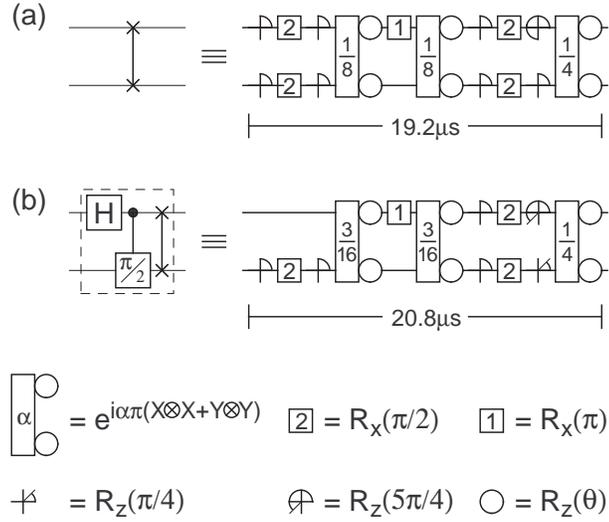}
\end{center}
\caption{(a) Swap gate expressed as a sequence of physical
operations via the canonical decomposition. (b) Similarly
decomposed compound gate consisting of a Hadamard gate, controlled
phase rotation, and swap gate.  Note that the Kane architecture
has been used for illustrative purposes.}
\label{ShorLNN:figure:swap_hphases}
\end{figure}

Fig.~\ref{ShorLNN:figure:serial_parallel_qft_no_ket}a shows the
usual circuit design for an architecture that can interact
arbitrary pairs of qubits.
Fig.~\ref{ShorLNN:figure:serial_parallel_qft_no_ket}b shows the
same circuit rearranged with the aid of swap gates to allow it to
be implemented on an LNN architecture. Note that the general QFT
circuit inverts the most significant to least significant ordering
of the qubits whereas the LNN circuit does not. Dashed boxes
indicate compound gates implemented with the aid of the canonical
decomposition. To emphasis the advantage of using compound gates,
Fig.~\ref{ShorLNN:figure:swap_hphases} contains a comparison of a
single swap gate with a Hadamard gate followed by a controlled
phase rotation, followed by a swap gate.

Counting compound gates as one, the total number of gates required
to implement a QFT on $L$ qubits for both the general and LNN
architectures is $L(L-1)/2$.  Assuming gates can be implemented in
parallel, the minimum circuit depth for both is $2L-3$. Note that
for large L it is both necessary and possible for nearly all of
the exponentially small controlled rotation gates to be omitted
\cite{Copp94,Fowl03b}. Omitting these gates does not, however,
enable a reduction in the depth of the circuit. This point will be
discussed in detail in Chapter~\ref{ShorGates}. Furthermore, in
the LNN case the swap gates associated with omitted controlled
rotations must remain for the circuit to work so the gate count
also remains unchanged.

\section{Modular Addition}
\label{ShorLNN:section:mod_add}

\begin{figure}
\begin{center}
\includegraphics[width=12cm]{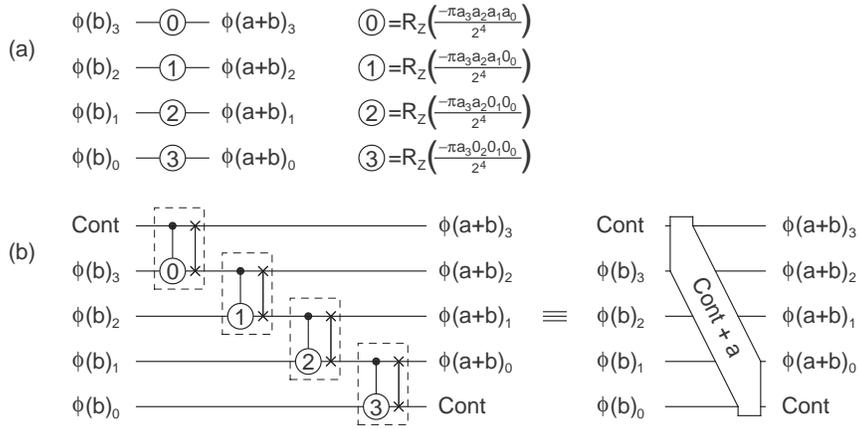}
\end{center}
\caption{(a) Quantum Fourier addition. (b) Controlled quantum
Fourier addition and its symbolic equivalent circuit. $Cont+a$
denotes the addition of $a$ if $Cont=1$.}
\label{ShorLNN:figure:fourier_add}
\end{figure}

Given a quantum register containing an arbitrary superposition of
binary numbers, there is a particularly easy way to add a binary
number to each number in the superposition \cite{Drap00,Beau03}.
By quantum Fourier transforming the superposition, the addition
can be performed simply by applying appropriate single-qubit
rotations as shown in Fig.~\ref{ShorLNN:figure:fourier_add}a. Such
an addition can also very easily be made dependant on a single
control qubit as shown in Fig.~\ref{ShorLNN:figure:fourier_add}b.

\begin{figure*}[p]
\begin{center}
\includegraphics[height=178mm]{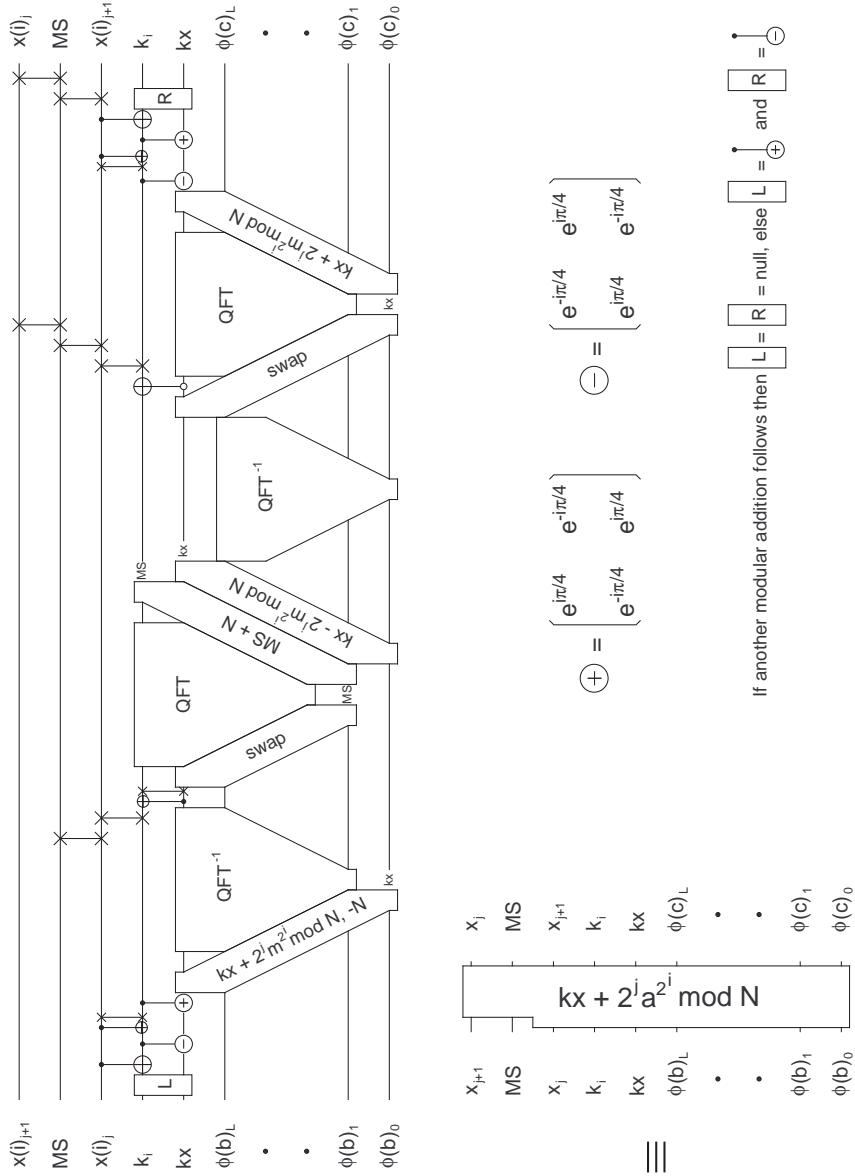}
\end{center}
\caption{Circuit to compute $c = (b + 2^{j}m^{2^{i}}) \bmod N$.
The diagonal circuit elements labelled swap represent a series of
2-qubit swap gates.  Small gates spaced close together represent
compound gates.  The qubits $x(i)$ are defined in
Eq.~\ref{ShorLNN:eq:xdef} and essentially store the current
partially calculated value of the modular exponentiation that
forms the heart of Shor's algorithm.  The $MS$ (Most Significant)
qubit is used to keep track of the sign of the partially
calculated modular addition result.  The $k_{i}$ qubit is the
$i$th bit of $k$ in Eq.~\ref{ShorLNN:eq:mult_series}.  The $kx$
qubit is set to 1 if and only if $x(i)_{j} = k_{i} = 1$. $kx\pm
2^{j}m^{2^{i}} \bmod N$ denotes modular addition (subtraction)
conditional on $kx=1$.} \label{ShorLNN:figure:mod_add}
\end{figure*}

Performing controlled modular addition is considerably more
complicated as shown in Fig.~\ref{ShorLNN:figure:mod_add}. This
circuit adds $2^{j}m^{2^{i}} \bmod N$ iff both $x(i)_{j}$ and
$k_{i}$ are 1 to the register containing $\phi(b)$ to obtain
$\phi(c)$ where $c=(b+2^{j}m^{2^{i}}) \bmod N$. Note that the
register containing $\phi(b)$ is $L+1$ qubits in length to prevent
overflow at any stage of the computation.

The first five gates comprise a Toffoli gate that sets $kx=1$ iff
$x(i)_{j}=k_{i}=1$.  $k_{i}$ and $x(i)_{j}$ are defined in
Eq.~(\ref{ShorLNN:eq:mult_series}) and
Eqs~(\ref{ShorLNN:eq:add_series_1}--\ref{ShorLNN:eq:add_series_2})
respectively. Note that the Beauregard circuit does not have a
$kx$ qubit, but without it the singly-controlled Fourier additions
become doubly-controlled and take four times as long.  The
calculations of the gate count and circuit depth of the Beauregard
circuit presented here have therefore been done with a $kx$ qubit
included.

The next circuit element firstly adds $2^{j}m^{2^{i}} \bmod N$ iff
$kx=1$ then subtracts $N$. If $b+(2^{j}m^{2^{i}} \bmod N) < N$,
subtracting $N$ will result in a negative number. In a binary
register, this means that the most significant bit will be 1.  The
next circuit element is an inverse QFT which takes the addition
result out of Fourier space and allows the most significant bit to
be accessed by the following \CNOT. The $MS$ (Most Significant)
qubit will now be 1 iff the addition result was negative. If
$b+(2^{j}m^{2^{i}} \bmod N)> N$, subtracting $N$ will yield the
positive number $(b+2^{j}m^{2^{i}}) \bmod N$ and the $MS$ qubit
will remain set to 0.

We now encounter the first circuit element that would not be
present if interactions between arbitrary pairs of qubits were
possible.  Note that while this ``long swap'' operation
technically consists of $L$ regular swap gates, it only increases
the depth of the circuit by 1. The subsequent QFT enables the $MS$
controlled Fourier addition of $N$ yielding the positive number
$(b+2^{j}m^{2^{i}}) \bmod N$ if $MS=1$ and leaving the already
correct result unchanged if $MS=0$.

While it might appear that we  are now done, the qubits $MS$ and
$kx$ must be reset so they can be reused. The next circuit element
subtracts $2^{j}m^{2^{i}} \bmod N$.  The result will be positive
and hence the most significant bit of the result equal to 0 iff
the very first addition $b+(2^{j}m^{2^{i}} \bmod N)$ gave a number
less than $N$.  This corresponds to the $MS=1$ case.  After
another inverse QFT to allow the most significant bit of the
result to be accessed, the $MS$ qubit is reset by a \CNOT\ gate
that flips the target qubit iff the control qubit is 0.  Note that
the long swap operation that occurs in the middle of all this to
move the $kx$ qubit to a more convenient location only increases
the depth of the circuit by 1.

After adding back $2^{j}m^{2^{i}} \bmod N$, the next few gates
form a Toffoli gate that resets $kx$.  The final two swap gates
move $x(i)_{j+1}$ into position ready for the next modular
addition.  Note that the $L$ and $R$ gates are inverses of one
another and hence not required if modular additions precede and
follow the circuit shown.  Only one of the final two swap gates
contributes to the overall depth of the circuit.

The total gate count of the LNN modular addition circuit is
$2L^{2}+8L+22$ and compares very favorably with the general
architecture gate count of $2L^{2}+6L+14$. Similarly, the LNN
depth is $8L+16$ versus the general depth of $8L+13$.

\section{Controlled swap}
\label{ShorLNN:section:cswap}

\begin{figure}
\begin{center}
\includegraphics[width=70mm]{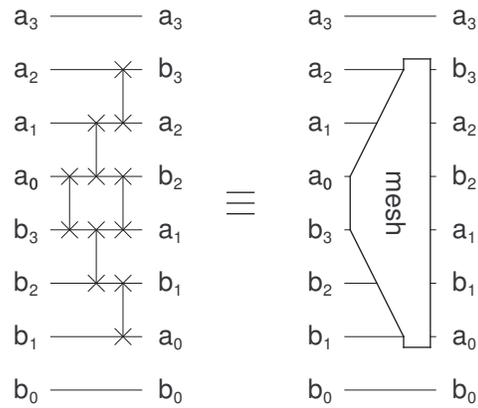}
\end{center}
\caption{Circuit designed to interleave two quantum registers.}
\label{ShorLNN:figure:mesh}
\end{figure}

Performing a controlled swap of two large registers is slightly
more difficult when only LNN interactions are available.  The two
registers need to be meshed so that pairs of equally significant
qubits can be controlled-swapped.  The mesh circuit is shown in
Fig.~\ref{ShorLNN:figure:mesh}.  This circuit element would not be
required in a general architecture.

\begin{figure}
\begin{center}
\includegraphics[width=70mm]{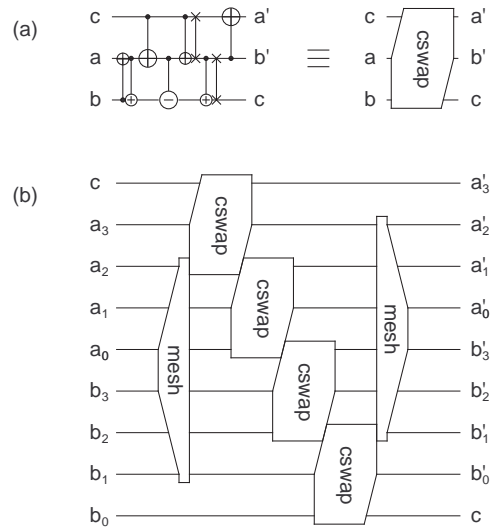}
\end{center}
\caption{(a) LNN circuit for the controlled swapping of two qubits
$|a\rangle$ and $|b\rangle$.  The qubits $|a'\rangle$ and
$|b'\rangle$ represent the potentially swapped states.  (b) LNN
circuit for the controlled swapping of two quantum registers. Note
that when chained together, the effective depth of the cswap gate
is 4.} \label{ShorLNN:figure:cswap}
\end{figure}

After the mesh circuit has been applied, the functional part of
the controlled swap circuit (Fig.~\ref{ShorLNN:figure:cswap}) can
be applied optimally with the control qubit moving from one end of
the meshed registers to the other.  The mesh circuit is then
applied in reverse to untangle the two registers.

The gate count and circuit depth of a mesh circuit is $L(L-1)/2$
and $L-1$ respectively.  The corresponding equations for a
complete LNN controlled swap are $L^{2}+5L$ and $6L$.  The general
controlled swap only requires $6L$ gates and can be implemented in
a circuit of depth $4L+2$.  The controlled swap is the only part
of this implementation of Shor's algorithm that is significantly
more difficult to implement on an LNN architecture.

\section{Modular Multiplication}
\label{ShorLNN:section:mod_mult}

\begin{figure*}[p]
\begin{center}
\includegraphics[height=185mm]{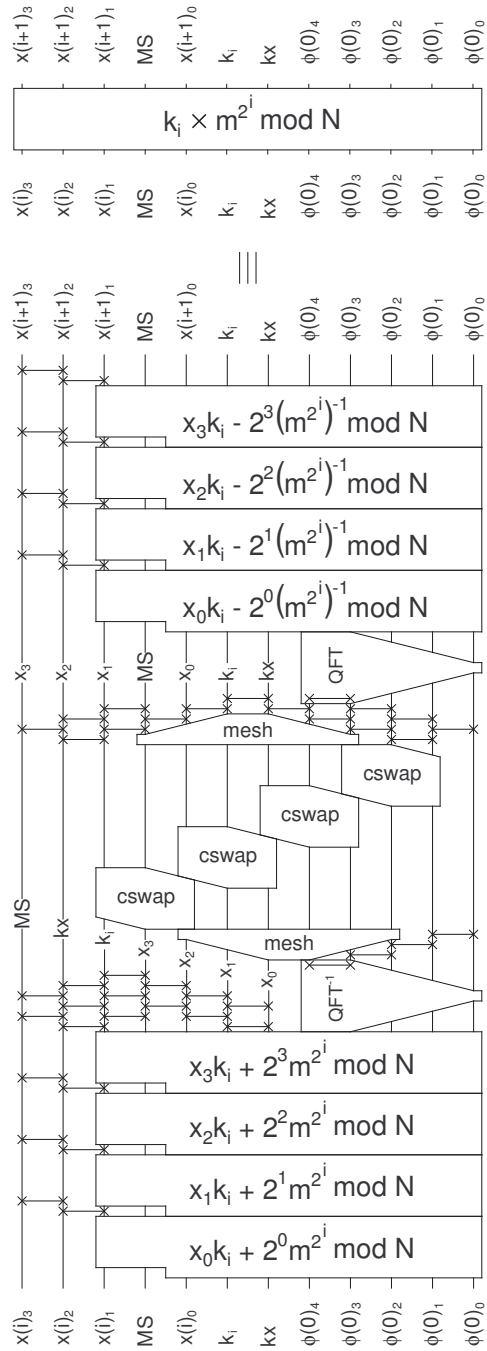}
\end{center}
\caption{Circuit designed to modularly multiply $x(i)$ by
$m^{2^{i}}$ if and only if $k_{i}=1$.  Note that for simplicity
the circuit for $L=4$ has been shown.  Note that the bottom $L+1$
qubits are ancilla and as such start and end in the
$|\phi(0)\rangle$ state.  The swap gates within the two QFT
structures represent compound gates. $k_{i}\times m^{2^{i}}\bmod
N$ denotes modular multiplication conditional on $k_{i}=1$.}
\label{ShorLNN:figure:multiply}
\end{figure*}

The ideas behind the modular multiplication circuit of
Fig.~\ref{ShorLNN:figure:multiply} were discussed in
Section~\ref{ShorLNN:section:shor_decomp}.  The first third
comprises a controlled modular multiply (via repeated addition)
with the result being stored in a temporary register. The middle
third implements a controlled swap of registers.  The final third
resets the temporary register.

Note that the main way in which the performance of the LNN circuit
differs from the ideal general case is due to the inclusion of the
two mesh circuits.  Nearly all of the remaining swaps shown in the
circuit do not contribute to the overall depth.  Note that the two
swaps drawn within the QFT and inverse QFT are intended to
indicate the appending of a swap gate to the first and last
compound gates in these circuits respectively.

The total gate count for the LNN modular multiplication circuit is
$4L^{3}+20L^{2}+58L-2$ versus the general gate count of
$4L^{3}+13L^{2}+35L+4$. The LNN depth is $16L^{2}+40L-7$ and the
general depth $16L^{2}+33L-6$.

\section{Complete Circuit}
\label{ShorLNN:section:comp_circ}

\begin{figure*}[p]
\begin{center}
\includegraphics[height=185mm]{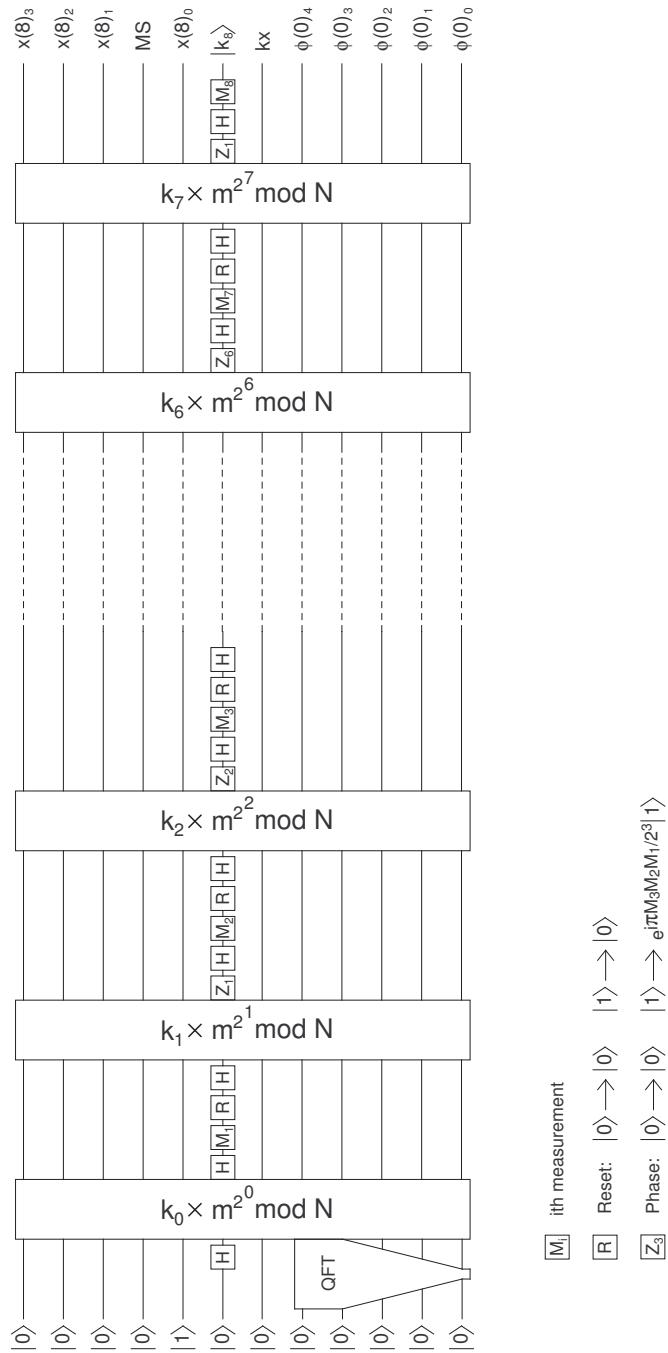}
\end{center}
\caption{Circuit implementing the quantum part of Shor's
algorithm. The single-qubit gates interleaved between the modular
multiplications comprise a QFT that has been decomposed by using
measurement gates to remove the need for controlled quantum phase
rotations.  Note that without these single-qubit gates the
remaining circuit is simply modular exponentiation.}
\label{ShorLNN:figure:modexp}
\end{figure*}

The complete circuit for Shor's algorithm
(Fig.~\ref{ShorLNN:figure:modexp}) can best be understood with
reference to Fig.~\ref{ShorLNN:figure:serial_parallel_qft_no_ket}a
and the four steps described in
Section~\ref{ShorLNN:section:shor_decomp}. The last two steps of
Shor's algorithm are a QFT and measurement of the qubits involved
in the QFT.  When a 2-qubit controlled quantum gate is followed by
measurement of the controlled qubit, it is equivalent to measure
the control qubit first and then apply a classically controlled
gate to the target qubit.  If this is done to every qubit in
Fig.~\ref{ShorLNN:figure:serial_parallel_qft_no_ket}a, it can be
seen that every qubit is decoupled.  Furthermore, since the QFT is
applied to the $k$ register and the $k$ register qubits are never
interacted with one another, it is possible to arrange the circuit
such that each qubit in the $k$ register is sequentially used to
control a modular multiplication, QFTed, then measured. Even
better, after the first quit of the $k$ register if manipulated in
this manner, it can be reset and used as the second qubit of the
$k$ register. This one qubit trick \cite{Park00} forms the basis
of Fig.~\ref{ShorLNN:figure:modexp}.

The total number of gates required in the LNN and general cases
are $8L^{4}+40L^{3}+116\frac{1}{2}L^{2}+4\frac{1}{2}L-2$ and
$8L^{4}+26L^{3}+70\frac{1}{2}L^{2}+8\frac{1}{2}L-1$ respectively.
The circuit depths are $32L^{3}+80L^{2}-4L-2$ and
$32L^{3}+66L^{2}-2L-1$ respectively.  The primary result of this
chapter is that the gate count and depth equations for both
architectures are identical to leading order.

\section{Conclusion}
\label{ShorLNN:section:conc}

We have presented a circuit implementing Shor's algorithm in a
manner appropriate for a linear nearest neighbor qubit array and
studied the number of extra gates and consequent increase in
circuit depth such a design entails.  To leading order our circuit
involves $8L^{4}$ gates arranged in a circuit of depth $32L^{3}$
on $2L+4$ qubits --- figures identical to that possible when
interactions between arbitrary pairs of qubits are allowed. Given
the importance of Shor's algorithm, this result supports the
widespread experimental study of linear nearest neighbor
architectures.

Simulations of the robustness of the circuit when subjected to
random discrete errors have been completed \cite{Devi04}, showing
extreme sensitivity to even small numbers of errors. Future
simulations will investigate the performance of the circuit when
protected by LNN quantum error correction.

\chapter{Shor's algorithm with a limited set of rotation gates}
\label{ShorGates}

Every circuit implementation of Shor's algorithm (see
Table~\ref{Shor:table:one}) ideally calls for controlled rotation
gates of magnitude $\pi/2^{2L}$ where $L$ is the binary length of
the integer $N$ to be factored. Such exponentially small rotations
are physically impossible to implement for large $L$. Prior work
by Coppersmith focusing solely on a quantum Fourier transform
suggested that it would be sufficient to implement controlled
$\pi/10^{6}$ rotations if integers thousands of bits long were
desired factored \cite{Copp94}. In this chapter, we study in
detail the complete Shor's algorithm using only controlled
$\pi/2^{d}$ rotation gates with $d$ less than or equal to some
$d_{\rm max}$. It is found that integers up to length $L_{\rm max}
= O(4^{d_{\rm max}})$ can be factored without significant
performance penalty. Consequently, we are able to show that
controlled rotation gates of magnitude $\pi/64$ are sufficient to
factor integers thousands of bits long.

The reader is assumed to be familiar with the description of
Shor's algorithm and notation as outlined in Chapter~\ref{Shor}.
In Section~\ref{ShorGates:section:background}, Coppersmith's
approximate quantum Fourier transform is introduced. In
Section~\ref{ShorGates:section:sVr}, we investigate the
relationship between the period $r$ of the function $f(k)$ given
as input to the quantum part of Shor's algorithm, and the
probability $s$ of obtaining useful output. In
Section~\ref{ShorGates:section:sVLd}, we study the relationship
between $s$ and both the length $L$ of the integer $N$ being
factored and the minimum angle controlled rotation $\pi/2^{d_{\rm
max}}$. This is then used to relate $L_{\rm max}$ to $d_{\rm
max}$. Section~\ref{ShorGates:section:conc} contains a summary of
results.

\section{Approximate quantum Fourier transform}
\label{ShorGates:section:background}

\begin{figure}
\begin{center}
\includegraphics[width=9cm]{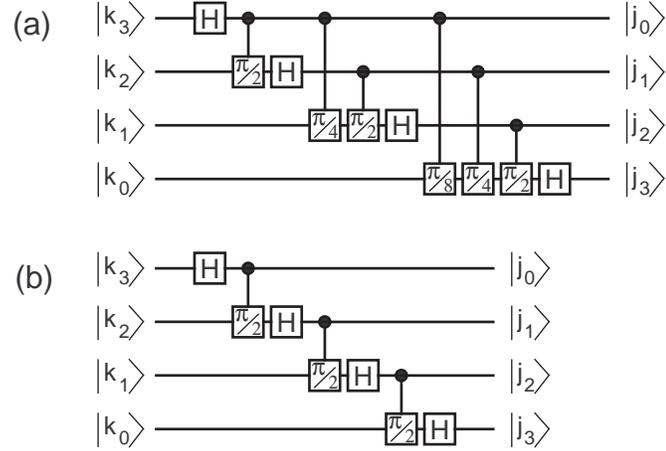}
\end{center}
\caption{Circuit for a 4-qubit (a) quantum Fourier transform and
(b) approximate quantum Fourier transform with $d_{\rm max}=1$.}
\label{ShorGates:figure:qftboth}
\end{figure}

Provided any circuit from Table~\ref{Shor:table:one} other than
Beauregard's is used to implement Shor's algorithm, exponentially
small rotations only occur in the one and only quantum Fourier
transform required just before measurement. The standard QFT
circuit is shown in Fig.~\ref{ShorGates:figure:qftboth}a. Note the
use of controlled rotations of magnitude $\pi/2^{d}$. In matrix
notation these 2-qubit operations correspond to
\begin{equation}
\label{ShorGates:eq:contphase} \left( \begin{array}{cccc}
1 & 0 & 0 & 0 \\
0 & 1 & 0 & 0 \\
0 & 0 & 1 & 0 \\
0 & 0 & 0 & e^{i\pi/2^{d}}
\end{array} \right).
\end{equation}

Coppersmith's approximate QFT (AQFT) circuit \cite{Copp94} is very
similar with just the deletion of rotation gates with $d$ greater
than some $d_{\rm max}$. For example,
Fig.~\ref{ShorGates:figure:qftboth}b shows an AQFT with $d_{\rm
max}=1$. Let $[j]_{m}$ denote the $m$th bit of $j$. The the action
of the AQFT on a computational basis state $|k\rangle$ is
\begin{equation}
\label{ShorGates:eq:aqfteq} |k\rangle \rightarrow
\frac{1}{\sqrt{2^{2L}}}\sum_{j=0}^{2^{2L}-1}|j\rangle
\mathrm{exp}\Big(\frac{2\pi i}{2^{2L}} {\textstyle
\tilde{\sum}_{mn}}[j]_{m}[k]_{n}2^{m+n}\Big)
\end{equation}
where $\tilde{\sum}_{mn}$ denotes a sum over all $m$, $n$ such
that $0\leq m,n<2L$ and $2L-d_{\rm max}+1\leq m+n<2L$. It has been
shown by Coppersmith that the AQFT is a good approximation of the
QFT \cite{Copp94} in the sense that the phase of individual
computational basis states in the output of the AQFT differ in
angle from those in the output of the QFT by at most $2\pi
L/2^{d_{\rm max}}$. The purpose of this chapter is to investigate
in detail the effect of using the AQFT in Shor's algorithm.

\section{Dependence of output reliability on period of $f(k)=m^{k}\bmod N$}
\label{ShorGates:section:sVr}

\begin{figure}
\psfrag{dmax}{$d_{\rm max}$}
\begin{center}
\includegraphics[width=78mm]{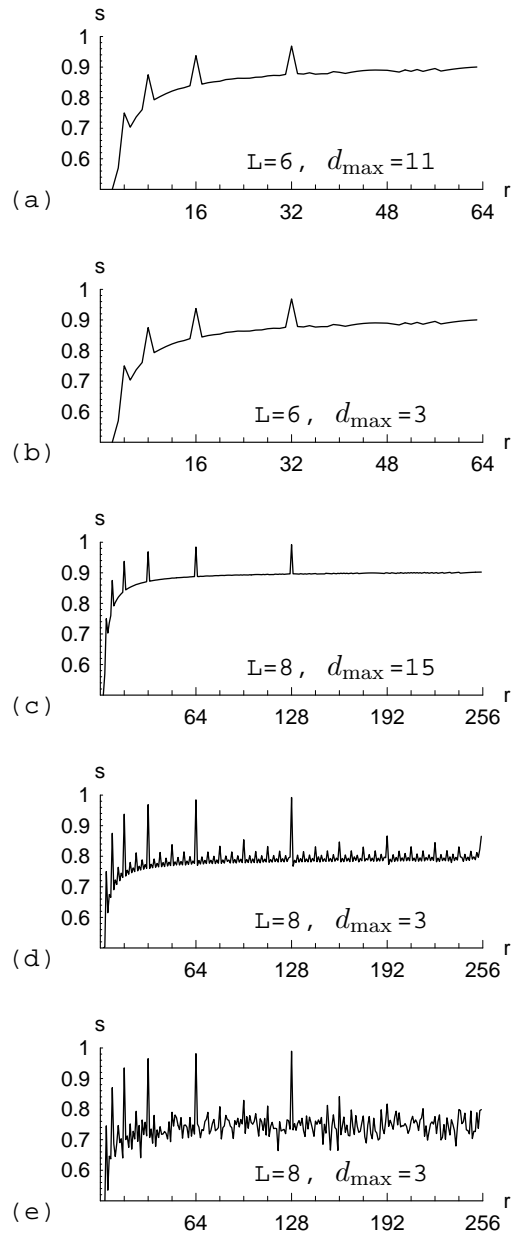}
\end{center}
\caption{Probability $s$ of obtaining useful output from quantum
period finding as a function of period $r$ for different integer
lengths $L$ and rotation gate restrictions $\pi/2^{d_{\rm max}}$.
The effect of using inaccurate controlled rotation gates
($\sigma=\pi/32$) is shown in (e).}
\label{ShorGates:figure:Ldarray}
\end{figure}

Different values of $r$ (the period of $f(k)=x^{k}\bmod N$) imply
different probabilities $s$ that the value $j$ measured at the end
of QPF will be useful. In particular, as discussed in
Chapter~\ref{Shor}, if $r$ is a power of 2 the probability of
useful output is much higher (Fig.~\ref{Shor:figure:period}). This
section investigates how sensitive $s$ is to variations in $r$.
Recall Eq.~(\ref{Shor:eq:prj}) for the probability of measuring a
given value of $j$. When the AQFT of
Eq.~(\ref{ShorGates:eq:aqfteq}) is used this becomes
\begin{equation}
\label{ShorGates:eq:aqftprj}
{\rm Pr}(j,r,L,d_{\rm max}) =
\left|\frac{\sqrt{r}}{2^{2L}}\sum_{p=0}^{
2^{2L}/r-1}\mathrm{exp}\Big(\frac{2\pi i}{2^{2L}}{\textstyle
\tilde{\sum}_{mn}[j]_{m}[pr]_{n}2^{m+n}}\Big)\right|^{2}
\end{equation}
The probability $s$ of useful output is thus
\begin{equation}
\label{ShorGates:eq:s} s(r,L,d_{\rm max})=\sum_{\{{\rm
useful}~j\}}{\rm Pr}(j,r,L,d_{\rm max})
\end{equation}
where $\{{\rm useful}~j\}$ denotes all $j=\lfloor
c2^{2L}/r\rfloor$ or $\lceil c2^{2L}/r\rceil$ such that $0<c<r$.
Fig.~\ref{ShorGates:figure:Ldarray} shows $s$ for $r$ ranging from
2 to $2^{L}-1$ and for various values of $L$ and $d_{\rm max}$.
The decrease in $s$ for small values of $r$ is more a result of
the definition of $\{{\rm useful}~j\}$ than an indication of poor
data. When $r$ is small there are few useful values of $j\simeq
c2^{2L}/r$, $0<c<r$ and a large range states likely to be observed
around each one resulting superficially in a low probability of
useful output $s$ as $s$ is the sum of the probabilities of
observing only values $j=\lfloor c2^{2L}/r\rfloor$ or $\lceil
c2^{2L}/r\rceil$, $0<c<r$. However, in practice values much
further from $j\simeq c2^{2L}/r$ can be used to obtain useful
output. For example if $r=4$ and $j=16400$ the correct output
value (4) can still be determined from the continued fraction
expansion of $16400/65536$ which is far from the ideal case of
$16384/65536$. To simplify subsequent analysis each pair $(L,
d_{\rm max})$ will from now on be associated with
$s(2^{L-1}+2,L,d_{\rm max})$ which corresponds to the minimum
value of $s$ to the right of the central peak. The choice of this
point as a meaningful characterization of the entire graph is
justified by the discussion above.

\begin{figure}
\begin{center}
\includegraphics[width=78mm]{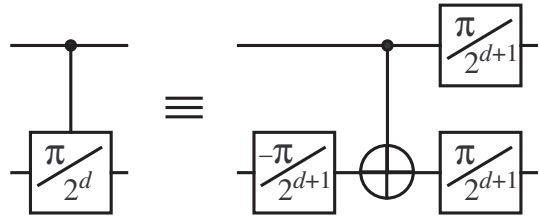}
\end{center}
\caption{Decomposition of a controlled phase gate into
single-qubit rotations and a \CNOT\ gate.}
\label{ShorGates:figure:phasedec}
\end{figure}

For completeness, Fig.~\ref{ShorGates:figure:Ldarray}e shows the
case of noisy controlled rotation gates of the form
\begin{equation}
\label{ShorGates:eq:contphase_delta} \left( \begin{array}{cccc}
1 & 0 & 0 & 0 \\
0 & 1 & 0 & 0 \\
0 & 0 & 1 & 0 \\
0 & 0 & 0 & e^{i(\pi/2^{d}+\delta)}
\end{array} \right).
\end{equation}
where $\delta$ is a normally distributed random variable of
standard deviation $\sigma$. This has been included to simulate
the effect of using approximate rotation gates built out of a
finite number of fault-tolerant gates.  The general form and
probability of successful output can be seen to be similar despite
$\sigma=\pi/32$. This $\sigma$ corresponds to $\pi/2^{d_{\rm
max}+2}$. For a controlled $\pi/64$ rotation, single-qubit
rotations of angle $\pi/128$ are required, as shown in
Fig.~\ref{ShorGates:figure:phasedec}.
Fig.~\ref{ShorGates:figure:Ldarray}e implies that it is acceptable
for these rotations to be implemented within $\pi/512$, implying
\begin{equation}
\label{ShorGates:eq:contphase_U} U = \left( \begin{array}{cccc}
1 & 0 \\
0 & e^{i(\pi/128+\pi/512)}
\end{array} \right)
\end{equation}
is an acceptable approximation of $R_{128}$. This point will be
developed further in Chapter~\ref{ShorGates}.

\section{Dependence of output usefulness on integer length and rotation gate
set}
\label{ShorGates:section:sVLd}

\begin{figure}
\begin{center}
\begin{tabular}{cc}
\includegraphics[width=6cm]{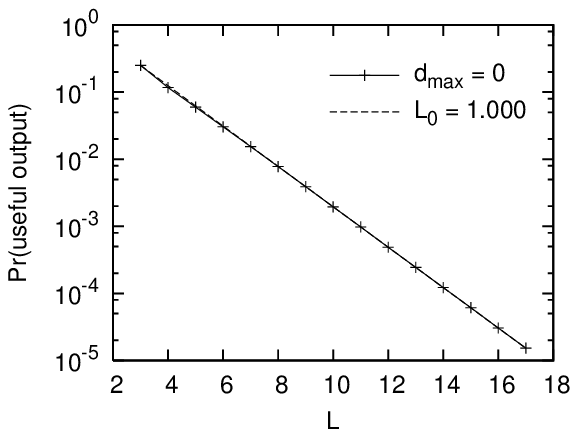} & \includegraphics[width=6cm]{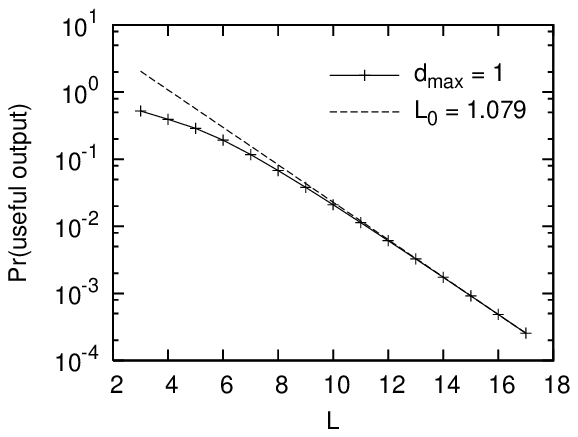}\\
\includegraphics[width=6cm]{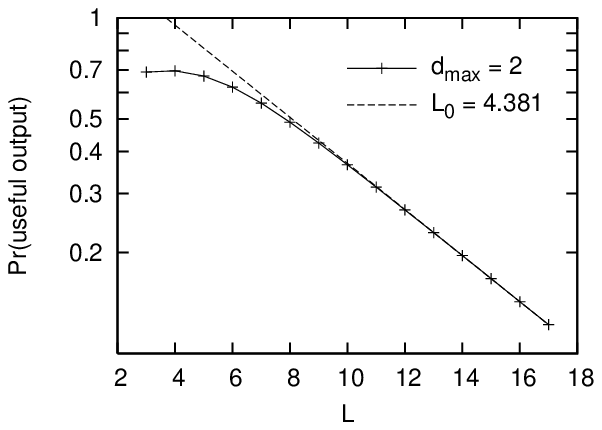} & \includegraphics[width=6cm]{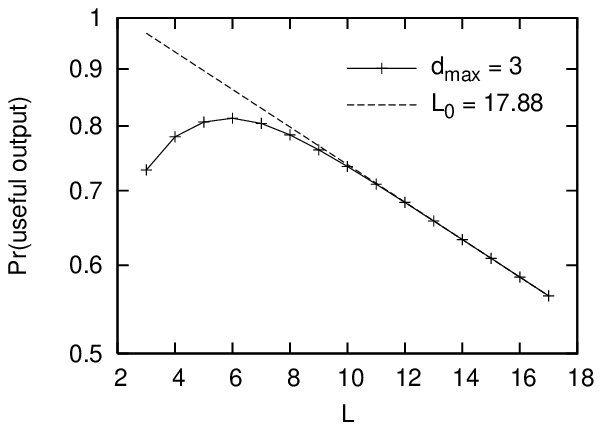}\\
\includegraphics[width=6cm]{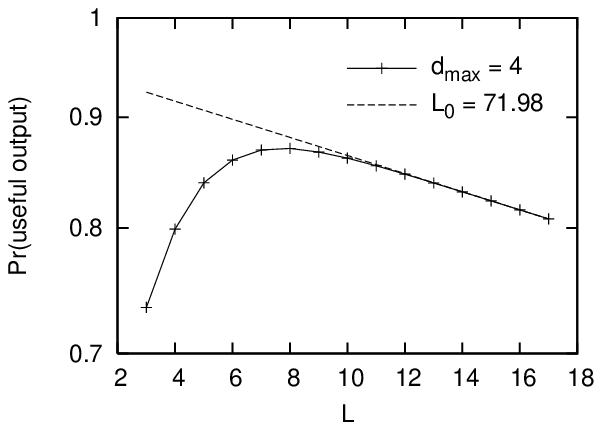} & \includegraphics[width=6cm]{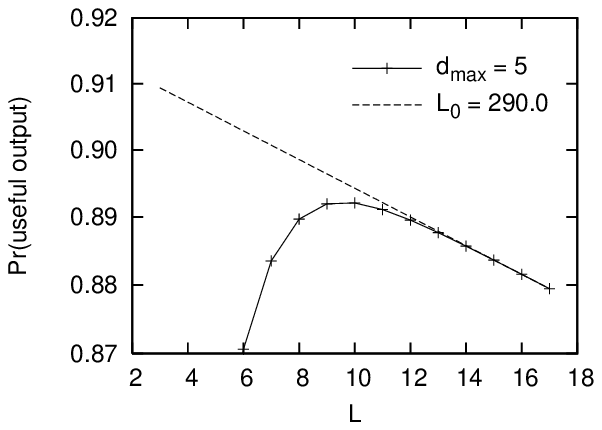}\\
\includegraphics[width=6cm]{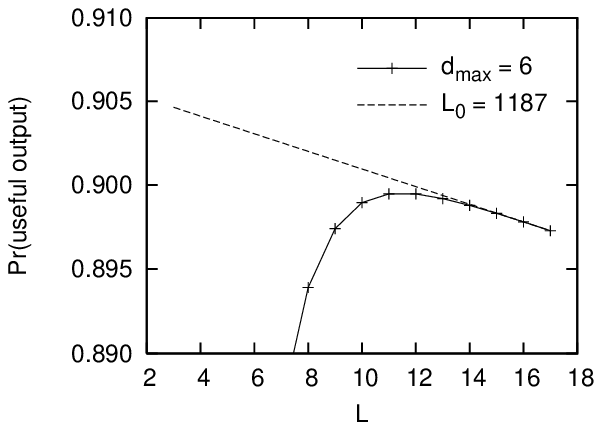} & \includegraphics[width=6cm]{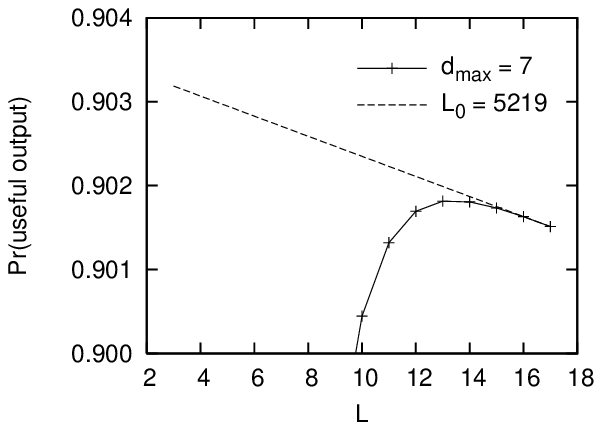}\\
\end{tabular}
\end{center}
\caption{Dependence of the probability of useful output from the
quantum part of Shor's algorithm on the length $L$ of the integer
being factored for different levels of restriction of controlled
rotation gates of angle $\pi/2^{d_{\rm max}}$. The parameter
$L_{0}$ characterizes lines of best fit of the form $s \propto
2^{-L/L_{0}}$.} \label{ShorGates:figure:darray}
\end{figure}

In order to determine how the probability of useful output $s$
depends on both the integer length $L$ and the minimum allowed
controlled rotation $\pi/2^{d_{\rm max}}$,
Eq.~(\ref{ShorGates:eq:s}) was solved with $r=2^{L-1}+2$ as
discussed in Section~\ref{ShorGates:section:sVr}.
Fig.~\ref{ShorGates:figure:darray} contains semilog plots of $s$
versus $L$ for different values of $d_{\rm max}$. Note that
Eq.~(\ref{ShorGates:eq:s}) grows exponentially more difficult to
evaluate as $L$ increases.

For $d_{\rm max}$ from 0 to 5, the exponential decrease of $s$
with increasing $L$ is clear.  Asymptotic lines of best fit of the
form
\begin{equation}
\label{ShorGates:eq:fit} s \propto 2^{-L/L_{0}}
\end{equation}
have been shown. Note that for $d_{\rm max}>0$, the value of
$L_{0}$ increases by greater than a factor of 4 when $d_{\rm max}$
increases by 1.  This enables one to generalize
Eq.~(\ref{ShorGates:eq:fit}) to an asymptotic lower bound valid
for all $d_{\rm max}>0$
\begin{equation}
\label{ShorGates:eq:genfit} s \propto 2^{-L/4^{d_{\rm max}-1}}
\end{equation}
with the constant of proportionality approximately equal to 1.

Keeping in mind that the required number of repetitions of QPF is
$O(1/s)$, one can relate $L_{\rm max}$ to $d_{\rm max}$ by
introducing an additional parameter $f_{\rm max}$ characterizing
the acceptable number of repetitions of QPF
\begin{equation}
\label{ShorGates:eq:Leq} L_{\rm max}\simeq 4^{d_{\rm
max}-1}\log_{2}f_{\rm max}.
\end{equation}

Available RSA \cite{Rive78} encryption programs such as PGP
typically use integers of length $L$ up to 4096. The circuit in
\cite{Vedr96} runs in $150L^3$ steps when an architecture that can
interact arbitrary pairs of qubits in parallel is assumed and
fault-tolerant gates are used. By virtue of the fact that this
circuit only interacts a few qubits at a time leaving the rest
idle, error correction can be easily built in without increasing
the circuit depth. Thus for $L=4096$, $\sim$$10^{13}$ steps are
required to perform a single run of QPF. On an electron spin or
charge quantum computer \cite{Burk00,Holl03} running at 10GHz this
corresponds to $\sim$$15$ minutes of computing. If we assume
$\sim$24 hours of computing is acceptable then $f_{\rm max}\sim
10^2$. Substituting these values of $L_{\rm max}$ and $f_{\rm
max}$ into Eq.~(\ref{ShorGates:eq:Leq}) gives $d_{\rm max}=6$
after rounding up. Thus provided controlled $\pi/64$ rotations can
be implemented accurately, implying the need to accurately
implement $\pi/128$ single-qubit rotations, it is conceivable that
a quantum computer could one day be used to break a 4096-bit RSA
encryption in a single day. With additional qubits, this time
could be reduced by several orders of magnitude by using one of
the circuits described in Ref.~\cite{VanM04}.

\section{Conclusion}
\label{ShorGates:section:conc}

We have demonstrated the robustness of Shor's algorithm when a
limited set of rotation gates is used. The length $L_{\rm max}$ of
the longest factorable integer can be related to the maximum
acceptable runs of quantum period finding $f_{\rm max}$ and the
smallest accurately implementable controlled rotation gate
$\pi/2^{d_{\rm max}}$ via $L_{\rm max}\sim 4^{d_{\rm
max}-1}\log_{2}f_{\rm max}$. Integers thousands of digits in
length can be factored provided controlled $\pi/64$ rotations can
be implemented with rotation angle accurate to $\pi/256$,
corresponding to single-qubit $\pi/128$ rotations implemented
within $\pi/512$. Sufficiently accurate fault-tolerant
approximations of such single-qubit rotation gates are presented
in Chapter~\ref{Solovay}.

\chapter{Constructing arbitrary single-qubit fault-tolerant gates}
\label{Solovay}

In large-scale quantum computation, every qubit of data is encoded
across multiple physical qubits to form a logical qubit permitting
quantum error correction and fault-tolerant computation.
Unfortunately, only very small sets of fault-tolerant gates
$\mathscr{G}$ can be applied simply and exactly to logical qubits,
where $\mathscr{G}$ depends on the number of logical qubits
considered, the code used, and the level of complexity one is
prepared to tolerate when implementing fault-tolerant gates. Gates
outside $\mathscr{G}$ must be approximated with sequences of gates
in $\mathscr{G}$. The existence of efficient approximating
sequences has been established by the Solovay-Kitaev theorem and
subsequent work \cite{Solo95,Kita97,Niel00,Harr02}. In this
chapter, we describe a numerical procedure taking a universal gate
set $\mathscr{G}$, gate $U$, and integer $l$ and outputting an
optimal approximation of $U$ using at most $l$ gates from
$\mathscr{G}$. This procedure is used to explore the properties of
approximations of the single-qubit phase rotation gates built out
of fault-tolerant gates that can be applied to a single Steane
code logical qubit. The average rate of convergence of Steane code
fault-tolerant approximations to arbitrary single-qubit gates is
also obtained.

Section~\ref{Solovay:section:opt_approx} describes the basics of
the numerical procedure used to find optimal gate sequences
approximating a given gate. A universal set of 24 gates that can
be applied fault-tolerantly to a single Steane code logical qubit
is given in Section~\ref{Solovay:section:gates}, along with most
of their quantum circuits. The complicated circuits comprising the
T-gate, which is part of this universal set, are described
separately in Section~\ref{Solovay:section:Tgate}.
Section~\ref{Solovay:section:phase} contains a discussion of
single-qubit phase rotations and their fault-tolerant
approximations, followed by approximations of arbitrary gates in
Section~\ref{Solovay:section:arbitrary}.
Section~\ref{Solovay:section:conc} summarizes the results of this
chapter and their implications, and points to further work.

\section{Finding optimal approximations}
\label{Solovay:section:opt_approx}

Let $U(m)$ denote the $m$-dimensional unitary group.  In this
section, we outline a numerical procedure that takes a finite gate
set $\mathscr{G} \subset U(m)$ that generates $U(m)$, a gate $U
\in U(m)$, and an integer $l$ and outputs an optimal sequence
$U_{l}$ of at most $l$ gates from $\mathscr{G}$ minimizing the
metric
\begin{equation}
\label{Solovay:eq:dist}
{\rm dist}(U,U_{l}) = \sqrt{\frac{m-|{\rm tr}(U^{\dag}U_{l})|}{m}}.
\end{equation}
The rationale of Eq.~(\ref{Solovay:eq:dist}) is that if $U$ and
$U_{l}$ are similar, $U^{\dag}U_{l}$ will be close to the identity
matrix (possibly up to some global phase) and the absolute value
of the trace will be close to $m$. By subtracting this absolute
value from $m$ and dividing by $m$ a number between 0 and 1 is
obtained. The overall square root is required to ensure that the
triangle inequality
\begin{equation}
\label{Solovay:eq:triangle}
{\rm dist}(U,W) \leq {\rm dist}(U,V)+{\rm dist}(V,W)
\end{equation}
is satisfied. This metric has been used in preference to the trace
distance used in the Solovay-Kitaev theorem \cite{Kita97,Niel00},
as the trace distance does not ignore global phase, and hence
leads to unnecessarily long phase correct approximating sequences.

Finding optimal gate sequences is a difficult task, and the
run-time of the numerical procedure presented here scales
exponentially with $l$.  Nevertheless, as we shall see in
Section~\ref{Solovay:section:phase}, gate sequences of sufficient
length for practical purposes can be obtained.

For a set $\mathscr{G}$ of size $g=|\mathscr{G}|$ and a maximum
sequence length of $l$, the size of the set of all possible gate
sequences of length up to $l$ is approximately $g^{l}$. For even
moderate $g$ and $l$, this set cannot be searched exhaustively. To
describe the basics of the actual method used, a few more
definitions are required. Let $G$ denote a gate in $\mathscr{G}$.
Order $\mathscr{G}$, and denote the $i$th gate by $G_{i}$. Let $S$
denote a sequence of gates in $\mathscr{G}$. Order the possible
gate sequences in the obvious manner $G_{1}, \ldots, G_{g},
G_{1}G_{1}, \ldots, G_{1}G_{g}, G_{2}G_{1}, \ldots$, and let
$S_{n}$ denote the $n$th sequence in this ordering. Let
$\{S\}_{l}$ denote all sequences with length less than or equal to
$l$. Let $\{Q\}_{l'}, l'<l$ denote the set of unique sequences of
length at most $l'$. Naively, $\{Q\}_{l'}$ can be constructed by
starting with the set containing the identity matrix, sequentially
testing whether $S_{n}\in \{S\}_{l'}$ satisfies ${\rm
dist}(S_{n},Q)>0$ for all $Q\in \{Q\}_{l'}$, and adding $S_{n}$ to
$\{Q\}_{l'}$ if it does. A search for an optimal approximation of
$U$ using gates in $\mathscr{G}$ begins with the construction of a
very large set of unique sequences $\{Q\}_{l'}$.

The utility of $\{Q\}_{l'}$ lies in its ability to predict which
sequences in $\{S\}_{l}, l>l'$ do not need to be compared with $U$
to determine whether they are good approximations, and what the
next sequence worth comparing is. To be more precise, assume every
sequence up to $S_{n-1}$ has been compared with $U$. Let
$\{S_{n-1}\}$ denote this set of compared sequences. Consider
subsequences of $S_{n}$ of length $l'$. If any subsequence is not
in $\{Q\}_{l'}$, there exists a sequence in $\{S_{n-1}\}$
equivalent to $S_{n}$. In other words, a sequence equivalent to
$S_{n}$ has already been compared with $U$, and $S_{n}$ can be
skipped. Furthermore, let
\begin{equation}
S_{n}=G_{i_{N}}\ldots G_{i_{k+l'+1}}G_{i_{k+l'}}\ldots
G_{i_{k+1}}G_{i_{k}}\ldots G_{i_{1}},
\end{equation}
where $G_{i_{k+l'}}\ldots G_{i_{k+1}}$ is the subsequence not in
$\{Q\}_{l'}$.  Let $Q(G_{i_{k+l'}}\ldots G_{i_{k+1}})$ denote the
next sequence in $\{Q\}_{l'}$ after $G_{i_{k+l'}}\ldots
G_{i_{k+1}}$. The next sequence with the potential to not be
equivalent to a sequence in $\{S_{n-1}\}$ is
\begin{equation}
G_{i_{N}}\ldots G_{i_{k+l'+1}}Q(G_{i_{k+l'}}\ldots
G_{i_{k+1}})G_{1}\ldots G_{1}.
\end{equation}
The process of checking subsequences is then repeated on this new
sequence. Skipping sequences in this manner is vastly better than
an exhaustive search, and enables optimal sequences of interesting
length to be obtained. It should be stressed, however, that the
runtime is still exponentially in $l$.

Highly non-optimal but polynomial runtime sequence finding
techniques do exist \cite{Kita97,Niel00,Kita02,Daws04} but will
not be discussed here.

\section{Simple Steane code single-qubit gates}
\label{Solovay:section:gates}

\begin{figure}
\begin{center}
\includegraphics[width=12cm]{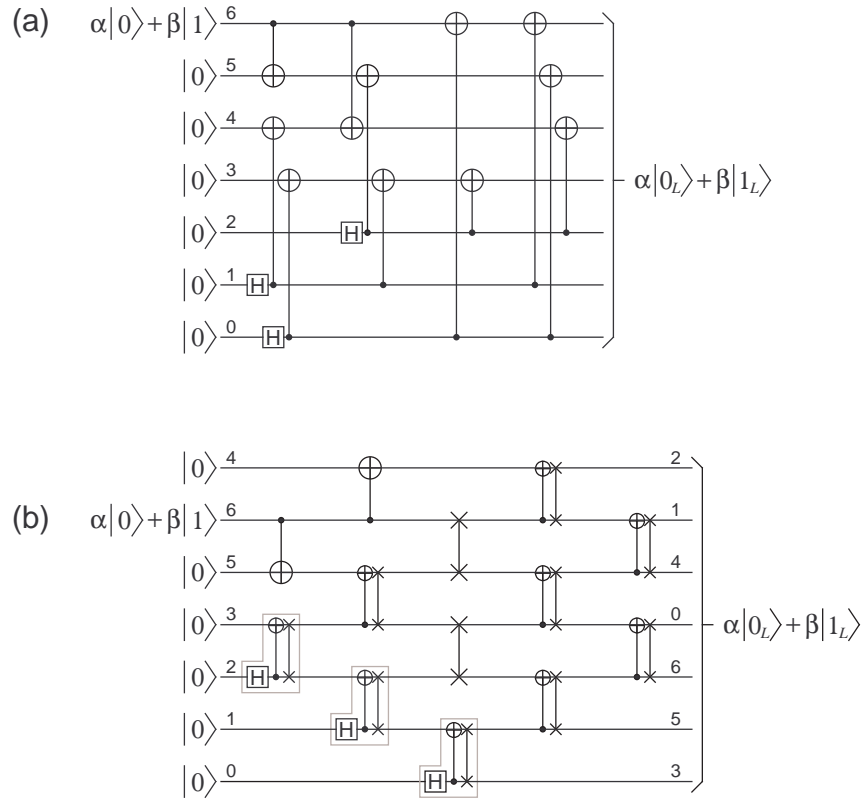}
\end{center}
\caption{Non-fault-tolerant 7-qubit Steane code encoding circuits
taking an arbitrary state $\alpha|0\rangle + \beta|1\rangle$ and
producing $\alpha|0_{L}\rangle + \beta|1_{L}\rangle$. (a) Depth 4
circuit for an architecture able to interact arbitrary pairs of
qubits. (b) Depth 5 circuit for a linear nearest neighbor
architecture.} \label{Solovay:figure:7qec_encode_both}
\end{figure}

For the remainder of the chapter we will restrict our attention to
fault-tolerant single-qubit gates that can be applied to the
7-qubit Steane code. The Steane code representation of states
$|0\rangle$ and $|1\rangle$ is \cite{Stea96}
\begin{eqnarray}
\label{Solovay:eq:zero_L} |0_{L}\rangle &=&
\frac{1}{\sqrt{8}}(|0000000\rangle+
|1010101\rangle+ |0110011\rangle+ |1100110\rangle \nonumber \\
&&\quad +|0001111\rangle+ |1011010\rangle+ |0111100\rangle+
|1101001\rangle), \\
\label{Solovay:eq:one_L} |1_{L}\rangle &=&
\frac{1}{\sqrt{8}}(|1111111\rangle+
|0101010\rangle+ |1001100\rangle+ |0011001\rangle \nonumber \\
&&\quad +|1110000\rangle+ |0100101\rangle+ |1000011\rangle+
|0010110\rangle).
\end{eqnarray}
An equivalent description of this code can be given in terms of
stabilizers \cite{Gott97} which are operators that map the logical
states $|0_{L}\rangle$ and $|1_{L}\rangle$ to themselves.
\begin{eqnarray}
\label{Solovay:eq:stabiliser1} \texttt{IIIXXXX} \\
\label{Solovay:eq:stabiliser2} \texttt{IXXIIXX} \\
\label{Solovay:eq:stabiliser3} \texttt{XIXIXIX} \\
\label{Solovay:eq:stabiliser4} \texttt{IIIZZZZ} \\
\label{Solovay:eq:stabiliser5} \texttt{IZZIIZZ} \\
\label{Solovay:eq:stabiliser6} \texttt{ZIZIZIZ}
\end{eqnarray}
States $|0_{L}\rangle$ and $|1_{L}\rangle$ are the only two that
are simultaneously stabilized by
Eqs~(\ref{Solovay:eq:stabiliser1}--\ref{Solovay:eq:stabiliser6}).
Non-fault-tolerant circuits for both a general and LNN
architecture that take an arbitrary state
$\alpha|0\rangle+\beta|1\rangle$ and produce
$\alpha|0_{L}\rangle+\beta|1_{L}\rangle$ are shown in
Fig.~\ref{Solovay:figure:7qec_encode_both}. The fault-tolerant
preparation of logical states is more complicated, and will be
discussed in the context of $T$-gate ancilla state preparation in
Section~\ref{Solovay:section:Tgate}.

The minimal universal set of single-qubit fault-tolerant gates
that can be applied to a Steane code logical qubit consists of
just the Hadamard gate and the $T$-gate
\begin{equation}
\label{Solovay:eq:Tgate}
T = \left(\begin{array}{cc}
1 & 0 \\
0 & e^{i\pi/4} \\
\end{array} \right).
\end{equation}
For practical purposes, the gates $X$, $Z$, $S$, $S^{\dag}$ should
be added to this set, where
\begin{equation}
\label{Solovay:eq:Sgate}
S = \left( \begin{array}{cc}
1 & 0 \\
0 & i \\
\end{array} \right),
\end{equation}
along with all gates generated by $H$, $X$, $Z$, $S$, $S^{\dag}$.
The complete list of gates that we shall consider is shown in
Eq.~(\ref{Solovay:eq:gate_set}). This is our set $\mathscr{G}$.
Note that gates $\{I,G_{1},\ldots,G_{23}\}$ form a group under
multiplication. Appendix~\ref{ap1} contains the multiplication
table of this group.

\begin{equation}
\begin{tabular}{rclcrcl}
$G_{1}$ &=& $H$ &\qquad\qquad\qquad& $G_{13}$ &=& $HS$ \\
$G_{2}$ &=& $X$ &\qquad\qquad\qquad& $G_{14}$ &=& $HS^{\dag}$ \\
$G_{3}$ &=& $Z$ &\qquad\qquad\qquad& $G_{15}$ &=& $ZXH$ \\
$G_{4}$ &=& $S$ &\qquad\qquad\qquad& $G_{16}$ &=& $SXH$ \\
$G_{5}$ &=& $S^{\dag}$ &\qquad\qquad\qquad& $G_{17}$ &=& $S^{\dag}XH$ \\
$G_{6}$ &=& $XH$ &\qquad\qquad\qquad& $G_{18}$ &=& $HSH$ \\
$G_{7}$ &=& $ZH$ &\qquad\qquad\qquad& $G_{19}$ &=& $HS^{\dag}H$ \\
$G_{8}$ &=& $SH$ &\qquad\qquad\qquad& $G_{20}$ &=& $HSX$ \\
$G_{9}$ &=& $S^{\dag}H$ &\qquad\qquad\qquad& $G_{21}$ &=& $HS^{\dag}X$ \\
$G_{10}$ &=& $ZX$ &\qquad\qquad\qquad& $G_{22}$ &=& $S^{\dag}HS$ \\
$G_{11}$ &=& $SX$ &\qquad\qquad\qquad& $G_{23}$ &=& $SHS^{\dag}$ \\
$G_{12}$ &=& $S^{\dag}X$ &\qquad\qquad\qquad& $G_{24}$ &=& $T$
\end{tabular}
\label{Solovay:eq:gate_set}
\end{equation}

\begin{figure}
\begin{center}
\includegraphics[width=12cm]{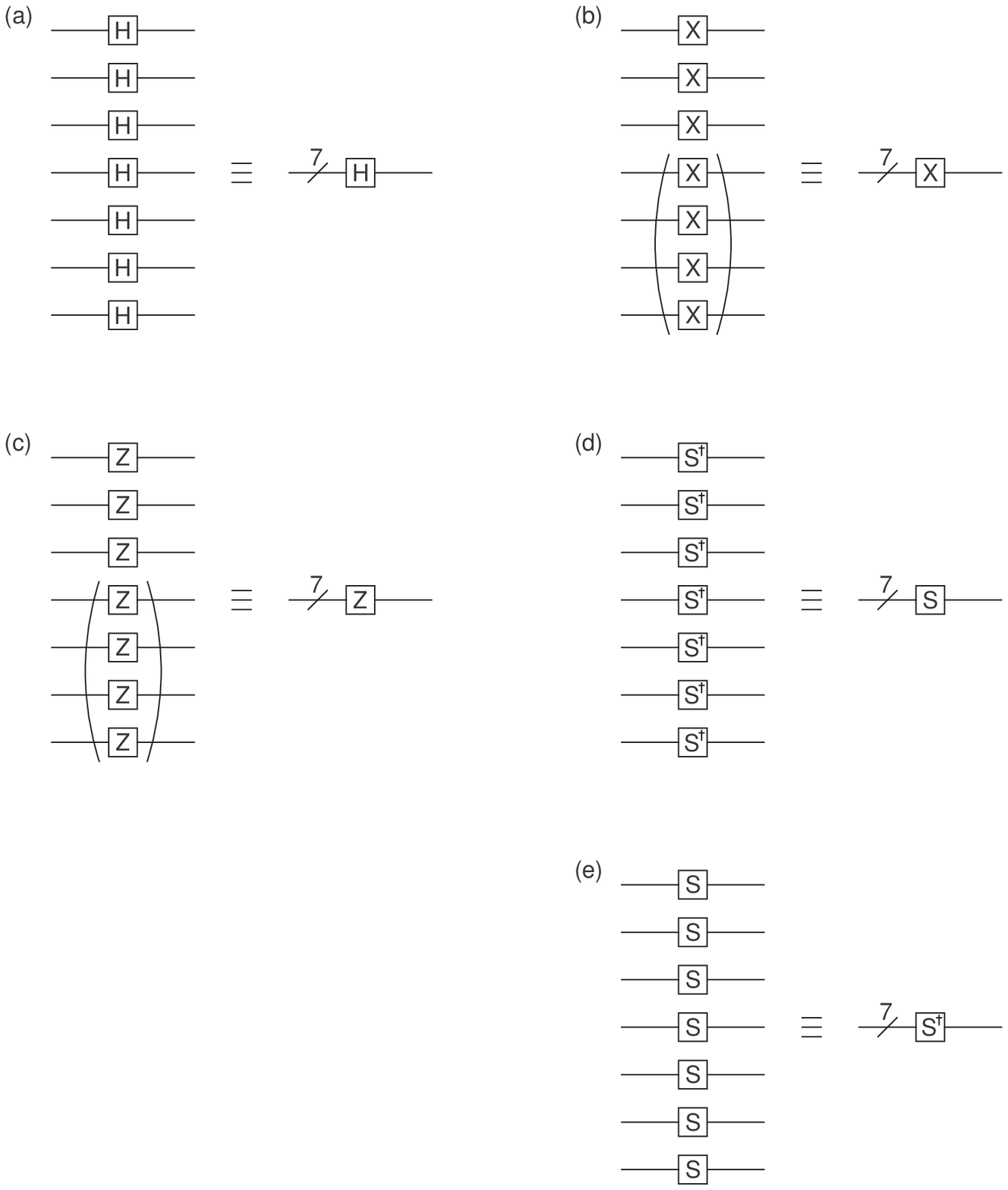}
\end{center}
\caption{Circuits fault-tolerantly applying common single-qubit
gates to Steane code logical qubits.  Gates in brackets are
optional as they implement stabilizers.}
\label{Solovay:figure:single}
\end{figure}

To justify the use of such a large set $\mathscr{G}$, consider the
transversal circuits shown in Fig.~\ref{Solovay:figure:single}
implementing $H$, $X$, $Z$, $S$ and $S^{\dag}$. By combination, it
can be seen that gates $\{G_{6},\ldots,G_{23}\}$ can also be
implemented with simple transversal applications of single qubit
gates. As we shall see in Section~\ref{Solovay:section:Tgate}, by
comparison the $T$-gate is extremely complicated to implement.
Since we are interested in minimal complexity as well as minimum
length sequences of gates in $\mathscr{G}$, it would be
unreasonable to count $G_{23}$ as three gates when in reality it
can be implemented as easily as any other gate
$\{G_{1},\ldots,G_{22}\}$. Since $\{I,G_{1},\ldots,G_{23}\}$ is a
group under multiplication, minimum length sequences of gates
approximating some $U$ outside $\mathscr{G}$ will alternate
between an element of $\{G_{1},\ldots,G_{23}\}$ and a $T$-gate.
Note that the $T^{\dag}$-gate is not required in $\mathscr{G}$ for
universality or efficiency as, in gate sequences of length $l\geq
2$, it is equally efficient to use $S^{\dag}T$ or $TS^{\dag}$. The
extra $S^{\dag}$-gate is absorbed into neighboring $G_{i}$-gates,
$i < 24$.

\section{The fault-tolerant T-gate}
\label{Solovay:section:Tgate}

\begin{figure}
\begin{center}
\includegraphics[width=9cm]{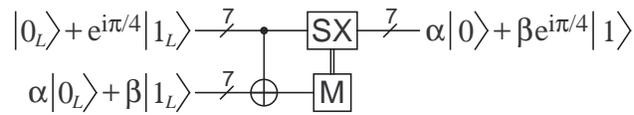}
\end{center}
\caption{High-level representation of the circuit implementing the
$T$-gate on a Steane code logical qubit. Input
$\alpha|0_{L}\rangle + \beta|1_{L}\rangle$ is transformed into
$\alpha|0_{L}\rangle + \beta e^{i\pi/4}|1_{L}\rangle$.}
\label{Solovay:figure:Tgate_simp}
\end{figure}

Moving on to implementing the fault-tolerant $T$-gate
\cite{Niel00}, the basic idea is to prepare an ancilla state
$|0_{L}\rangle+e^{i\pi/4}|1_{L}\rangle$ then apply the circuit
shown in Fig.~\ref{Solovay:figure:Tgate_simp}. Tracing the action
of Fig.~\ref{Solovay:figure:Tgate_simp}, we initially have
\begin{equation}
(|0_{L}\rangle + e^{i\pi/4}|1_{L}\rangle)(\alpha|0_{L}\rangle + \beta|1_{L}\rangle).
\end{equation}
After applying the \CNOT\ we obtain
\begin{equation}
\begin{tabular}{rl}
&$\alpha|0_{L}\rangle|0_{L}\rangle+\beta|0_{L}\rangle|1_{L}\rangle
+\alpha e^{i\pi/4}|1_{L}\rangle|1_{L}\rangle+\beta
e^{i\pi/4}|1_{L}\rangle|0_{L}\rangle$ \\
=&$(\alpha|0_{L}\rangle + \beta e^{i\pi/4}|1_{L}\rangle)|0_{L}\rangle
+(\beta|0_{L}\rangle + \alpha e^{i\pi/4}|1_{L}\rangle)|1_{L}\rangle$.
\end{tabular}
\end{equation}
After measuring the lower logical qubit, if $|0_{L}\rangle$ is
observed (meaning one of the eight bit strings shown in
Eq.~(\ref{Solovay:eq:zero_L}) or a bit string a single bit
different from one of these eight), no further action is required.
If $|1_{L}\rangle$ is observed, applying the logical gate $SX$ to
the top qubit will yield the desired state up to an irrelevant
global phase. Note that the measurement step and subsequent
classical processing allows the correction of a single bit-flip
error and is insensitive to phase errors.

\begin{figure}
\begin{center}
\includegraphics[width=7.5cm]{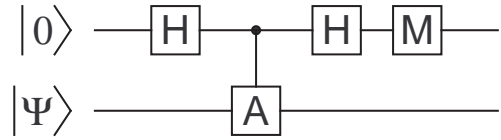}
\end{center}
\caption{Circuit measuring whether $|\Psi\rangle$ is in the $+1$
or $-1$ eigenstate of $A$.} \label{Solovay:figure:nft_op_meas}
\end{figure}

To fault-tolerantly prepare the ancilla state, we first need to be
able to fault-tolerantly prepare the state $|0_{L}\rangle$. As we
shall see, to do this, we need to be able to fault-tolerantly
determine whether a state $|\Psi\rangle$ is in the $+1$ or $-1$
eigenstate of a self-inverse operator $A$ ($A^{2}=I$). A
non-fault-tolerant circuit doing this is shown in
Fig.~\ref{Solovay:figure:nft_op_meas}. It is instructive to trace
the action of the circuit. The initial state is
$|0\rangle|\Psi\rangle$, which after the first Hadamard becomes
$(|0\rangle + |1\rangle)|\Psi\rangle$. After the controlled-$A$
the state becomes $|0\rangle|\Psi\rangle + |1\rangle
A|\Psi\rangle$. After the second Hadamard
\begin{equation}
\begin{tabular}{rl}
& $(|0\rangle + |1\rangle)|\Psi\rangle + (|0\rangle - |1\rangle)A|\Psi\rangle $ \\
= & $|0\rangle(|\Psi\rangle +
A|\Psi\rangle)+|1\rangle(|\Psi\rangle - A|\Psi\rangle)$.
\end{tabular}
\end{equation}
If a zero is measured, the lower qubit will be in the $+1$
eigenstate $|\Psi\rangle + A|\Psi\rangle$.  Conversely, if one is
measured, the lower qubit will be in the $-1$ eigenstate
$|\Psi\rangle - A|\Psi\rangle$.

\begin{figure}
\begin{center}
\includegraphics[width=12cm]{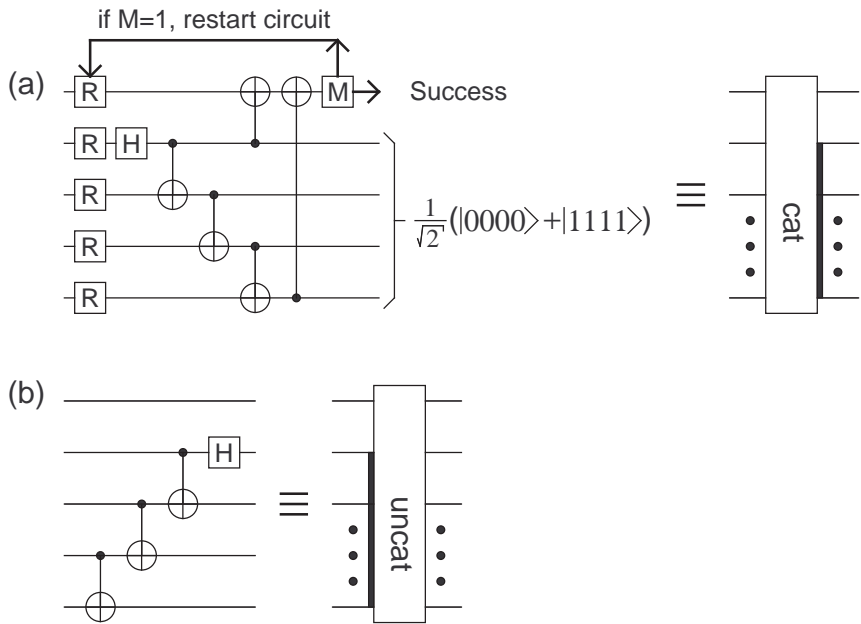}
\end{center}
\caption{(a) Simple circuit fault-tolerantly preparing a cat
state. (b) Circuit undoing the preparation of a cat state.}
\label{Solovay:figure:cat_1_meas}
\end{figure}

\begin{figure}
\begin{center}
\includegraphics[width=12cm]{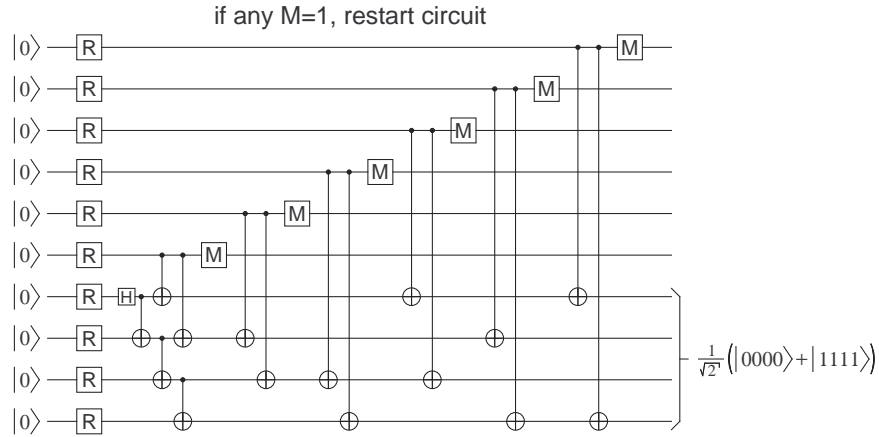}
\end{center}
\caption{Typical, but unnecessarily complicated circuit
fault-tolerantly preparing a cat state.}
\label{Solovay:figure:cat_6_meas}
\end{figure}

The specific self-inverse operators we wish to measure are the
stabilizers
Eqs~(\ref{Solovay:eq:stabiliser1}--\ref{Solovay:eq:stabiliser3}).
To build a fault-tolerant circuit measuring these multiple qubit
operators, the control qubit shown in
Fig.~\ref{Solovay:figure:nft_op_meas} must be replaced by a cat
state so that each qubit modified by the stabilizer is controlled
by a different qubit in the cat state. This is necessary to
prevent a single error in a control qubit propagating to multiple
target qubits. This in turn necessitates fault-tolerant cat state
preparation which is shown in
Fig.~\ref{Solovay:figure:cat_1_meas}a \cite{Pres97}. A single bit-
or phase-flip anywhere in this circuit causes at most one error in
the final state. This circuit is significantly simpler, and no
less robust than the fault-tolerant cat state preparation circuit
suggested in \cite{Niel00} (Fig.~\ref{Solovay:figure:cat_6_meas}).
The uncat circuit of Fig.~\ref{Solovay:figure:cat_1_meas}b is
fault-tolerant purely because its output is a single qubit and by
definition a single error can cause of most one error in the
output.

\begin{figure}
\begin{center}
\includegraphics[height=6cm]{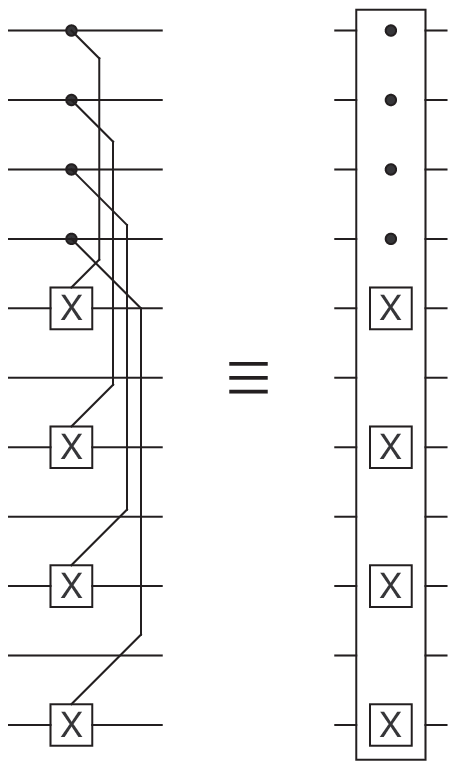}
\end{center}
\caption{Symbolic representation of transversal controlled
operations.} \label{Solovay:figure:tangle}
\end{figure}

\begin{figure}
\begin{center}
\includegraphics[width=12cm]{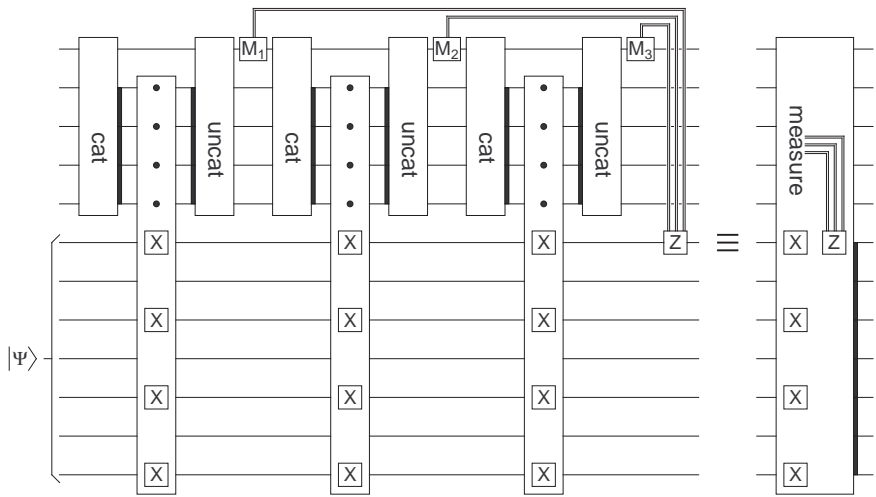}
\end{center}
\caption{Circuit fault-tolerantly projecting $|\Psi\rangle$ onto
the $\pm 1$ eigenstates of $\texttt{XIXIXIX}$, then converting
$-1$ eigenstate's into $+1$ eigenstates. The third measurement
structure can be omitted if $M_{1}=M_{2}$.}
\label{Solovay:figure:measure_plus_classical}
\end{figure}

Using the circuit notation shown in
Fig.~\ref{Solovay:figure:tangle}, the complete circuit for
fault-tolerantly measuring a stabilizer is shown in
Fig.~\ref{Solovay:figure:measure_plus_classical}. Note that the
basic stabilizer measurement circuit appears three times since a
single error in a cat state block, while not propagating to
multiple qubits in the logical state block, almost always causes
an incorrect measurement.  To ensure a probability $O(p^{2})$ of
incorrect measurement, the process must be repeated up to three
times.  The third measurement structure can be omitted if the
first two measurements are the same. The final triply controlled
$Z$-gate is only applied if the majority of the measurements are
one. Note that this assumes fast and reliable classical processing
is available. The final $Z$-gate converts a $-1$ eigenstate of
$XIXIXIX$ into a $+1$ eigenstate. Thus the output of
Fig.~\ref{Solovay:figure:measure_plus_classical} is the $+1$
eigenstate of $XIXIXIX$ with probability $O(p^{2})$ of failure
(i.e. more than one incorrect output qubit).

\begin{figure}
\begin{center}
\includegraphics[width=12cm]{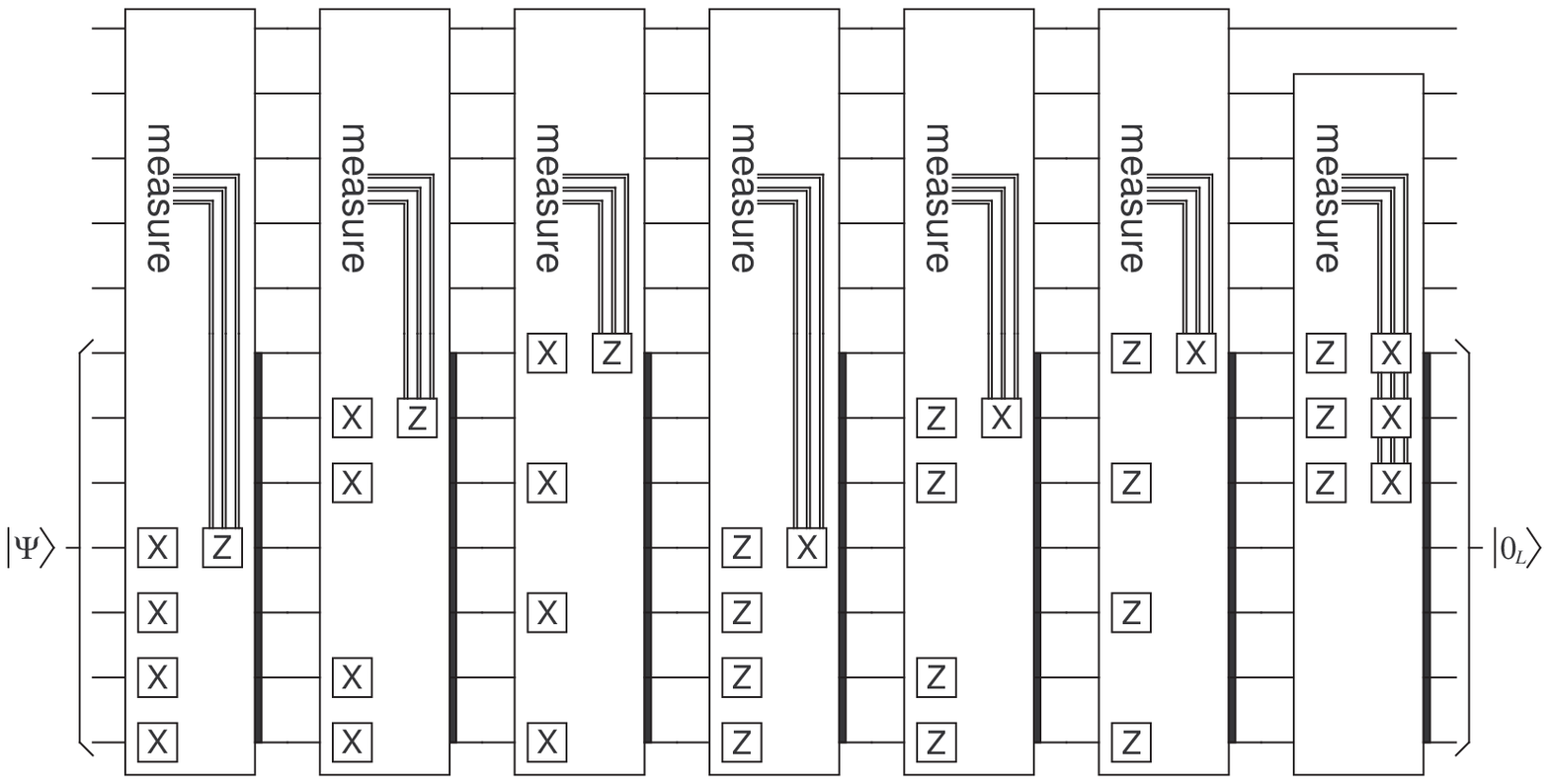}
\end{center}
\caption{Circuit taking arbitrary input and producing
$|0_{L}\rangle$ by repeated stabilizer measurement.}
\label{Solovay:figure:initialise_7}
\end{figure}

We now have the necessary tools to fault-tolerantly prepare
$|0_{L}\rangle$. Recall that $|0_{L}\rangle$ and $|1_{L}\rangle$
are the only two states simultaneously stabilized by all of
Eqs~(\ref{Solovay:eq:stabiliser1}--\ref{Solovay:eq:stabiliser6}).
If we include the logical $Z$ operator, $|0_{L}\rangle$ is the
unique state stabilized by $Z$ and all six stabilizers. The state
$|0_{L}\rangle$ could thus be created using the circuit of
Fig.~\ref{Solovay:figure:initialise_7} which outputs
$|0_{L}\rangle$ for arbitrary input \cite{Niel00}.

\begin{figure}
\begin{center}
\includegraphics[width=9cm]{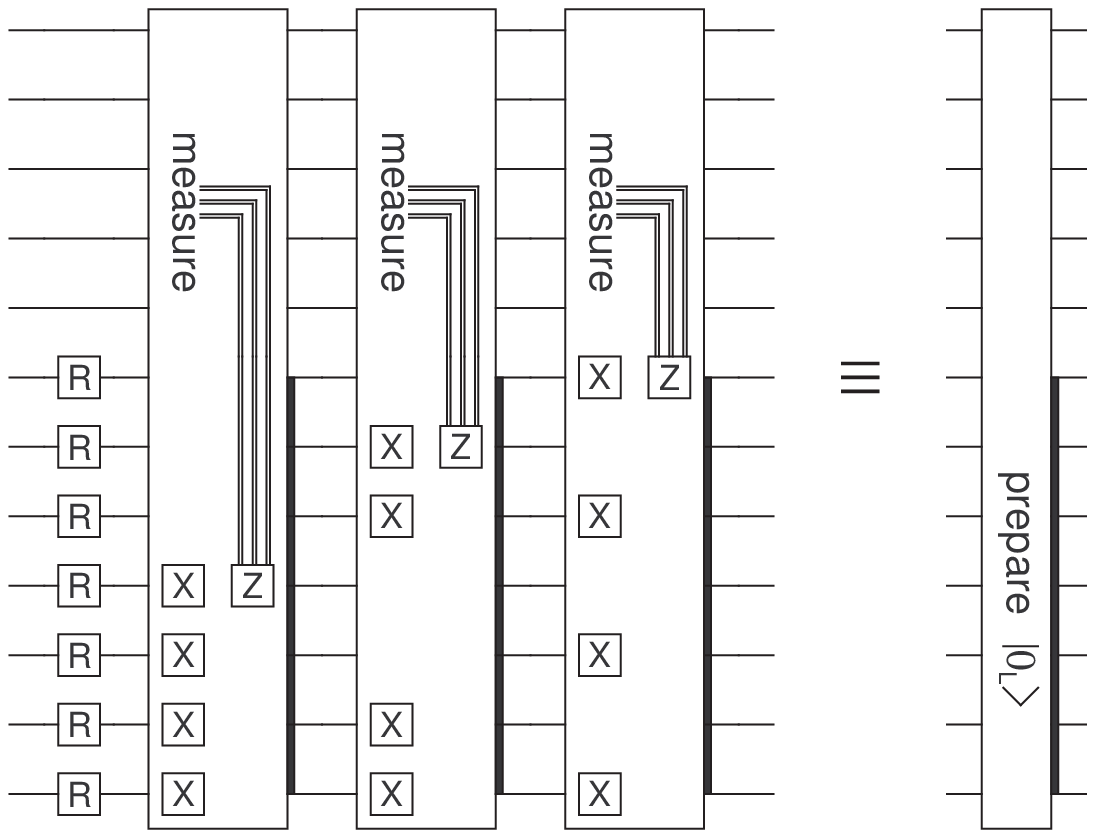}
\end{center}
\caption{Circuit taking arbitrary input and producing
$|0_{L}\rangle$ via physical resetting and just three stabilizer
measurements.} \label{Solovay:figure:initialise}
\end{figure}

A better way of obtaining $|0_{L}\rangle$, is to start with the
state $|0000000\rangle$ which is physically accessible in a
quantum computer architecture either via some form of special
reset operation, or measurement possibly followed by an $X$-gate.
State $|0000000\rangle$ is a $+1$ eigenstate of logical $Z$ and
Eqs~(\ref{Solovay:eq:stabiliser4}--\ref{Solovay:eq:stabiliser6}),
therefore only stabilizers
Eqs~(\ref{Solovay:eq:stabiliser1}--\ref{Solovay:eq:stabiliser3})
need to be measured (Fig.~\ref{Solovay:figure:initialise}).

\begin{figure}
\begin{center}
\includegraphics[width=12cm]{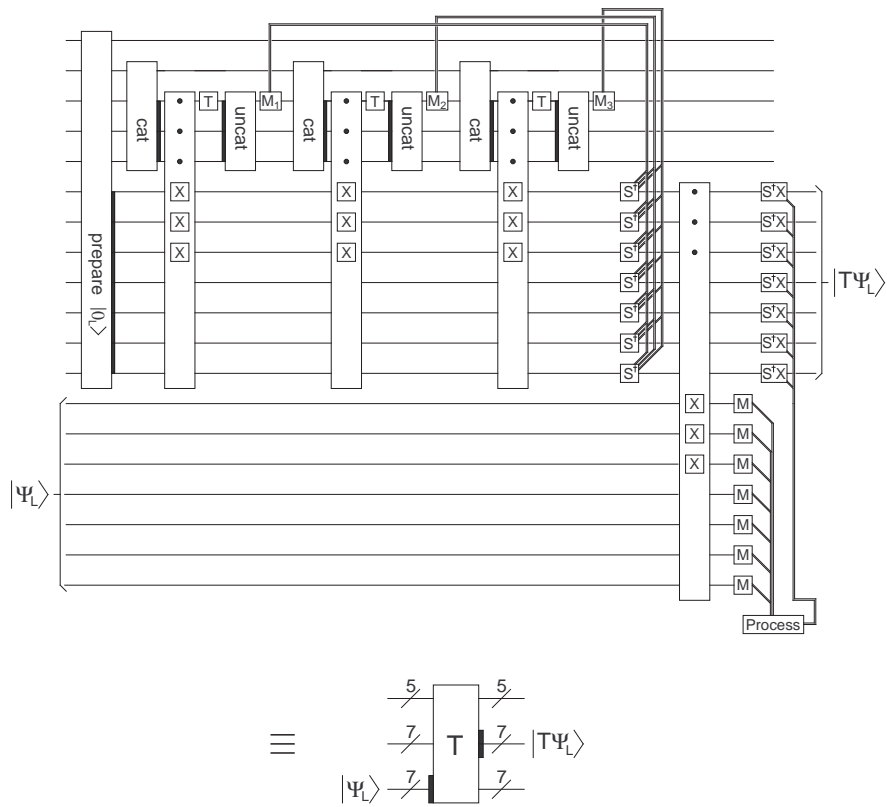}
\end{center}
\caption{Complete circuit implementing the $T$-gate on a Steane
code logical qubit.} \label{Solovay:figure:Tgate}
\end{figure}

To complete the construction of the ancilla state, and hence the
$T$-gate (Fig.~\ref{Solovay:figure:Tgate}), the operator
$e^{i\pi/4}X$ is measured. Note that $e^{i\pi/4}X$ is not
self-inverse, but nevertheless the circuit works as required.
Specifically, before cat state preparation we have
$|000\rangle|0_{L}\rangle$. After cat state creation this becomes
\begin{equation}
\label{Solovay:eq:cat7}
\frac{1}{\sqrt{2}}(|000\rangle+|111\rangle)|0_{L}\rangle.
\end{equation}
After transversal \CNOT\ we have
\begin{equation}
\frac{1}{\sqrt{2}}(|000\rangle|0_{L}\rangle+|111\rangle|1_{L}\rangle).
\end{equation}
Note that only three physical \CNOT\ gates are required to
implement a logical \CNOT\ gate on the Steane code due to its
stabilizer structure. After the single-qubit $T$-gate we have
\begin{equation}
\label{Solovay:eq:single_Tgate}
\frac{1}{\sqrt{2}}(|000\rangle|0_{L}\rangle+e^{i\pi/4}|111\rangle|1_{L}\rangle).
\end{equation}
After uncat we have
\begin{equation}
\label{Solovay:eq:premeas}
|0\rangle\frac{1}{\sqrt{2}}(|0_{L}\rangle+e^{i\pi/4}|1_{L}\rangle)
+|1\rangle\frac{1}{\sqrt{2}}(|0_{L}\rangle-e^{i\pi/4}|1_{L}\rangle),
\end{equation}
resulting in the state
$(|0_{L}\rangle+e^{i\pi/4}|1_{L}\rangle)/\sqrt{2}$ if zero is
measured, and $(|0_{L}\rangle-e^{i\pi/4}|1_{L}\rangle)/\sqrt{2}$
if one is measured.  Note that the steps shown in
Eqs~(\ref{Solovay:eq:cat7}--\ref{Solovay:eq:premeas}) must be
repeated up to three times to be able to say that a 0 or 1 has
been measured with probability of error $O(p^{2})$. The final
logical $S$-gate converts
$(|0_{L}\rangle-e^{i\pi/4}|1_{L}\rangle)/\sqrt{2}$ into
$(|0_{L}\rangle+e^{i\pi/4}|1_{L}\rangle)/\sqrt{2}$. Under the
assumptions that 2-qubit gates, measurement, reset and classical
processing each have depth 1, single-qubit gates have depth zero
and do not contribute to the gate count, and arbitrary disjoint
2-qubit gates can be implemented in parallel,
Table~\ref{Solovay:table:Tgate} summaries the best case complexity
of the $T$-gate.

\begin{table}
  \centering
  \begin{tabular}{c|c}
    Circuit Element & Count \\
    \hline
    Qubits & 19 \\
    Gates & 93 \\
    Resets & 45 \\
    Measurements & 17 \\
    Depth & 92 \\
  \end{tabular}
  \caption{Best case complexity of the T-gate.}\label{Solovay:table:Tgate}
\end{table}

\section{Approximations of phase gates}
\label{Solovay:section:phase}

We now use the machinery described in this chapter to construct
optimal fault-tolerant approximations of single-qubit phase
rotation gates
\begin{equation}
R_{2^{d}}= \left(
\begin{array}{cc}
1 & 0 \\
0 & e^{i\pi/2^{d}}
\end{array}
\right).
\end{equation}
Gates $R_{2^{d}}$ are examples of gates used in the single-qubit
quantum Fourier transform that forms part of the Shor circuits
described in Chapters~\ref{ShorLNN}--\ref{ShorGates}. Note that
phase rotations of angle $2\pi x/2^{d}$, where $x$ is a $d$-digit
binary number, are also required, but the properties of
fault-tolerant approximations of such gates can be inferred from
$R_{2^{d}}$.

\begin{figure}
\begin{center}
\begin{tabular}{cc}
\includegraphics[width=60mm]{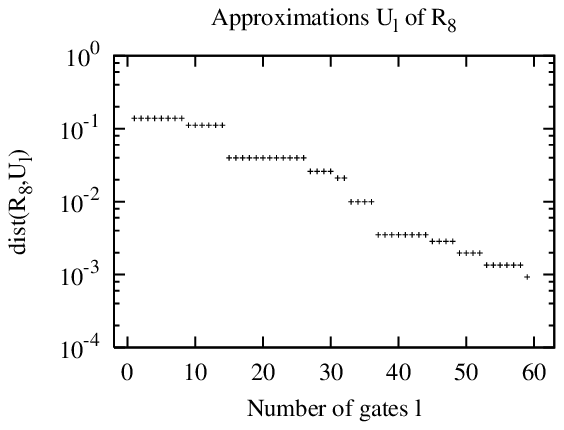} & \includegraphics[width=60mm]{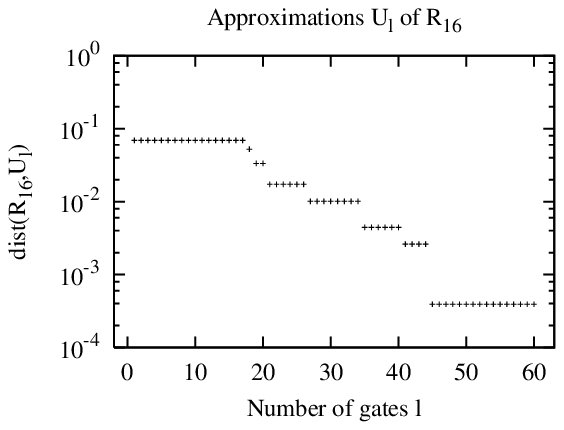}\\
\includegraphics[width=60mm]{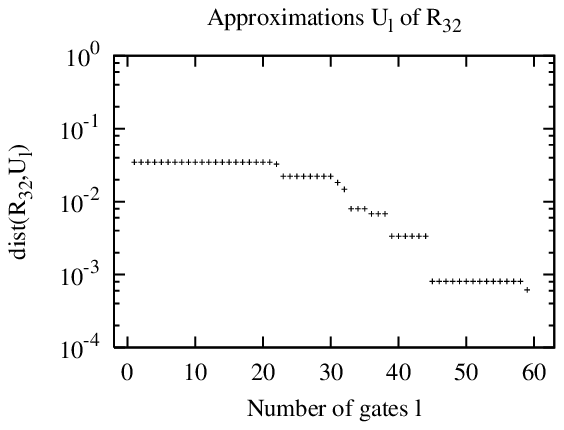} & \includegraphics[width=60mm]{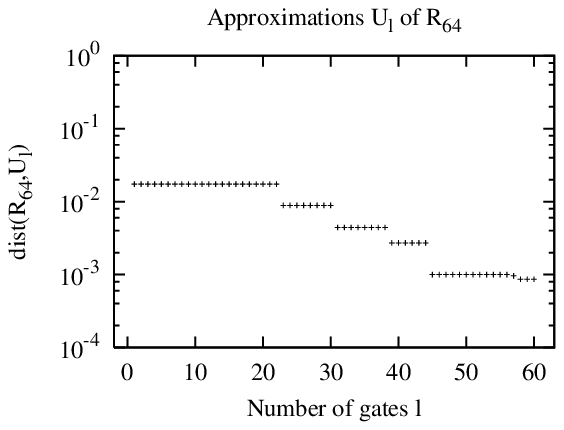}\\
\multicolumn{2}{c}{\includegraphics[width=60mm]{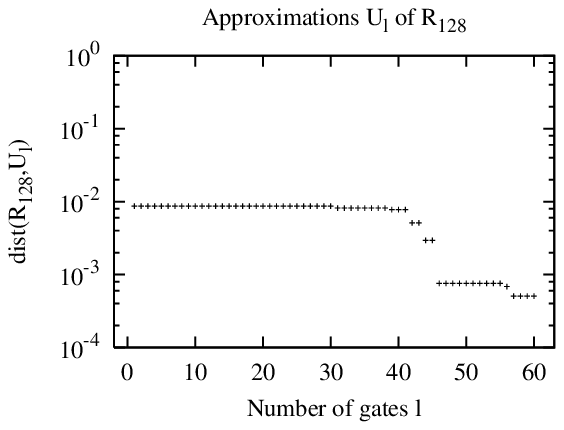}} \\
\end{tabular}
\end{center}
\caption{Optimal fault-tolerant approximations $U_{l}$ of phase
rotation gates $R_{2^{d}}$, for $R_{8}$ to $R_{128}$.}
\label{Solovay:figure:constructions}
\end{figure}

For a given $R_{2^{d}}$, and maximum number of gates $l$ in
$\mathscr{G}$, Fig.~\ref{Solovay:figure:constructions} shows ${\rm
dist}(R_{2^{d}},U_{l})$ where $U_{l}$ is an optimal sequence of at
most $l$ gates in $\mathscr{G}$ minimizing ${\rm
dist}(R_{2^{d}},U_{l})$. For $d\geq 3$, $U_{1}$ is equivalent to
the identity. Note that as $d$ increases, $R_{2^{d}}$ becomes
closer and closer to the identity, lowering the value of ${\rm
dist}(R_{2^{d}},U_{1})$, and increasing the value of $l$ required
to obtain an approximation $U_{l}$ that is closer to $R_{2^{d}}$
than the identity. In fact, for $R_{128}$ the shortest sequence of
gates that provides a better approximation of $R_{128}$ than the
identity has length $l=31$. There are a very large number of
optimal sequences of this length. An example of one with a minimal
number of $T$-gates is
\begin{equation}
\label{Solovay:eq:U31}
\begin{tabular}{rcl}
$U_{31}$&=&$HTHT(SH)T(SH)T(SH)THTHT(SH)$ \\
       &&$THTHT(SH)THTHTHT(SH)T(S^{\dag}H)$
\end{tabular}
\end{equation}
The parentheses group standard gates into elements of the set
$\mathscr{G}$ defined in Eq.~(\ref{Solovay:eq:gate_set}). Note
that ${\rm dist}(R_{128},I)= 8.7\times 10^{-3}$ whereas ${\rm
dist}(R_{128},U_{33})= 8.1\times 10^{-3}$. In other words
Eq.~(\ref{Solovay:eq:U31}) is only slightly better than the
identity. This immediately raises the question of how many gates
are required to construct a sufficiently good approximation.

\begin{figure}
\begin{center}
\includegraphics[width=12cm]{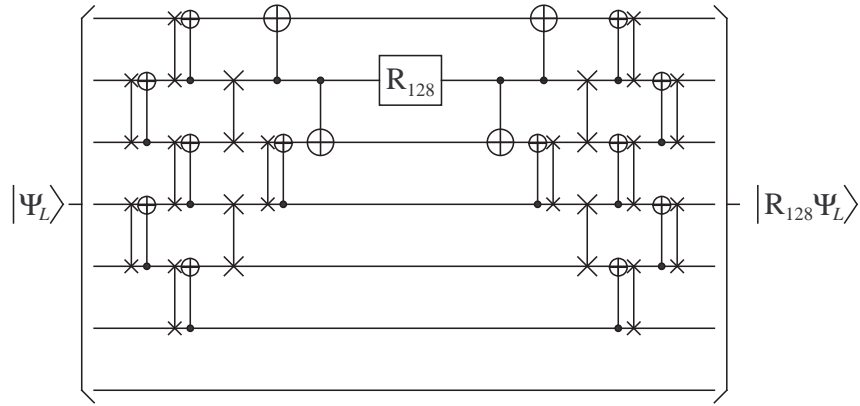}
\end{center}
\caption{Non-fault-tolerant circuit exactly implementing $R_{128}$
by first decoding the logical qubit and re-encoding after
application of $R_{128}$.} \label{Solovay:figure:nft_rotate}
\end{figure}

In Chapter~\ref{ShorGates}, it was shown that
\begin{equation}
U=\left(
\begin{array}{cc}
  1 & 0 \\
  0 & e^{i(\pi/128 + \pi/512)}  \\
\end{array}
\right)
\end{equation}
was sufficiently close to $R_{128}$.  This is, of course, only a
property of Shor's algorithm, not a universal property of quantum
circuits. Given ${\rm dist}(R_{128},U)=2.2\times 10^{-3}$, a
sufficiently accurate fault-tolerant approximation $U_{l}$ of
$R_{128}$ must therefore satisfy ${\rm
dist}(R_{128},U_{l})<2.2\times 10^{-3}$. The smallest value of $l$
for which this is true is 46, and one of the many optimal gate
sequences satisfying ${\rm dist}(R_{128},U_{46})=7.5\times
10^{-4}$ is
\begin{equation}
\label{Solovay:eq:U46}
\begin{tabular}{rcl}
$U_{46}$&=&$HTHTHT(SH)THT(SH)T(SH)T(SH)THT$ \\
       &&$(SH)T(SH)THTHT(SH)T(SH)THT(SH)T$ \\
       &&$(SH)T(SH)THT(SH)THT(HS^{\dag})T$
\end{tabular}
\end{equation}
Parentheses again group standard gates into elements of
$\mathscr{G}$. Now that we have a minimal complexity circuit
sufficiently close to $R_{128}$, the immediate question is whether
it is practical. An alternative to Eq.~(\ref{Solovay:eq:U46}) is
shown in Fig.~\ref{Solovay:figure:nft_rotate} which simply decodes
the logical qubit, applies $R_{128}$, then re-encodes. This simple
non-fault-tolerant circuit will fail (generate more than one error
in the output logical qubit) if a single error occurs almost
anywhere in the top six qubits. Given there are $11\times 6=66$
possible error locations, the probability of no errors in the top
six qubits is $(1-p)^{66}$. This is the worst-case reliability of
the circuit.

\begin{figure}
\begin{center}
\includegraphics[width=12cm]{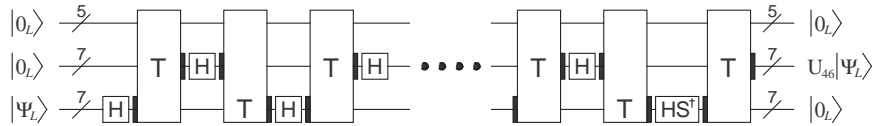}
\end{center}
\caption{Schematic of a minimum complexity, sufficiently accurate
fault-tolerant approximation of $R_{128}$ given in full by
Eq.~(\ref{Solovay:eq:U46}).} \label{Solovay:figure:U46}
\end{figure}

A partial schematic of the circuit corresponding to
Eq.~\ref{Solovay:eq:U46} is shown in
Fig.~\ref{Solovay:figure:U46}. As the circuit is fault-tolerant,
it only fails if at least two errors occur within the circuit. Any
analysis of the reliability of the circuit is complicated by the
fact that the $T$-gates that comprise the bulk of the circuit have
error correction built in at a number of places. Furthermore, when
errors are detected and corrected, the circuit typically increases
in depth. Referring back to Fig.~\ref{Solovay:figure:Tgate}, we
shall assume that the $T$-gate is only sensitive to errors in the
lower 14 qubits and that the depth of the circuit is never
increased by errors. From Table~\ref{Solovay:table:Tgate}, the
best case depth of the $T$-gate is 92. This implies an area
sensitive to errors of approximately 1300. Given there are 23
$T$-gates in Fig.~\ref{Solovay:figure:U46}, the total area
sensitive to errors is approximately 30000. The reliability of
Fig.~\ref{Solovay:figure:U46} is thus approximately
$(1-p)^{30000}+30000p(1-p)^{29999}$, which is only greater than
the non-fault-tolerant circuit for $p<1.4\times 10^{-7}$.

\begin{figure}
\begin{center}
\includegraphics[width=12cm]{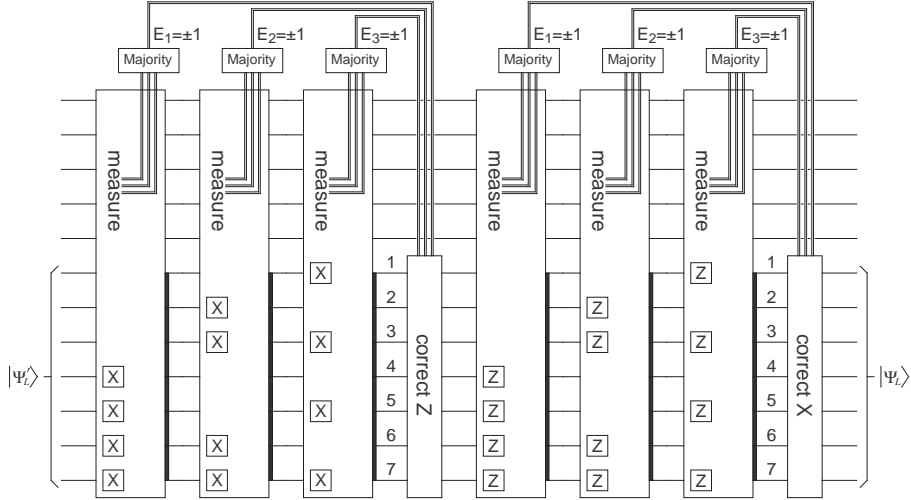}
\end{center}
\caption{Circuit fault-tolerantly correcting an arbitrary single
error within the logical qubit. The qubit acted on by the correct
$Z$/$X$ boxes is described by Table~\ref{Solovay:table:select}.}
\label{Solovay:figure:ft_correct}
\end{figure}

\begin{figure}
\begin{center}
\includegraphics[width=12cm]{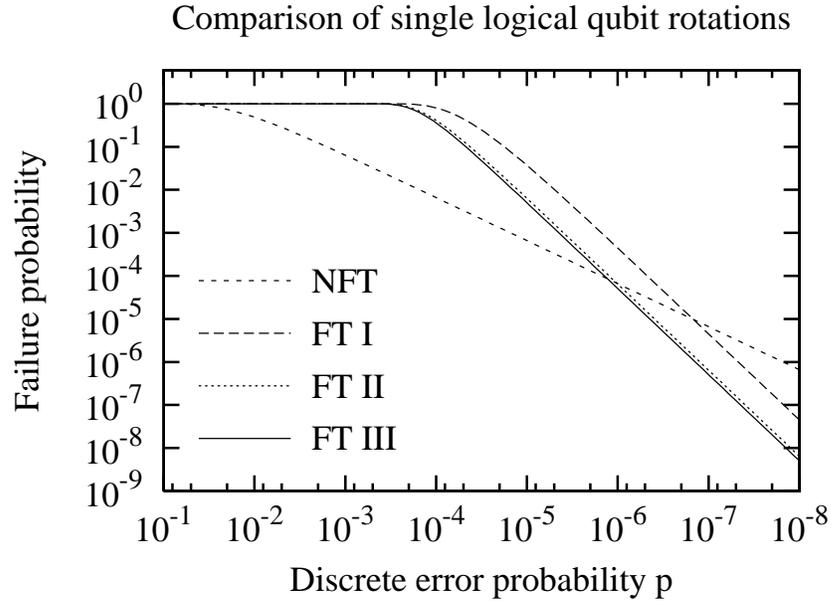}
\end{center}
\caption{Approximate probability of more than one error in the
output logical qubit versus probability per qubit per time step of
discrete error for different circuits implementing a $R_{128}$
phase rotation gate. (a) NFT: non-fault-tolerant circuit from
Fig.~\ref{Solovay:figure:nft_rotate}, (b) FT I: fault-tolerant
circuit from Fig.~\ref{Solovay:figure:U46}, (c) FT II: as above
but with Fig.~\ref{Solovay:figure:ft_correct} error correction
after every second T-gate, (d) FT III: as above but with error
correction after every T-gate. Note that all fault-tolerant
results are for the 7-qubit Steane code without concatenation.}
\label{Solovay:figure:comparison}
\end{figure}

A fault-tolerant circuit correcting an arbitrary single error in a
Steane logical qubit is shown in
Fig.~\ref{Solovay:figure:ft_correct}. Consider the first half of
the circuit. Given eigenvalue measurements $E_{1}$, $E_{2}$,
$E_{3}$, the appropriate qubit to correct is shown in
Table~\ref{Solovay:table:select}. Note that slightly less complex
circuits exist that use more qubits \cite{Stea02}, but our choice
of circuit will not play a significant role in the following
analysis. By applying error correction to the logical qubit after
each $T$-gate in Fig.~\ref{Solovay:figure:U46}, the reliability of
the circuit can be increased. The best case depth of
Fig.~\ref{Solovay:figure:ft_correct} is 120, and if we assume that
the circuit is only sensitive to errors in the lower seven qubits,
the total area sensitive to errors is approximately 800. The error
corrected $U_{46}$ circuit will only fail if two errors occur
within a single $T$-gate and error correction block. The
reliability of a single block is $(1-p)^{2100}+2100p(1-p)^{2099}$.
The failure probability of the non-fault-tolerant circuit, the
fault-tolerant circuit without correction, with correction after
every second $T$-gate, and with correction after every $T$-gate is
compared in Fig.~\ref{Solovay:figure:comparison}.

\begin{table}
\begin{center}
\begin{tabular}{ccc|c}
$E_{1}$ & $E_{2}$ & $E_{3}$ & Qubit \\
\hline
\rule[-1ex]{0pt}{3.7ex}%
$+1$ & $+1$ & $+1$ & no error \\
$+1$ & $+1$ & $-1$ & 1 \\
$+1$ & $-1$ & $+1$ & 2 \\
$+1$ & $-1$ & $-1$ & 3 \\
$-1$ & $+1$ & $+1$ & 4 \\
$-1$ & $+1$ & $-1$ & 5 \\
$-1$ & $-1$ & $+1$ & 6 \\
$-1$ & $-1$ & $-1$ & 7
\end{tabular}
\caption{Qubit to correct given certain sequence of eigenvalue
measurements in both halves of
Fig.~\ref{Solovay:figure:ft_correct}.}
\label{Solovay:table:select}
\end{center}
\end{table}

Of the fault-tolerant circuits, the one with error correction
after every $T$-gate performs best. Nevertheless, this circuit is
still only more reliable than the non-fault-tolerant circuit for
$p<1.3\times 10^{-6}$. Given that $p\sim 10^{-6}$ is likely to be
very difficult to achieve in practice, longer error correction
code words or concatenation would be required to make the
fault-tolerant circuit practical \cite{Stea03b}. Given that
Fig.~\ref{Solovay:figure:U46} is both extremely complex and the
simplest fault-tolerant circuit sufficiently close to $R_{128}$,
for practical computation non-fault-tolerant circuits similar to
Fig.~\ref{Solovay:figure:nft_rotate} are likely to remain the best
way to implement arbitrary rotations for the foreseeable future.

In Shor's algorithm, the use of non-fault-tolerant rotations would
be acceptable as only $2L$ such gates are used to factorize an
$L$-bit number $N$. Furthermore, only half of
Fig.~\ref{Solovay:figure:nft_rotate} would be required as these
gates immediately precede measurement, and there is no point
re-encoding before measurement. In a 4096 bit factorization, the
total area of non-fault-tolerant circuit would be approximately
$2\times 10^{5}$. Assuming the rest of the Shor circuit uses
sufficient error correction to be reliable, if $p\sim 10^{-5}$,
the average number of errors in the non-fault-tolerant part of the
circuit would be two --- completely manageable with just a few
repetitions of the entire circuit or minimal classical processing.

\section{Approximations of arbitrary gates}
\label{Solovay:section:arbitrary}

\begin{figure}
\begin{center}
\includegraphics[width=10cm]{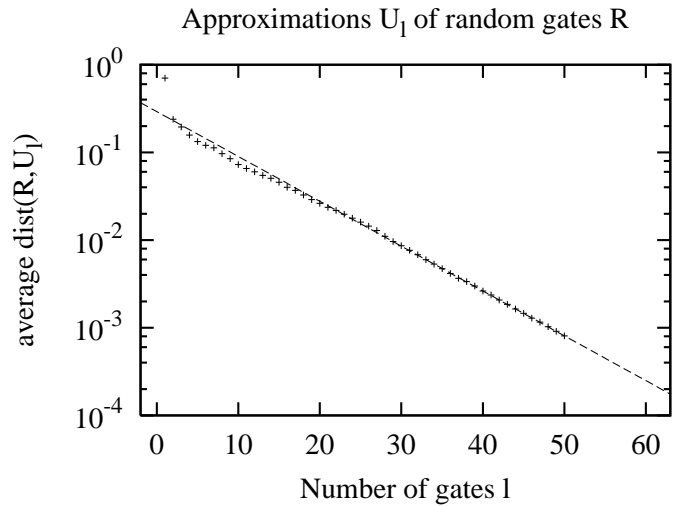}
\end{center}
\caption{Average accuracy of optimal fault-tolerant gate sequence
approximations of length $l$.} \label{Solovay:figure:average}
\end{figure}

In this section, we investigate the properties of fault-tolerant
approximations of arbitrary single-qubit gates
\begin{equation}
\label{Solovay:eq:arbitrary} U=\left(
\begin{array}{cc}
\cos(\theta/2)e^{i(\alpha+\beta)/2} & \sin(\theta/2)e^{i(\alpha-\beta)/2} \\
-\sin(\theta/2)e^{i(-\alpha+\beta)/2} &
\cos(\theta/2)e^{i(-\alpha-\beta)/2}
\end{array}
\right).
\end{equation}
Consider Fig.~\ref{Solovay:figure:average}.  This was constructed
using 1000 random matrices $U$ of the form
Eq.~\ref{Solovay:eq:arbitrary} with $\alpha,\beta,\theta$
uniformly distributed in $[0,2\pi)$.  Optimal fault-tolerant
approximations $U_{l}$ were constructed of each, with the average
${\rm dist}(U,U_{l})$ plotted for each $l$.  The indicated line of
best fit has the form
\begin{equation}
\label{Solovay:eq:av_scaling} \delta=0.292\times 10^{-0.0511l}.
\end{equation}
This equation characterizes the average number $l$ of Steane code
single-qubit fault-tolerant gates required to obtain a
fault-tolerant approximation $U_{l}$ of an arbitrary single-qubit
gate $U$ to within $\delta={\rm dist}(U,U_{l})$.

An important point to note is that even with unlimited resources
Eq.~(\ref{Solovay:eq:av_scaling}) does not provide a pathway to
construct arbitrarily accurate gates. The accuracy of the
fault-tolerant T-gate described in
Section~\ref{Solovay:section:Tgate} depends critically on the
accuracy of a single physical T-gate
(Eq.~(\ref{Solovay:eq:single_Tgate})). Any over or under rotation
at this point will be directly reflected in the output state of
the logical qubit. Since half the gates in an optimal
fault-tolerant approximation are T-gates, as the number of gates
increases rotation errors will inevitably accumulate.

Consider the over-rotation gate
\begin{equation}
\label{Solovay:eq:over_rotation}
I_{\theta}=\left(
\begin{array}{cc}
  1 & 0 \\
  0 & e^{i\theta} \\
\end{array}
\right),
\end{equation}
where $\theta\ll 1$. For sufficiently small $\theta$, $\delta={\rm
dist}(I,I_{\theta})=\sqrt{3/8}\theta$. Note that any metric on
$U(2)$ modulo global phase must have the property
$\delta\propto\theta$ and hence these results are not expected to
depend on the precise metric used. In the logical T-gate, even if
there is systematic over rotation in the single physical T-gate,
the stochastic nature of Eq.~(\ref{Solovay:eq:premeas}) ensures
that the final logical state will be out by a random angle
$\pm\theta$. This implies that a fault-tolerant approximation
involving $l/2$ T-gates will be uncertain by an amount
$\delta=\sqrt{3l/16}\theta$. The inequality
$\sqrt{3l/16}\theta<0.292\times 10^{-0.0511l}$ therefore sets the
maximum number of gates that can meaningfully be included in a
fault-tolerant approximation.  Note that even if $\theta\sim
10^{-4}$, $l_{\rm max}$ is only 60 -- a number of gates accessible
using the algorithm described in this chapter.

\section{Conclusion}
\label{Solovay:section:conc}

We have described an algorithm enabling the optimal approximation
of arbitrary unitary matrices given a discrete universal gate set.
We have used this algorithm to investigate the properties of
fault-tolerant approximations of arbitrary single-qubit gates
using the gates that can be applied to a single Steane code
logical qubit and found that on average an $l$ gate approximation
can be found within $\delta=0.292\times 10^{-0.0511l}$ of the
ideal gate. We have considered the specific case of the phase
rotation gates used in Shor's algorithm and found that even the
minimal complexity fault-tolerant circuits obtained are still so
large that they are outperformed by non-fault-tolerant
equivalents. The work here suggests that practical quantum
algorithms should avoid, where possible, logical gates that must
be implemented using an approximate sequence of fault-tolerant
gates. An important extension of this work would be to similarly
examine the properties of fault-tolerant approximations of
multiple-qubit gates and larger circuits.

\chapter{Concluding remarks}
\label{remarks}

Neglecting Chapters~\ref{intro}, \ref{canonical} and \ref{Shor}
which contain review material only, in this thesis we
\begin{enumerate}
\item (Chapter~\ref{CNOT}) performed simulations of the adiabatic
Kane $^{31}$P in $^{28}$Si \CNOT\ gate suggesting that achieving a
probability of error less than $10^{-4}$ is possible provided the
presence of the silicon dioxide layer, gate electrodes, and
control circuitry do not reduce the experimentally measured
coherence times of the $^{31}$P donor electron and nucleus, which
were obtained in bulk $^{28}$Si, by more than a factor of 6.

\item (Chapter~\ref{readout}) performed simulations of the
adiabatic Kane $^{31}$P in $^{28}$Si readout operation suggesting
that the fidelity, stability, and accessibility of the states
required to transfer nuclear spin information onto the donor
electrons is insufficient to permit readout. We briefly outlined
an alternative readout scheme based on resonant fields.

\item (Chapter~\ref{5QEC}) presented a simple 5-qubit quantum
error correction (QEC) scheme designed for a linear nearest
neighbor (LNN) architecture, simulating its performance when
subjected to both discrete and continuous errors. Threshold error
rates, at which a QEC scheme provides precisely no reduction in
error, were obtained for both discrete ($p=1.6\times 10^{-3}$) and
continuous ($\sigma=4.7\times 10^{-2}$) errors.

\item (Chapter~\ref{5QECNM}) showed that it is possible to remove
the need for measurement in the QEC scheme of Chapter~\ref{5QEC}
and, with the addition of four qubits, make do with slow
resetting. Discrete and continuous threshold error rates were
reduced to $p=3.7\times 10^{-4}$ and $\sigma=3.1\times 10^{-2}$
respectively.

\item (Chapter~\ref{ShorLNN}) presented a minimal qubit count LNN
circuit implementing the quantum part of Shor's $L$-bit integer
factoring algorithm. Achieved circuit depth and gate count
identical to leading order in $L$ to that possible when long-range
interactions are available.

\item (Chapter~\ref{ShorGates}) showed that Shor's algorithm can
be used to factor integers $O(4^{d})$ bits long, provided
single-qubit phase rotations of angle $\pi/2^{d}$ can be
implemented. Specifically, with sufficient qubits and error
correction, we showed that a 4096 bit integer could conceivably be
factored in a single day provided single-qubit phase rotations of
angle $\pi/128 \pm \pi/512$ could be implemented.

\item (Chapter~\ref{Solovay}) presented a numerical algorithm
capable of obtaining optimal fault-tolerant approximations of
arbitrary single-qubit gates. Used this algorithm to assess the
properties of fault-tolerant approximations of single-qubit phase
rotations with the conclusion that it is better to use simple
non-fault-tolerant circuits to implement phase rotations in Shor's
algorithm.
\end{enumerate}

Significant further work is planned in three broad areas. Firstly,
overcoming or coping with the lack of long-range communication in
the solid-state. For example, teleportation has been proposed as a
possible long-range communication technique \cite{Cops03,Oski03}
but the details of how this would be implemented in practice have
yet to be worked out. A threshold gate error rate $p=2.4\times
10^{-6}$ below which an arbitrarily large quantum computation can
be performed on an LNN architecture has been shown to exist for a
simple discrete error model \cite{Szko04}, but it is highly
desirable that better LNN circuits with a higher threshold are
found, and that an analysis using a more physical model of errors
is performed.

Secondly, many quantum algorithms have now been proposed but very
few have been analyzed for practicality, especially when only a
limited set of quantum gates is available. A similar analysis to
that carried out for Shor's algorithm in this thesis could be
applied to quantum algorithms dealing with Poincar\'e recurrences
and periodic orbits (classical dynamics) \cite{Geor04}, eigenvalue
calculation \cite{Jaks03}, pattern recognition \cite{Schu03},
Schur and Clebsch-Gordon transforms (technically not algorithms in
their own right) \cite{Baco04}, numerical integrals and stochastic
processes \cite{Abra99}, black box function determination
\cite{Du00}, Jones polynomials (knot theory) \cite{Subr02}, a vast
array of problems in the field of quantum system simulation
\cite{Lloy96,Zalk98b}, a somewhat controversial algorithm related
to the classically uncomputable halting problem \cite{Kieu04}, and
a number of promising algorithms based on adiabatic quantum
computation \cite{Farh01,Chil00}. Explicit techniques for the
translation of adiabatic quantum algorithms into quantum circuits
also need to be developed.

Thirdly and finally, while we have developed a quantum compiler
capable of optimally approximating an arbitrary single-qubit gate,
it would be extremely interesting to look at optimal
approximations of larger circuits. Exact quantum compilation has
received a great deal of attention \cite{Tucc04,Shen05,Berg04} but
in the worst-case results in circuits containing a number of gates
that scales exponentially with the number of qubits. By contrast,
the number of gates needed to approximate an arbitrary computation
scales logarithmically with the required accuracy, and even if the
required accuracy increases exponentially with the number of
qubits, this implies a polynomial growth of gate count.

\partfont{\huge}
\appendix                      % Appendices are supposed to go at end, but bib style dominates
\chapter{Simple Steane code gates}
\label{ap1}

This appendix contains the complete multiplication tables of the
largest possible group of fault-tolerant Steane code single
logical qubit gates that can be implemented as products of single
physical qubit rotations.  These gates were introduced in
Chapter~\ref{Solovay}. For convenience, we list these gates again
below.
\begin{equation}
\begin{tabular}{rclcrcl}
$G_{0}$ &=& $I$ &\qquad\qquad\qquad& $G_{12}$ &=& $S^{\dag}X$ \\
$G_{1}$ &=& $H$ &\qquad\qquad\qquad& $G_{13}$ &=& $HS$ \\
$G_{2}$ &=& $X$ &\qquad\qquad\qquad& $G_{14}$ &=& $HS^{\dag}$ \\
$G_{3}$ &=& $Z$ &\qquad\qquad\qquad& $G_{15}$ &=& $ZXH$ \\
$G_{4}$ &=& $S$ &\qquad\qquad\qquad& $G_{16}$ &=& $SXH$ \\
$G_{5}$ &=& $S^{\dag}$ &\qquad\qquad\qquad& $G_{17}$ &=& $S^{\dag}XH$ \\
$G_{6}$ &=& $XH$ &\qquad\qquad\qquad& $G_{18}$ &=& $HSH$ \\
$G_{7}$ &=& $ZH$ &\qquad\qquad\qquad& $G_{19}$ &=& $HS^{\dag}H$ \\
$G_{8}$ &=& $SH$ &\qquad\qquad\qquad& $G_{20}$ &=& $HSX$ \\
$G_{9}$ &=& $S^{\dag}H$ &\qquad\qquad\qquad& $G_{21}$ &=& $HS^{\dag}X$ \\
$G_{10}$ &=& $ZX$ &\qquad\qquad\qquad& $G_{22}$ &=& $S^{\dag}HS$ \\
$G_{11}$ &=& $SX$ &\qquad\qquad\qquad& $G_{23}$ &=& $SHS^{\dag}$ \\
\end{tabular}
\label{ap1:eq:gate_set}
\end{equation}
The tables on the following pages show $G_{i}G_{j}=G_{k}$ with $i$
the vertical index and $j$ the horizontal index.
\begin{table}
  \centering\small
\begin{tabular}{c|c|c|c|c|c|c|c|c|c|c|c|c|}
  \multicolumn{1}{c}{} & \multicolumn{1}{c}{\raisebox{-1.4mm}{$G_{1}$}} & \multicolumn{1}{c}{\raisebox{-1.4mm}{$G_{2}$}} & \multicolumn{1}{c}{\raisebox{-1.4mm}{$G_{3}$}} & \multicolumn{1}{c}{\raisebox{-1.4mm}{$G_{4}$}} & \multicolumn{1}{c}{\raisebox{-1.4mm}{$G_{5}$}} & \multicolumn{1}{c}{\raisebox{-1.4mm}{$G_{6}$}} & \multicolumn{1}{c}{\raisebox{-1.4mm}{$G_{7}$}} & \multicolumn{1}{c}{\raisebox{-1.4mm}{$G_{8}$}} & \multicolumn{1}{c}{\raisebox{-1.4mm}{$G_{9}$}} & \multicolumn{1}{c}{\raisebox{-1.4mm}{$G_{10}$}} & \multicolumn{1}{c}{\raisebox{-1.4mm}{$G_{11}$}} & \multicolumn{1}{c}{\raisebox{-1.4mm}{$G_{12}$}} \\[2mm]
\cline{2-13}
  \raisebox{-1.4mm}{$G_{1}$} & \raisebox{-1.4mm}{$G_{0}$} & \raisebox{-1.4mm}{$G_{7}$} & \raisebox{-1.4mm}{$G_{6}$} & \raisebox{-1.4mm}{$G_{13}$} & \raisebox{-1.4mm}{$G_{14}$} & \raisebox{-1.4mm}{$G_{3}$} & \raisebox{-1.4mm}{$G_{2}$} & \raisebox{-1.4mm}{$G_{18}$} & \raisebox{-1.4mm}{$G_{19}$} & \raisebox{-1.4mm}{$G_{15}$} & \raisebox{-1.4mm}{$G_{20}$} & \raisebox{-1.4mm}{$G_{21}$}\\[2mm]
\cline{2-13}
  \raisebox{-1.4mm}{$G_{2}$} & \raisebox{-1.4mm}{$G_{6}$} & \raisebox{-1.4mm}{$G_{0}$} & \raisebox{-1.4mm}{$G_{10}$} & \raisebox{-1.4mm}{$G_{12}$} & \raisebox{-1.4mm}{$G_{11}$} & \raisebox{-1.4mm}{$G_{1}$} & \raisebox{-1.4mm}{$G_{15}$} & \raisebox{-1.4mm}{$G_{17}$} & \raisebox{-1.4mm}{$G_{16}$} & \raisebox{-1.4mm}{$G_{3}$} & \raisebox{-1.4mm}{$G_{5}$} & \raisebox{-1.4mm}{$G_{4}$}\\[2mm]
\cline{2-13}
  \raisebox{-1.4mm}{$G_{3}$} & \raisebox{-1.4mm}{$G_{7}$} & \raisebox{-1.4mm}{$G_{10}$} & \raisebox{-1.4mm}{$G_{0}$} & \raisebox{-1.4mm}{$G_{5}$} & \raisebox{-1.4mm}{$G_{4}$} & \raisebox{-1.4mm}{$G_{15}$} & \raisebox{-1.4mm}{$G_{1}$} & \raisebox{-1.4mm}{$G_{9}$} & \raisebox{-1.4mm}{$G_{8}$} & \raisebox{-1.4mm}{$G_{2}$} & \raisebox{-1.4mm}{$G_{12}$} & \raisebox{-1.4mm}{$G_{11}$}\\[2mm]
\cline{2-13}
  \raisebox{-1.4mm}{$G_{4}$} & \raisebox{-1.4mm}{$G_{8}$} & \raisebox{-1.4mm}{$G_{11}$} & \raisebox{-1.4mm}{$G_{5}$} & \raisebox{-1.4mm}{$G_{3}$} & \raisebox{-1.4mm}{$G_{0}$} & \raisebox{-1.4mm}{$G_{16}$} & \raisebox{-1.4mm}{$G_{9}$} & \raisebox{-1.4mm}{$G_{7}$} & \raisebox{-1.4mm}{$G_{1}$} & \raisebox{-1.4mm}{$G_{12}$} & \raisebox{-1.4mm}{$G_{10}$} & \raisebox{-1.4mm}{$G_{2}$}\\[2mm]
\cline{2-13}
  \raisebox{-1.4mm}{$G_{5}$} & \raisebox{-1.4mm}{$G_{9}$} & \raisebox{-1.4mm}{$G_{12}$} & \raisebox{-1.4mm}{$G_{4}$} & \raisebox{-1.4mm}{$G_{0}$} & \raisebox{-1.4mm}{$G_{3}$} & \raisebox{-1.4mm}{$G_{17}$} & \raisebox{-1.4mm}{$G_{8}$} & \raisebox{-1.4mm}{$G_{1}$} & \raisebox{-1.4mm}{$G_{7}$} & \raisebox{-1.4mm}{$G_{11}$} & \raisebox{-1.4mm}{$G_{2}$} & \raisebox{-1.4mm}{$G_{10}$}\\[2mm]
\cline{2-13}
  \raisebox{-1.4mm}{$G_{6}$} & \raisebox{-1.4mm}{$G_{2}$} & \raisebox{-1.4mm}{$G_{15}$} & \raisebox{-1.4mm}{$G_{1}$} & \raisebox{-1.4mm}{$G_{14}$} & \raisebox{-1.4mm}{$G_{13}$} & \raisebox{-1.4mm}{$G_{10}$} & \raisebox{-1.4mm}{$G_{0}$} & \raisebox{-1.4mm}{$G_{19}$} & \raisebox{-1.4mm}{$G_{18}$} & \raisebox{-1.4mm}{$G_{7}$} & \raisebox{-1.4mm}{$G_{21}$} & \raisebox{-1.4mm}{$G_{20}$}\\[2mm]
\cline{2-13}
  \raisebox{-1.4mm}{$G_{7}$} & \raisebox{-1.4mm}{$G_{3}$} & \raisebox{-1.4mm}{$G_{1}$} & \raisebox{-1.4mm}{$G_{15}$} & \raisebox{-1.4mm}{$G_{21}$} & \raisebox{-1.4mm}{$G_{20}$} & \raisebox{-1.4mm}{$G_{0}$} & \raisebox{-1.4mm}{$G_{10}$} & \raisebox{-1.4mm}{$G_{23}$} & \raisebox{-1.4mm}{$G_{22}$} & \raisebox{-1.4mm}{$G_{6}$} & \raisebox{-1.4mm}{$G_{14}$} & \raisebox{-1.4mm}{$G_{13}$}\\[2mm]
\cline{2-13}
  \raisebox{-1.4mm}{$G_{8}$} & \raisebox{-1.4mm}{$G_{4}$} & \raisebox{-1.4mm}{$G_{9}$} & \raisebox{-1.4mm}{$G_{16}$} & \raisebox{-1.4mm}{$G_{19}$} & \raisebox{-1.4mm}{$G_{23}$} & \raisebox{-1.4mm}{$G_{5}$} & \raisebox{-1.4mm}{$G_{11}$} & \raisebox{-1.4mm}{$G_{14}$} & \raisebox{-1.4mm}{$G_{21}$} & \raisebox{-1.4mm}{$G_{17}$} & \raisebox{-1.4mm}{$G_{18}$} & \raisebox{-1.4mm}{$G_{22}$}\\[2mm]
\cline{2-13}
  \raisebox{-1.4mm}{$G_{9}$} & \raisebox{-1.4mm}{$G_{5}$} & \raisebox{-1.4mm}{$G_{8}$} & \raisebox{-1.4mm}{$G_{17}$} & \raisebox{-1.4mm}{$G_{22}$} & \raisebox{-1.4mm}{$G_{18}$} & \raisebox{-1.4mm}{$G_{4}$} & \raisebox{-1.4mm}{$G_{12}$} & \raisebox{-1.4mm}{$G_{20}$} & \raisebox{-1.4mm}{$G_{13}$} & \raisebox{-1.4mm}{$G_{16}$} & \raisebox{-1.4mm}{$G_{23}$} & \raisebox{-1.4mm}{$G_{19}$}\\[2mm]
\cline{2-13}
  \raisebox{-1.4mm}{$G_{10}$} & \raisebox{-1.4mm}{$G_{15}$} & \raisebox{-1.4mm}{$G_{3}$} & \raisebox{-1.4mm}{$G_{2}$} & \raisebox{-1.4mm}{$G_{11}$} & \raisebox{-1.4mm}{$G_{12}$} & \raisebox{-1.4mm}{$G_{7}$} & \raisebox{-1.4mm}{$G_{6}$} & \raisebox{-1.4mm}{$G_{16}$} & \raisebox{-1.4mm}{$G_{17}$} & \raisebox{-1.4mm}{$G_{0}$} & \raisebox{-1.4mm}{$G_{4}$} & \raisebox{-1.4mm}{$G_{5}$}\\[2mm]
\cline{2-13}
  \raisebox{-1.4mm}{$G_{11}$} & \raisebox{-1.4mm}{$G_{16}$} & \raisebox{-1.4mm}{$G_{4}$} & \raisebox{-1.4mm}{$G_{12}$} & \raisebox{-1.4mm}{$G_{2}$} & \raisebox{-1.4mm}{$G_{10}$} & \raisebox{-1.4mm}{$G_{8}$} & \raisebox{-1.4mm}{$G_{17}$} & \raisebox{-1.4mm}{$G_{6}$} & \raisebox{-1.4mm}{$G_{15}$} & \raisebox{-1.4mm}{$G_{5}$} & \raisebox{-1.4mm}{$G_{0}$} & \raisebox{-1.4mm}{$G_{3}$}\\[2mm]
\cline{2-13}
  \raisebox{-1.4mm}{$G_{12}$} & \raisebox{-1.4mm}{$G_{17}$} & \raisebox{-1.4mm}{$G_{5}$} & \raisebox{-1.4mm}{$G_{11}$} & \raisebox{-1.4mm}{$G_{10}$} & \raisebox{-1.4mm}{$G_{2}$} & \raisebox{-1.4mm}{$G_{9}$} & \raisebox{-1.4mm}{$G_{16}$} & \raisebox{-1.4mm}{$G_{15}$} & \raisebox{-1.4mm}{$G_{6}$} & \raisebox{-1.4mm}{$G_{4}$} & \raisebox{-1.4mm}{$G_{3}$} & \raisebox{-1.4mm}{$G_{0}$}\\[2mm]
\cline{2-13}
\end{tabular}
\end{table}

\begin{table}
  \centering\small
\begin{tabular}{c|c|c|c|c|c|c|c|c|c|c|c|c|}
  \multicolumn{1}{c}{} & \multicolumn{1}{c}{\raisebox{-1.4mm}{$G_{13}$}} & \multicolumn{1}{c}{\raisebox{-1.4mm}{$G_{14}$}} & \multicolumn{1}{c}{\raisebox{-1.4mm}{$G_{15}$}} & \multicolumn{1}{c}{\raisebox{-1.4mm}{$G_{16}$}} & \multicolumn{1}{c}{\raisebox{-1.4mm}{$G_{17}$}} & \multicolumn{1}{c}{\raisebox{-1.4mm}{$G_{18}$}} & \multicolumn{1}{c}{\raisebox{-1.4mm}{$G_{19}$}} & \multicolumn{1}{c}{\raisebox{-1.4mm}{$G_{20}$}} & \multicolumn{1}{c}{\raisebox{-1.4mm}{$G_{21}$}} & \multicolumn{1}{c}{\raisebox{-1.4mm}{$G_{22}$}} & \multicolumn{1}{c}{\raisebox{-1.4mm}{$G_{23}$}} \\[2mm]
\cline{2-12}
  \raisebox{-1.4mm}{$G_{1}$} & \raisebox{-1.4mm}{$G_{4}$} & \raisebox{-1.4mm}{$G_{5}$} & \raisebox{-1.4mm}{$G_{10}$} & \raisebox{-1.4mm}{$G_{22}$} & \raisebox{-1.4mm}{$G_{23}$} & \raisebox{-1.4mm}{$G_{8}$} & \raisebox{-1.4mm}{$G_{9}$} & \raisebox{-1.4mm}{$G_{11}$} & \raisebox{-1.4mm}{$G_{12}$} & \raisebox{-1.4mm}{$G_{16}$} & \raisebox{-1.4mm}{$G_{17}$}\\[2mm]
\cline{2-12}
  \raisebox{-1.4mm}{$G_{2}$} & \raisebox{-1.4mm}{$G_{14}$} & \raisebox{-1.4mm}{$G_{13}$} & \raisebox{-1.4mm}{$G_{7}$} & \raisebox{-1.4mm}{$G_{9}$} & \raisebox{-1.4mm}{$G_{8}$} & \raisebox{-1.4mm}{$G_{19}$} & \raisebox{-1.4mm}{$G_{18}$} & \raisebox{-1.4mm}{$G_{21}$} & \raisebox{-1.4mm}{$G_{20}$} & \raisebox{-1.4mm}{$G_{23}$} & \raisebox{-1.4mm}{$G_{22}$}\\[2mm]
\cline{2-12}
  \raisebox{-1.4mm}{$G_{3}$} & \raisebox{-1.4mm}{$G_{21}$} & \raisebox{-1.4mm}{$G_{20}$} & \raisebox{-1.4mm}{$G_{6}$} & \raisebox{-1.4mm}{$G_{17}$} & \raisebox{-1.4mm}{$G_{16}$} & \raisebox{-1.4mm}{$G_{23}$} & \raisebox{-1.4mm}{$G_{22}$} & \raisebox{-1.4mm}{$G_{14}$} & \raisebox{-1.4mm}{$G_{13}$} & \raisebox{-1.4mm}{$G_{19}$} & \raisebox{-1.4mm}{$G_{18}$}\\[2mm]
\cline{2-12}
  \raisebox{-1.4mm}{$G_{4}$} & \raisebox{-1.4mm}{$G_{19}$} & \raisebox{-1.4mm}{$G_{23}$} & \raisebox{-1.4mm}{$G_{17}$} & \raisebox{-1.4mm}{$G_{15}$} & \raisebox{-1.4mm}{$G_{6}$} & \raisebox{-1.4mm}{$G_{14}$} & \raisebox{-1.4mm}{$G_{21}$} & \raisebox{-1.4mm}{$G_{18}$} & \raisebox{-1.4mm}{$G_{22}$} & \raisebox{-1.4mm}{$G_{13}$} & \raisebox{-1.4mm}{$G_{20}$}\\[2mm]
\cline{2-12}
  \raisebox{-1.4mm}{$G_{5}$} & \raisebox{-1.4mm}{$G_{22}$} & \raisebox{-1.4mm}{$G_{18}$} & \raisebox{-1.4mm}{$G_{16}$} & \raisebox{-1.4mm}{$G_{6}$} & \raisebox{-1.4mm}{$G_{15}$} & \raisebox{-1.4mm}{$G_{20}$} & \raisebox{-1.4mm}{$G_{13}$} & \raisebox{-1.4mm}{$G_{23}$} & \raisebox{-1.4mm}{$G_{19}$} & \raisebox{-1.4mm}{$G_{21}$} & \raisebox{-1.4mm}{$G_{14}$}\\[2mm]
\cline{2-12}
  \raisebox{-1.4mm}{$G_{6}$} & \raisebox{-1.4mm}{$G_{12}$} & \raisebox{-1.4mm}{$G_{11}$} & \raisebox{-1.4mm}{$G_{3}$} & \raisebox{-1.4mm}{$G_{23}$} & \raisebox{-1.4mm}{$G_{22}$} & \raisebox{-1.4mm}{$G_{17}$} & \raisebox{-1.4mm}{$G_{16}$} & \raisebox{-1.4mm}{$G_{5}$} & \raisebox{-1.4mm}{$G_{4}$} & \raisebox{-1.4mm}{$G_{9}$} & \raisebox{-1.4mm}{$G_{8}$}\\[2mm]
\cline{2-12}
  \raisebox{-1.4mm}{$G_{7}$} & \raisebox{-1.4mm}{$G_{5}$} & \raisebox{-1.4mm}{$G_{4}$} & \raisebox{-1.4mm}{$G_{2}$} & \raisebox{-1.4mm}{$G_{19}$} & \raisebox{-1.4mm}{$G_{18}$} & \raisebox{-1.4mm}{$G_{9}$} & \raisebox{-1.4mm}{$G_{8}$} & \raisebox{-1.4mm}{$G_{12}$} & \raisebox{-1.4mm}{$G_{11}$} & \raisebox{-1.4mm}{$G_{17}$} & \raisebox{-1.4mm}{$G_{16}$}\\[2mm]
\cline{2-12}
  \raisebox{-1.4mm}{$G_{8}$} & \raisebox{-1.4mm}{$G_{3}$} & \raisebox{-1.4mm}{$G_{0}$} & \raisebox{-1.4mm}{$G_{12}$} & \raisebox{-1.4mm}{$G_{13}$} & \raisebox{-1.4mm}{$G_{20}$} & \raisebox{-1.4mm}{$G_{7}$} & \raisebox{-1.4mm}{$G_{1}$} & \raisebox{-1.4mm}{$G_{10}$} & \raisebox{-1.4mm}{$G_{2}$} & \raisebox{-1.4mm}{$G_{15}$} & \raisebox{-1.4mm}{$G_{6}$}\\[2mm]
\cline{2-12}
  \raisebox{-1.4mm}{$G_{9}$} & \raisebox{-1.4mm}{$G_{0}$} & \raisebox{-1.4mm}{$G_{3}$} & \raisebox{-1.4mm}{$G_{11}$} & \raisebox{-1.4mm}{$G_{21}$} & \raisebox{-1.4mm}{$G_{14}$} & \raisebox{-1.4mm}{$G_{1}$} & \raisebox{-1.4mm}{$G_{7}$} & \raisebox{-1.4mm}{$G_{2}$} & \raisebox{-1.4mm}{$G_{10}$} & \raisebox{-1.4mm}{$G_{6}$} & \raisebox{-1.4mm}{$G_{15}$}\\[2mm]
\cline{2-12}
  \raisebox{-1.4mm}{$G_{10}$} & \raisebox{-1.4mm}{$G_{20}$} & \raisebox{-1.4mm}{$G_{21}$} & \raisebox{-1.4mm}{$G_{1}$} & \raisebox{-1.4mm}{$G_{8}$} & \raisebox{-1.4mm}{$G_{9}$} & \raisebox{-1.4mm}{$G_{22}$} & \raisebox{-1.4mm}{$G_{23}$} & \raisebox{-1.4mm}{$G_{13}$} & \raisebox{-1.4mm}{$G_{14}$} & \raisebox{-1.4mm}{$G_{18}$} & \raisebox{-1.4mm}{$G_{19}$}\\[2mm]
\cline{2-12}
  \raisebox{-1.4mm}{$G_{11}$} & \raisebox{-1.4mm}{$G_{23}$} & \raisebox{-1.4mm}{$G_{19}$} & \raisebox{-1.4mm}{$G_{9}$} & \raisebox{-1.4mm}{$G_{1}$} & \raisebox{-1.4mm}{$G_{7}$} & \raisebox{-1.4mm}{$G_{21}$} & \raisebox{-1.4mm}{$G_{14}$} & \raisebox{-1.4mm}{$G_{22}$} & \raisebox{-1.4mm}{$G_{18}$} & \raisebox{-1.4mm}{$G_{20}$} & \raisebox{-1.4mm}{$G_{13}$}\\[2mm]
\cline{2-12}
  \raisebox{-1.4mm}{$G_{12}$} & \raisebox{-1.4mm}{$G_{18}$} & \raisebox{-1.4mm}{$G_{22}$} & \raisebox{-1.4mm}{$G_{8}$} & \raisebox{-1.4mm}{$G_{7}$} & \raisebox{-1.4mm}{$G_{1}$} & \raisebox{-1.4mm}{$G_{13}$} & \raisebox{-1.4mm}{$G_{20}$} & \raisebox{-1.4mm}{$G_{19}$} & \raisebox{-1.4mm}{$G_{23}$} & \raisebox{-1.4mm}{$G_{14}$} & \raisebox{-1.4mm}{$G_{21}$}\\[2mm]
\cline{2-12}
\end{tabular}
\end{table}

\begin{table}
  \centering\small
\begin{tabular}{c|c|c|c|c|c|c|c|c|c|c|c|c|}
  \multicolumn{1}{c}{} & \multicolumn{1}{c}{\raisebox{-1.4mm}{$G_{1}$}} & \multicolumn{1}{c}{\raisebox{-1.4mm}{$G_{2}$}} & \multicolumn{1}{c}{\raisebox{-1.4mm}{$G_{3}$}} & \multicolumn{1}{c}{\raisebox{-1.4mm}{$G_{4}$}} & \multicolumn{1}{c}{\raisebox{-1.4mm}{$G_{5}$}} & \multicolumn{1}{c}{\raisebox{-1.4mm}{$G_{6}$}} & \multicolumn{1}{c}{\raisebox{-1.4mm}{$G_{7}$}} & \multicolumn{1}{c}{\raisebox{-1.4mm}{$G_{8}$}} & \multicolumn{1}{c}{\raisebox{-1.4mm}{$G_{9}$}} & \multicolumn{1}{c}{\raisebox{-1.4mm}{$G_{10}$}} & \multicolumn{1}{c}{\raisebox{-1.4mm}{$G_{11}$}} & \multicolumn{1}{c}{\raisebox{-1.4mm}{$G_{12}$}} \\[2mm]
\cline{2-13}
  \raisebox{-1.4mm}{$G_{13}$} & \raisebox{-1.4mm}{$G_{18}$} & \raisebox{-1.4mm}{$G_{20}$} & \raisebox{-1.4mm}{$G_{14}$} & \raisebox{-1.4mm}{$G_{6}$} & \raisebox{-1.4mm}{$G_{1}$} & \raisebox{-1.4mm}{$G_{22}$} & \raisebox{-1.4mm}{$G_{19}$} & \raisebox{-1.4mm}{$G_{2}$} & \raisebox{-1.4mm}{$G_{0}$} & \raisebox{-1.4mm}{$G_{21}$} & \raisebox{-1.4mm}{$G_{15}$} & \raisebox{-1.4mm}{$G_{7}$}\\[2mm]
\cline{2-13}
  \raisebox{-1.4mm}{$G_{14}$} & \raisebox{-1.4mm}{$G_{19}$} & \raisebox{-1.4mm}{$G_{21}$} & \raisebox{-1.4mm}{$G_{13}$} & \raisebox{-1.4mm}{$G_{1}$} & \raisebox{-1.4mm}{$G_{6}$} & \raisebox{-1.4mm}{$G_{23}$} & \raisebox{-1.4mm}{$G_{18}$} & \raisebox{-1.4mm}{$G_{0}$} & \raisebox{-1.4mm}{$G_{2}$} & \raisebox{-1.4mm}{$G_{20}$} & \raisebox{-1.4mm}{$G_{7}$} & \raisebox{-1.4mm}{$G_{15}$}\\[2mm]
\cline{2-13}
  \raisebox{-1.4mm}{$G_{15}$} & \raisebox{-1.4mm}{$G_{10}$} & \raisebox{-1.4mm}{$G_{6}$} & \raisebox{-1.4mm}{$G_{7}$} & \raisebox{-1.4mm}{$G_{20}$} & \raisebox{-1.4mm}{$G_{21}$} & \raisebox{-1.4mm}{$G_{2}$} & \raisebox{-1.4mm}{$G_{3}$} & \raisebox{-1.4mm}{$G_{22}$} & \raisebox{-1.4mm}{$G_{23}$} & \raisebox{-1.4mm}{$G_{1}$} & \raisebox{-1.4mm}{$G_{13}$} & \raisebox{-1.4mm}{$G_{14}$}\\[2mm]
\cline{2-13}
  \raisebox{-1.4mm}{$G_{16}$} & \raisebox{-1.4mm}{$G_{11}$} & \raisebox{-1.4mm}{$G_{17}$} & \raisebox{-1.4mm}{$G_{8}$} & \raisebox{-1.4mm}{$G_{23}$} & \raisebox{-1.4mm}{$G_{19}$} & \raisebox{-1.4mm}{$G_{12}$} & \raisebox{-1.4mm}{$G_{4}$} & \raisebox{-1.4mm}{$G_{21}$} & \raisebox{-1.4mm}{$G_{14}$} & \raisebox{-1.4mm}{$G_{9}$} & \raisebox{-1.4mm}{$G_{22}$} & \raisebox{-1.4mm}{$G_{18}$}\\[2mm]
\cline{2-13}
  \raisebox{-1.4mm}{$G_{17}$} & \raisebox{-1.4mm}{$G_{12}$} & \raisebox{-1.4mm}{$G_{16}$} & \raisebox{-1.4mm}{$G_{9}$} & \raisebox{-1.4mm}{$G_{18}$} & \raisebox{-1.4mm}{$G_{22}$} & \raisebox{-1.4mm}{$G_{11}$} & \raisebox{-1.4mm}{$G_{5}$} & \raisebox{-1.4mm}{$G_{13}$} & \raisebox{-1.4mm}{$G_{20}$} & \raisebox{-1.4mm}{$G_{8}$} & \raisebox{-1.4mm}{$G_{19}$} & \raisebox{-1.4mm}{$G_{23}$}\\[2mm]
\cline{2-13}
  \raisebox{-1.4mm}{$G_{18}$} & \raisebox{-1.4mm}{$G_{13}$} & \raisebox{-1.4mm}{$G_{19}$} & \raisebox{-1.4mm}{$G_{22}$} & \raisebox{-1.4mm}{$G_{9}$} & \raisebox{-1.4mm}{$G_{17}$} & \raisebox{-1.4mm}{$G_{14}$} & \raisebox{-1.4mm}{$G_{20}$} & \raisebox{-1.4mm}{$G_{5}$} & \raisebox{-1.4mm}{$G_{12}$} & \raisebox{-1.4mm}{$G_{23}$} & \raisebox{-1.4mm}{$G_{8}$} & \raisebox{-1.4mm}{$G_{16}$}\\[2mm]
\cline{2-13}
  \raisebox{-1.4mm}{$G_{19}$} & \raisebox{-1.4mm}{$G_{14}$} & \raisebox{-1.4mm}{$G_{18}$} & \raisebox{-1.4mm}{$G_{23}$} & \raisebox{-1.4mm}{$G_{16}$} & \raisebox{-1.4mm}{$G_{8}$} & \raisebox{-1.4mm}{$G_{13}$} & \raisebox{-1.4mm}{$G_{21}$} & \raisebox{-1.4mm}{$G_{11}$} & \raisebox{-1.4mm}{$G_{4}$} & \raisebox{-1.4mm}{$G_{22}$} & \raisebox{-1.4mm}{$G_{17}$} & \raisebox{-1.4mm}{$G_{9}$}\\[2mm]
\cline{2-13}
  \raisebox{-1.4mm}{$G_{20}$} & \raisebox{-1.4mm}{$G_{22}$} & \raisebox{-1.4mm}{$G_{13}$} & \raisebox{-1.4mm}{$G_{21}$} & \raisebox{-1.4mm}{$G_{7}$} & \raisebox{-1.4mm}{$G_{15}$} & \raisebox{-1.4mm}{$G_{18}$} & \raisebox{-1.4mm}{$G_{23}$} & \raisebox{-1.4mm}{$G_{3}$} & \raisebox{-1.4mm}{$G_{10}$} & \raisebox{-1.4mm}{$G_{14}$} & \raisebox{-1.4mm}{$G_{1}$} & \raisebox{-1.4mm}{$G_{6}$}\\[2mm]
\cline{2-13}
  \raisebox{-1.4mm}{$G_{21}$} & \raisebox{-1.4mm}{$G_{23}$} & \raisebox{-1.4mm}{$G_{14}$} & \raisebox{-1.4mm}{$G_{20}$} & \raisebox{-1.4mm}{$G_{15}$} & \raisebox{-1.4mm}{$G_{7}$} & \raisebox{-1.4mm}{$G_{19}$} & \raisebox{-1.4mm}{$G_{22}$} & \raisebox{-1.4mm}{$G_{10}$} & \raisebox{-1.4mm}{$G_{3}$} & \raisebox{-1.4mm}{$G_{13}$} & \raisebox{-1.4mm}{$G_{6}$} & \raisebox{-1.4mm}{$G_{1}$}\\[2mm]
\cline{2-13}
  \raisebox{-1.4mm}{$G_{22}$} & \raisebox{-1.4mm}{$G_{20}$} & \raisebox{-1.4mm}{$G_{23}$} & \raisebox{-1.4mm}{$G_{18}$} & \raisebox{-1.4mm}{$G_{17}$} & \raisebox{-1.4mm}{$G_{9}$} & \raisebox{-1.4mm}{$G_{21}$} & \raisebox{-1.4mm}{$G_{13}$} & \raisebox{-1.4mm}{$G_{12}$} & \raisebox{-1.4mm}{$G_{5}$} & \raisebox{-1.4mm}{$G_{19}$} & \raisebox{-1.4mm}{$G_{16}$} & \raisebox{-1.4mm}{$G_{8}$}\\[2mm]
\cline{2-13}
  \raisebox{-1.4mm}{$G_{23}$} & \raisebox{-1.4mm}{$G_{21}$} & \raisebox{-1.4mm}{$G_{22}$} & \raisebox{-1.4mm}{$G_{19}$} & \raisebox{-1.4mm}{$G_{8}$} & \raisebox{-1.4mm}{$G_{16}$} & \raisebox{-1.4mm}{$G_{20}$} & \raisebox{-1.4mm}{$G_{14}$} & \raisebox{-1.4mm}{$G_{4}$} & \raisebox{-1.4mm}{$G_{11}$} & \raisebox{-1.4mm}{$G_{18}$} & \raisebox{-1.4mm}{$G_{9}$} & \raisebox{-1.4mm}{$G_{17}$}\\[2mm]
\cline{2-13}
\end{tabular}
\end{table}

\begin{table}
  \centering\small
\begin{tabular}{c|c|c|c|c|c|c|c|c|c|c|c|c|}
  \multicolumn{1}{c}{} & \multicolumn{1}{c}{\raisebox{-1.4mm}{$G_{13}$}} & \multicolumn{1}{c}{\raisebox{-1.4mm}{$G_{14}$}} & \multicolumn{1}{c}{\raisebox{-1.4mm}{$G_{15}$}} & \multicolumn{1}{c}{\raisebox{-1.4mm}{$G_{16}$}} & \multicolumn{1}{c}{\raisebox{-1.4mm}{$G_{17}$}} & \multicolumn{1}{c}{\raisebox{-1.4mm}{$G_{18}$}} & \multicolumn{1}{c}{\raisebox{-1.4mm}{$G_{19}$}} & \multicolumn{1}{c}{\raisebox{-1.4mm}{$G_{20}$}} & \multicolumn{1}{c}{\raisebox{-1.4mm}{$G_{21}$}} & \multicolumn{1}{c}{\raisebox{-1.4mm}{$G_{22}$}} & \multicolumn{1}{c}{\raisebox{-1.4mm}{$G_{23}$}} \\[2mm]
\cline{2-12}
  \raisebox{-1.4mm}{$G_{13}$} & \raisebox{-1.4mm}{$G_{9}$} & \raisebox{-1.4mm}{$G_{17}$} & \raisebox{-1.4mm}{$G_{23}$} & \raisebox{-1.4mm}{$G_{10}$} & \raisebox{-1.4mm}{$G_{3}$} & \raisebox{-1.4mm}{$G_{5}$} & \raisebox{-1.4mm}{$G_{12}$} & \raisebox{-1.4mm}{$G_{8}$} & \raisebox{-1.4mm}{$G_{16}$} & \raisebox{-1.4mm}{$G_{4}$} & \raisebox{-1.4mm}{$G_{11}$}\\[2mm]
\cline{2-12}
  \raisebox{-1.4mm}{$G_{14}$} & \raisebox{-1.4mm}{$G_{16}$} & \raisebox{-1.4mm}{$G_{8}$} & \raisebox{-1.4mm}{$G_{22}$} & \raisebox{-1.4mm}{$G_{3}$} & \raisebox{-1.4mm}{$G_{10}$} & \raisebox{-1.4mm}{$G_{11}$} & \raisebox{-1.4mm}{$G_{4}$} & \raisebox{-1.4mm}{$G_{17}$} & \raisebox{-1.4mm}{$G_{9}$} & \raisebox{-1.4mm}{$G_{12}$} & \raisebox{-1.4mm}{$G_{5}$}\\[2mm]
\cline{2-12}
  \raisebox{-1.4mm}{$G_{15}$} & \raisebox{-1.4mm}{$G_{11}$} & \raisebox{-1.4mm}{$G_{12}$} & \raisebox{-1.4mm}{$G_{0}$} & \raisebox{-1.4mm}{$G_{18}$} & \raisebox{-1.4mm}{$G_{19}$} & \raisebox{-1.4mm}{$G_{16}$} & \raisebox{-1.4mm}{$G_{17}$} & \raisebox{-1.4mm}{$G_{4}$} & \raisebox{-1.4mm}{$G_{5}$} & \raisebox{-1.4mm}{$G_{8}$} & \raisebox{-1.4mm}{$G_{9}$}\\[2mm]
\cline{2-12}
  \raisebox{-1.4mm}{$G_{16}$} & \raisebox{-1.4mm}{$G_{2}$} & \raisebox{-1.4mm}{$G_{10}$} & \raisebox{-1.4mm}{$G_{5}$} & \raisebox{-1.4mm}{$G_{20}$} & \raisebox{-1.4mm}{$G_{13}$} & \raisebox{-1.4mm}{$G_{6}$} & \raisebox{-1.4mm}{$G_{15}$} & \raisebox{-1.4mm}{$G_{0}$} & \raisebox{-1.4mm}{$G_{3}$} & \raisebox{-1.4mm}{$G_{1}$} & \raisebox{-1.4mm}{$G_{7}$}\\[2mm]
\cline{2-12}
  \raisebox{-1.4mm}{$G_{17}$} & \raisebox{-1.4mm}{$G_{10}$} & \raisebox{-1.4mm}{$G_{2}$} & \raisebox{-1.4mm}{$G_{4}$} & \raisebox{-1.4mm}{$G_{14}$} & \raisebox{-1.4mm}{$G_{21}$} & \raisebox{-1.4mm}{$G_{15}$} & \raisebox{-1.4mm}{$G_{6}$} & \raisebox{-1.4mm}{$G_{3}$} & \raisebox{-1.4mm}{$G_{0}$} & \raisebox{-1.4mm}{$G_{7}$} & \raisebox{-1.4mm}{$G_{1}$}\\[2mm]
\cline{2-12}
  \raisebox{-1.4mm}{$G_{18}$} & \raisebox{-1.4mm}{$G_{6}$} & \raisebox{-1.4mm}{$G_{1}$} & \raisebox{-1.4mm}{$G_{21}$} & \raisebox{-1.4mm}{$G_{4}$} & \raisebox{-1.4mm}{$G_{11}$} & \raisebox{-1.4mm}{$G_{2}$} & \raisebox{-1.4mm}{$G_{0}$} & \raisebox{-1.4mm}{$G_{15}$} & \raisebox{-1.4mm}{$G_{7}$} & \raisebox{-1.4mm}{$G_{10}$} & \raisebox{-1.4mm}{$G_{3}$}\\[2mm]
\cline{2-12}
  \raisebox{-1.4mm}{$G_{19}$} & \raisebox{-1.4mm}{$G_{1}$} & \raisebox{-1.4mm}{$G_{6}$} & \raisebox{-1.4mm}{$G_{20}$} & \raisebox{-1.4mm}{$G_{12}$} & \raisebox{-1.4mm}{$G_{5}$} & \raisebox{-1.4mm}{$G_{0}$} & \raisebox{-1.4mm}{$G_{2}$} & \raisebox{-1.4mm}{$G_{7}$} & \raisebox{-1.4mm}{$G_{15}$} & \raisebox{-1.4mm}{$G_{3}$} & \raisebox{-1.4mm}{$G_{10}$}\\[2mm]
\cline{2-12}
  \raisebox{-1.4mm}{$G_{20}$} & \raisebox{-1.4mm}{$G_{17}$} & \raisebox{-1.4mm}{$G_{9}$} & \raisebox{-1.4mm}{$G_{19}$} & \raisebox{-1.4mm}{$G_{0}$} & \raisebox{-1.4mm}{$G_{2}$} & \raisebox{-1.4mm}{$G_{12}$} & \raisebox{-1.4mm}{$G_{5}$} & \raisebox{-1.4mm}{$G_{16}$} & \raisebox{-1.4mm}{$G_{8}$} & \raisebox{-1.4mm}{$G_{11}$} & \raisebox{-1.4mm}{$G_{4}$}\\[2mm]
\cline{2-12}
  \raisebox{-1.4mm}{$G_{21}$} & \raisebox{-1.4mm}{$G_{8}$} & \raisebox{-1.4mm}{$G_{16}$} & \raisebox{-1.4mm}{$G_{18}$} & \raisebox{-1.4mm}{$G_{2}$} & \raisebox{-1.4mm}{$G_{0}$} & \raisebox{-1.4mm}{$G_{4}$} & \raisebox{-1.4mm}{$G_{11}$} & \raisebox{-1.4mm}{$G_{9}$} & \raisebox{-1.4mm}{$G_{17}$} & \raisebox{-1.4mm}{$G_{5}$} & \raisebox{-1.4mm}{$G_{12}$}\\[2mm]
\cline{2-12}
  \raisebox{-1.4mm}{$G_{22}$} & \raisebox{-1.4mm}{$G_{7}$} & \raisebox{-1.4mm}{$G_{15}$} & \raisebox{-1.4mm}{$G_{14}$} & \raisebox{-1.4mm}{$G_{11}$} & \raisebox{-1.4mm}{$G_{4}$} & \raisebox{-1.4mm}{$G_{3}$} & \raisebox{-1.4mm}{$G_{10}$} & \raisebox{-1.4mm}{$G_{1}$} & \raisebox{-1.4mm}{$G_{6}$} & \raisebox{-1.4mm}{$G_{0}$} & \raisebox{-1.4mm}{$G_{2}$}\\[2mm]
\cline{2-12}
  \raisebox{-1.4mm}{$G_{23}$} & \raisebox{-1.4mm}{$G_{15}$} & \raisebox{-1.4mm}{$G_{7}$} & \raisebox{-1.4mm}{$G_{13}$} & \raisebox{-1.4mm}{$G_{5}$} & \raisebox{-1.4mm}{$G_{12}$} & \raisebox{-1.4mm}{$G_{10}$} & \raisebox{-1.4mm}{$G_{3}$} & \raisebox{-1.4mm}{$G_{6}$} & \raisebox{-1.4mm}{$G_{1}$} & \raisebox{-1.4mm}{$G_{2}$} & \raisebox{-1.4mm}{$G_{0}$}\\[2mm]
\cline{2-12}
\end{tabular}
\end{table}

\bibliographystyle{unsrt}
\bibliography{References}

\end{document}